\documentclass[12pt,oneside]{extbook}
\usepackage{arxiv}

\usepackage[utf8]{inputenc}
\usepackage[hidelinks]{hyperref}       
\usepackage{url}            
\usepackage{booktabs}       
\usepackage{amsfonts}       
\usepackage{nicefrac}       
\usepackage{microtype}      
\usepackage{graphicx}
\usepackage[numbers]{natbib}
\usepackage{doi}
\usepackage{subfiles}
\usepackage[acronym]{glossaries}
\usepackage{subcaption}

\title{The security of the Coordicide: the implementation and analysis of possible attack vectors}

\date{March, 2022}	

\author{\href{https://orcid.org/0000-0002-2303-5607}{\includegraphics[scale=0.06]{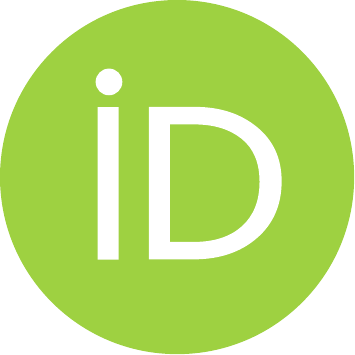}\hspace{1mm}Daria Dziubałtowska}\\
    The Wrocław School of Information Technology, Wroclaw, Poland\\
	IOTA Foundation, Berlin, Germany\\
	\texttt{daria.dziubaltowska@iota.org} \\
}

\renewcommand{\headeright}{Master Thesis}
\renewcommand{\undertitle}{Master Thesis}
\renewcommand{\shorttitle}{}

\hypersetup{
pdftitle={The Security of the Coordicide},
pdfsubject={},
pdfauthor={Daria Dziubałtow},
pdfkeywords={DLT, Iota, Tangle, orphanage, attack vectors},
}

\makeglossaries

\begin{document}

\maketitle

{\normalsize\itshape A thesis submitted in partial fulfilment of the requirements for the degree of Master in Computer Science}
\vspace{0.4cm}

{\normalsize   Developed under supervision of \hfill and industrial supervision of\\}
\vspace{0.4cm}
{\normalsize \itshape  Paweł Morawiec, PhD \hfill Andreas Penzkofer, PhD}

\vspace{2cm}

\begin{abstract}
The goal of the thesis is to study and perform an analysis of the possible attack vectors on the Iota network 2.0 version of the protocol. 
In this work existing attack vectors on Distributed Ledger Technologies are studied and their applicability to the Iota 2.0 protocol discussed. A specific attack that targets the capability of honest participants to write to the ledger is presented and analysed in a network of nodes that run a full node software version.
\end{abstract}

\vfill
\frontmatter
\mainmatter
\tableofcontents

\newacronym{mcmc}{MCMC}{Markov Chain Monte Carlo}
\newacronym{tsa}{TSA}{Tip Selection Algorithm}
\newacronym{dlt}{DLT}{Distributed Ledger Technology}
\newacronym{urts}{URTS}{Uniform Random Tip Selection}
\newacronym{rurts}{R-URTS}{Restricted Uniform Random Tip Selection}
\newacronym{otv}{OTV}{On Tangle Voting}
\newacronym{fpc}{FPC}{Fast Probabilistic Consensus}
\newacronym{fpcs}{FPCS}{Fast Probabilistic Consensus on a Set}
\newacronym{flp}{FLP}{The impossibility theory}
\newacronym{cap}{CAP}{CAP theorem - consistency, availability, partition tolerance}
\newacronym{dos}{DOS}{Denial of Service}
\newacronym{aw}{AW}{Approval Weight}
\newacronym{pow}{PoW}{Proof of Work}
\newacronym{pos}{PoS}{Proof of Stake}
\newacronym{dos}{DoS}{Denial-of-service}
\newacronym{ddos}{DDoS}{Distributed-denial-of-service}
\newacronym{podc}{PODC}{Conference on Principles of Distributed Computing}
\newacronym{gst}{GST}{Global Stabilization Time}
\newacronym{mps}{mps}{message per second}
\newacronym{pc}{PC}{Parasite Chain}
\newacronym{poh}{PoH}{Proof of History}
\newacronym{dpos}{DPoS}{Delegated Proof of Stake}
\newacronym{api}{API}{Application Programming Interface}
\newacronym{bft}{BFT}{Byzantine Fault Tolerance}
\newacronym{pbft}{pBFT}{Practical Byzantine Fault Tolerance}
\newacronym{dag}{DAG}{Directed Acyclic Graph}
\newacronym{dapp}{DApp}{decentralized application}
\newacronym{nft}{NFT}{non-fungible token}
\newacronym{defi}{DeFi}{Decentralized Finance}
\newacronym{fba}{FBA}{Federated Byzantine agreement}
\printglossaries

{\backmatter \chapter{Introduction}}

The main goal of the thesis is to reason about the security of the IOTA protocol with an emphasis on the orphanage problem. The main focus will be put on the newest prototype version of the node software, commonly referred to as the  "Coordicide" solution, which is still in the research phase. 

The topic of distributed systems is one of the oldest branches of computer science. Many famous insights and proofs introduced in the early days of computers are still being used on a wide scale. Many families of protocols designed to ensure an agreement among many distinct processes or network participants have their beginning in the previous century. A brilliant example is the Byzantine Failure protocols which constitute a large part of many currently developed blockchain projects. 
With the emergence of blockchain technology, distributed systems have experienced a revitalization in interest. Nakamoto's consensus introduced in 2007 awakened scientists and engineers to look for new ways to redesign the web again, to create a space in which system users have full control over their data. All of this is happening with the usage of cutting-edge technology that feeds from many fields of study, such as cryptography, distributed systems, finance, and many others. The success of early cryptocurrency projects has brought blockchain technology to the attention of many investors, developers, and innovators and has enabled an incredible amount of development and research in this field. We are in the moment where blockchain is no longer only an investment, hype, or buzzword, as there are thousands of projects, platforms, and applications currently developed that are focusing on security and decentralization. With increased usage it becomes harder to follow all the changes, to validate new ideas, thus many new design ideas are subject to activities performed by dishonest parties, trying to make a profit by exploiting network vulnerabilities and finding gaps in the protocols' design.
All of this contributes to the importance of carrying out the research and testing before a system is brought into production.

One such innovative project is a cryptocurrency called IOTA, which has a quite distinctive approach compared to other DLTs.
Firstly, the IOTA protocol is being designed with specific principles in mind to address non-trivial and novel challenges, such as solving scaling issues of blockchain and creating a feeless \Gls{dlt}. All of this resulted in a unique design compared to other projects. As a consequence, a completely new range of studies and research questions need to be solved.
Secondly, IOTA is still under development and its current official production version of the network, called \textit{mainnet}, is employing the IOTA 1.5 Chrysalis version \footnote{\url{https://wiki.iota.org/chrysalis-docs/introduction/what_is_chrysalis}}. However, this version of the protocol will not be the main point of this thesis, as the focus will be laid on the in-development IOTA 2.0 version of the protocol, which is referred to as Coordicide. The rationale behind choosing the version that is still under development is as follows: the new version of the protocol is going to bring a vast array of changes, such that the majority of attacks and malicious behaviors researched for the current stable version would no longer be applicable after the protocol upgrade. Furthermore, since the upgrade of Coordicide will replace the current mainnet software, the current version forms only a transition state to a more decentralized version. It is, therefore, reasonable to focus on the analysis of and reasoning about the solutions in the current prototype software. Additionally, to the theoretical analysis and review of currently known problems, the major contribution of this work will be the implementation of selected attack scenarios and performance studies in the presence of these. This is done under controlled, real network conditions, to provide reproducible results about the impact it will have on the performance and robustness of the IOTA network. 

\subsection*{Structure of the thesis}

In order to provide the reader with a sufficient background needed for the understanding of security issues that might be present in this protocol, the first part of the thesis, Chapter \ref{ch:theory} provides an overview of distributed systems and introduces some of the most known problems that need to be considered during the design phase of such a system. In Section \ref{sec:iota} the IOTA protocol is portrayed along with a brief history of how this technology has changed through the recent years. In Chapter \ref{ch:attacks}  a  review of existing research about known attack vectors and vulnerabilities of IOTA is performed. Furthermore, existing problems in other blockchain-based solutions are examined to provide context. The main point is presented in Section \ref{sec:orphanage} where the orphanage problem is introduced. In Chapter \ref{ch:experiment} a practical demonstration of the discussed orphanage attack is presented by means of a simulation in a real network setting.  Results and a summary is provided in Chapter \ref{ch:conclusion}.

\chapter{Theory and technical introduction}

\label{ch:theory}
\section{Distributed Ledger Technology}

Today's world became data-driven -- each day an enormous number of data is sent across the Internet. All services that we use are backed up with millions of servers, machines, storage devices, and other systems that silently support processes for exchanging information and control billions of requests sent across all over the world. There is always a need for more computing power, resiliency, and security, and that is precisely where distributed systems find their applications. The existence of a collection of machines working independently of each other, but sharing the same goal and processes, might lead to multiple benefits, such as sharing resources for efficient utilization or increased resiliency and robustness by eliminating the single point of failure.
This vast group of systems differs in their functionality. Different implementations may vary due to their openness and transparency on who is allowed to access the system, and if users need special permissions. They may allow for different levels of concurrency, processing certain functionalities at the same time. Another important feature is the scalability of the system if it is able to grow and expand along with an increasing number of system users. How are they distributed, is the network a collection of a few big hubs or each machine is an equal participant? And finally, their fault tolerance: how many single entities can fail before the whole system will fail? 

In the next few sections, the most useful and well-known characteristics of distributed systems in the literature will be presented to provide a background for understanding the more specific security problems existing in \Gls{dlt}. In particular, in DAG-based systems, which include IOTA \cite{popov2018tangle} -- the technology that is at the focus of this thesis.

\subsection{Eight fallacies of distributed systems}
In computer science, there is a generally known list of \textit{The Eight Fallacies} for distributed systems introduced by L. Peter Deutcha and James A. Gosling \cite{fallacies}, which is pointing out the most common false assumption made by engineers new to distributed applications. Reviewing them might help to better understand the problems one might face during a new system design, which is spread across many network instances. 

\begin{enumerate}
    \item \textbf{Reliability.} The network is reliable - requests sent might fail and packets might be lost. The system needs to be prepared for network outages, be able to handle failed requests and multiple requests regarding the same action. This is especially important for financial systems. Hardware and software might fail due to bugs, updates, or faulty patches even in today's world, for which the recent outage in the Facebook network is a great example \cite{fboutage}.
    \item \textbf{Latency is zero.} - processing of database requests can be costly and takes several orders of magnitude longer than processing a local request, which is often forgotten. Correctly setting timeouts in communication with the database and requesting only the essential data helps to avoid unnecessary latency in the program. All network calls shall be separated from in-memory calls.
    \item \textbf{Bandwidth is infinite.} A bandwidth is a limit on how much data a network can process at a given time. It can be controlled by the characteristic called network throughput. It shows how much data was transferred during a given time range. In this case, the bandwidth of a network will be the maximum possible network throughput. This problem becomes especially apparent when a single component is responsible for all the processing in a certain part of the system. Thus, it can impose constraints on the entire system. A limited bandwidth might also be a result of the design decisions, e.g the bottleneck in blockchain-based systems explained in Section \ref{sec:dag}.
    \item \textbf{The network is secure.} This false assumption might lead to underestimation of the existing threats and vulnerabilities present in the network and leave gaps in the overall security of the system unnoticed. Therefore, it is reasonable to assume that some malicious third party is always present, wanting to steal or alter the exchanged data. This is the topic that resonates with the public the most, as it might lead to huge system and financial failures.
    \item \textbf{Topology does not change.} It assumes that all data sources will be always available, that there are no network partitions. Unfortunately in the real world, servers might go down due to many different reasons. The more distributed system is, the more likely this will occur. This problem is partially present in the \gls{cap} theorem, which will be covered in section \ref{sec:cap}.
    \item \textbf{There is one administrator.} The fallacy of this belief becomes apparent when a failure occurs in the system and identifying the responsible ones is not an easy task.  Moreover, if there is more than one administrator, introducing changes or collecting permissions might be problematic. There is no single control entry that could force other network participants to perform a task. 
    \item \textbf{Transport cost is zero.} This fallacy is related to the cost of maintenance and transport over the network. That includes the cost of devices, access to the network, computational resources consumed by the infrastructure, and the people responsible for the system.
    \item \textbf{The network is homogeneous.} In the modern world it became clear that people might use a variety of different devices to participate in the network, not all machines could have the same processing power or similar speed of the network connection. 
\end{enumerate}

Remembering the characteristics listed above can help in finding security flaws and points of failures of distributed systems. Negligence in any of those areas can open the possibility for an attacker to take control over the network or at least disturb its continuous operation.

\subsection{Asynchronicity and partial synchronicity of the system}
\label{sec:synchronicity}
One of the core characteristics of distributed systems is a lack of a global clock. There is no single truth source about the time of the events happening in the system. All system components exchange messages and there is always some non-zero network delay between issuing a message and receiving it. 
Three main models for distributed systems can be distinguished:
\begin{itemize}
    \item synchronous
    \item asynchronous
    \item partial synchronous
\end{itemize}

The synchronous model assumes that any message sent over the network will be delivered within some known finite time $\Delta$. However, using this kind of model would be problematic for two reasons. Firstly, if we want to be sure that message will be delivered we would need to agree on some fixed timeout. If the value would be too large it will affect the performance of the system. On the other hand, choosing too small $\Delta$ could lead to safety violations.

The opposite approach is taken in the asynchronous model, in which there are no assumptions about the delay of message delivery. However, the message will always be delivered eventually, although the delivery time is unbounded. Whereas this solves the above issues, this model makes the design of a robust system a much more complex task. Research in this field had shown many issues related to an agreement in the network, e.g.  \cite{fischer1985impossibility}, where it is proven that reaching consensus in an asynchronous system while one process fails unannounced would require an infinite execution time. Thus, in \cite{dwork1988consensus} the partially synchronous model was introduced. It is somewhere between the synchronous and asynchronous models. It is supposed to guarantee the safety of the protocol, without loosing liveness.

The partial synchronous model is defined as follows:
There exists a special event called \gls{gst}, which is characterized as:
\begin{itemize}
    \item The GST event will eventually happen after an unknown finite time
    \item any message must be delivered before time $\Delta + \max (x,\text{GST})\Delta + \max (x,\text{GST})$, where $x$ is the time of a message issuance
\end{itemize}

\subsection{CAP theorem}
\label{sec:cap}
In 2000 at \gls{podc} during his talk, E. Brewer introduced the CAP theorem, where C stands for consistency, A for availability, and P for partition tolerance \cite{brewer2000towards}, which was proved two years later by S.Gilbert and N.Lynch in \cite{gilbert2002brewer}. It stated that only two of the following requirements can be met for distributed systems:
\begin{itemize}
    \item \textbf{Consistency (C)} - in a consistent system operations are atomic and can either commit the change to the database answering the user's request or entirely fail, there is nothing in between. A consistent system is behaving as a single entity.
    \item \textbf{Availability (A)} - available systems always respond, even if the requested data will not be valid or most up-to-date. The main goal of the availability is that a service should be as available as their underlying network with no interruptions due to server issues.
    \item \textbf{Partition tolerance (P)} - is about the endurance of the system in a faulty environment, and still, be able to run even if certain parts of nodes will crush or a considerable amount of packets will be dropped.
\end{itemize}

After more than twenty years have passed since the introduction of the CAP theorem, the understanding of it and its implications have evolved along with an expansion of new technologies. Network partitions do not occur often, so in a healthy environment, a service can provide both availability and consistency without any trade-offs. Therefore, during network partitions, the choice between availability and consistency has to be made. This compromise is possible because in reality the system can be partially synchronous if the nodes have access to local clocks.

When taking into consideration the IOTA protocol, in the case of a network partition, the protocol should be able to recover from disturbances caused by partition. Also, there should be no agreement failure among the nodes. Data consistency is most essential to value-related messages or to any other data that requires full consistency and cannot be changed once considered final. In IOTA, each node keeps its local copy of the Tangle and protocol-related data structures, see also Section \ref{sec:iota-coordicide}. The data can be extracted from the node at any moment. However, there is no guarantee that all nodes in the network will have the same view of the ledger. Thus, regarding the local data, it is simply assumed through the partial synchronicity that each message eventually will be distributed among all network participants. This is enabled by a component, called a solidifier. It listens to all incoming messages and checks if the node has in its local database all parents referenced by this message. Whenever any parent message is missing, nodes request the missing message from their neighbors, thereby improving the consistency of its local database (see Section \ref{sec:solidification}). Nevertheless, there are parts of the protocol that need to have an undisturbed consistency, the most important of which are value transfers. There needs to be a certain point in time, after which we consider data as impossible to change and be fully consistent among all nodes. To achieve that the time for validation, voting in case of conflict, and other necessary steps are introduced, leading to the sacrifice of data availability for the sake of its consistency.

\subsection{Consensus}
\label{sec:consensus}
The consensus problem is the fundamental problem in the distributed systems field. Assume the system model in which processes are vertices of a connected graph and the links represent connections through which messages can be exchanged between processes. The variation between different network models depends on the characteristics of the links, whether connections are reliable (if a message can be lost), if processes might fail, and the time perception of each process. To model, the connections between vertices, the asynchronous, synchronous, and partial synchronous model described in Section \ref{sec:synchronicity} can be used.

In order to solve a consensus problem, the proposed consensus mechanism should fulfill the following properties:
\begin{itemize}
    \item Agreement. All the non-faulty processes need to decide on the same value.
    \item Termination. All decisions need to be made eventually.
    \item Validity. The decided value needs to be one of the initial values proposed by some process.
\end{itemize}

\paragraph{Consensus output.} 
The consensus problems may have different levels of difficulty, based on the assumptions and requirements imposed on the expected outcome. 
The consensus problem describes the situation in which multiple processes need to cooperate in order to make a decision. This decision might be to agree on certain single value \textit{(single-value consensus)}, to agree on a single-digit {0, 1} \textit{(binary consensus)}, or on series of values \textit{(multi-value consensus)}. An example of the decision might be a well-known problem of whether to commit transactions to the database \cite{dolev1982distributed, lampson1979crash}.

\paragraph{Faulty processes.}
Another assumption is about the existence of faulty processes. Different types of faults are distinguished. Firstly, the \textit{passive} fault, where the process is collecting information about the network but does not deviate from the protocol rules. The process that fails and stops working without notifying the rest of the system is a \textit{crush}. The next failure type is \textit{omission}, where the faulty process can omit some of the messages. A more difficult scenario would be with the existence of processes that do not follow the protocol, usually with an intention to damage the system. They are traditionally called \textit{byzantine} faults, the name was taken from the byzantine generals' problem \cite{lamport2019byzantine} which will be described in Section \ref{sec:byzantine}.

\paragraph{Resiliency.}
The model can be also characterized by its resiliency, which is the maximum number of faulty processes that can be tolerated in the system. 
Before the process of reaching the consensus starts (might be called \textit{voting}), each of the processes has the initial opinion, that is brought together from all of the processes for an initial configuration (state) of the system.

\begin{figure}
    \centering
    \includegraphics[width=.9\linewidth]{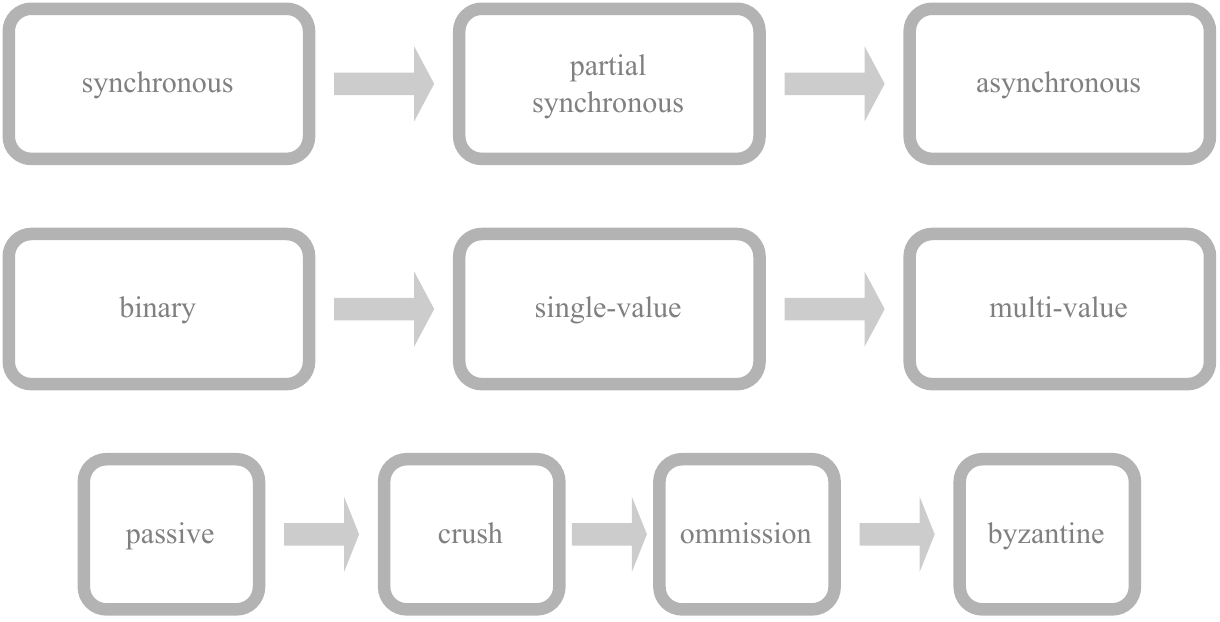}
    \caption{Characteristics of consensus, starting from the top: connection model, output decision, fault types. Each chain is ordered from the least to the most difficult for finding the solution.  }
    \label{img:consensus-diff}
\end{figure}

The specific system model is chosen based on the kind of characteristics desired for the system. The difficulty of finding a solution to a consensus problem varies depending on this choice. Designing protocol that is fault-tolerant in the synchronous system will be much easier than in partial synchronous, as, in the first one, each message delay is limited with known by each process bound. The same is for desired output decisions, the less complicated mechanisms will be required for binary consensus that multi-value. 
Thus, proving that a certain algorithm or mechanism works with stronger assumptions will usually be strong enough for a weaker consensus family. Diagram in Figure \ref{img:consensus-diff} shows the discussed characteristics ordered by their difficulty.

\subsubsection{Liveness and safety}
The liveness and safety terminology is usually used to describe how the system is affected and the mechanisms occurring during protocol execution. The meaning behind those terms can be read intuitively as:
\begin{itemize}
    \item \textbf{Liveness}. The 'good things' do happen, the decision will be made, the action will be taken.
    \item \textbf{Security}. The 'bad things' do not happen, careful decisions and the action is halted until the solution is correct.
\end{itemize}

 \paragraph{Liveness and security in a synchronous and asynchronous setting.}

It is not possible to assume network availability in the asynchronous setting. The reason is that even if all messages will be finally delivered, their delay is unbounded so that one never knows if a certain message was lost or has never been sent. The same issue relates to the liveness of the network, which can not be guaranteed without any stronger assumption about synchronicity. On the other hand, the synchronous model results in guaranteed liveness of the network, but the trade-off is the network security, as too strict assumptions on the message delivery time can be broken in real-world scenarios. The partial synchronous model can be seen as a golden mean between synchronous and asynchronous assumptions, allowing for both liveness and security control.

\paragraph{Liveness and safety in consensus}
The two properties are quite often used with regard to consensus protocols. They are both inseparable, any correct protocol aiming to solve a consensus problem should not leave out any of those two properties. Without a \textit{security} property, the consensus \textit{agreement} can not be reached. The same is for liveness, protocol that can not guarantee \textit{liveness} is violating the \textit{termination} rule.


\subsection{FLP impossibility theory}

As introduced in Section \ref{sec:consensus}, the assumptions about an underlying network model can have a great impact on what is theoretically possible in the network. There are many propositions on how to reach a consensus in the synchronous system. However, for many years the topic of reaching an agreement in an asynchronous setting was repeatedly discussed, without any solution. The \cite{fischer1985impossibility} brought this debate to an end. The proof introduced in this work is commonly referred as to FLP impossibility. In this paper, Fischer shows that reaching consensus in an asynchronous model is not possible. More specifically, given the solution to the consensus problem within an asynchronous model and assuming that it is tolerant to just a single fault, it was proven that the execution time can be infinite. In order to solve the consensus problem all three requirements: termination, agreement, and validity need to be met for all possible execution rounds. The FLP proof shows that agreement and validity were provided, however, termination was not fulfilled.  The proof was conducted for very weak assumptions, such as:
\begin{itemize}
    \item binary consensus
    \item crush faulty process
    \item agreement is reached when only some non-faulty nodes decides
\end{itemize}

Since the paper discusses a case with weak assumptions, the FLP impossibility covers also more complicated cases. As a consequence, it is known that there is no protocol that could deterministically solve the consensus with the asynchronous model, as there always could be one initial configuration that will run forever.

Nevertheless, systems in the real world are usually not fully asynchronous. Thus, after the introduction of the FLP impossibility, countless papers were presented,  aiming to find out how much the asynchronous model can be modified to solve the problem. 
In \cite{larrea2002impossibility} authors consider different classes of the weakest failure detectors that solve the reliable broadcast that was proven to work in an asynchronous system with a majority of correct processes \cite{kawazoe1999revisiting}. They show that none of those processes could be implemented in a partially synchronous model with only one failure.

\subsection{Byzantine Fault Tolerant Systems}
\label{sec:byzantine}
The term \gls{bft} took its name from the well known Lamport's paper \cite{lamport2019byzantine} where the concept of Byzantine failure is introduced based on an example of a thought experiment, where a group of Byzantine Generals besieging the enemy's camp needs to come to a consensus on whether to attack or retreat. Generals communicate via signed messages and among generals, there are traitors who are aiming to corrupt the plan by sending inconsistent messages and disrupting the honest generals' communication. The assumptions laid on messages are: messages are not lost, the issuer of each message is known, and generals know when a message was omitted by others. Thus the \textbf{Byzantine fault} is a fault in the system presenting different views to different observers and a \textbf{Byzantine failure} occurs when the system fails due to a Byzantine fault.

The outcome of the generals' problems was that for an operation to not fail $3m+1$ of generals are needed for $m$ traitors and that at least $m+1$ rounds of message exchange are needed.
Since the formation of the Byzantine General problem, it was a subject to many extensive studies that lead to many $BFT$ algorithms solving consensus under different conditions in a system with present failures.

In \cite{driscoll2003byzantine} it is shown that the Byzantine Generals problem is not only a theoretical problem, and there are many practical examples when it can be reflected in real systems. Therefore, either a system is designed without the need for reaching consensus, or if when it is not possible, an appropriate fault-tolerant consensus algorithm must be introduced to preserve system safety.

\paragraph{Practical Byzantine Fault Tolerance}
One of the most widely known \Gls{bft} algorithms is \gls{pbft} introduced in 1999 in \cite{castro1999practical}. The algorithm was designed with a state machines model with faulty nodes (replicas) present in the system. The algorithm utilizes the primary-backup technique \cite{oki1988viewstamped} and introduces the role of the primary node (leader). 
All nodes are sequentially ordered, and the rest of the nodes besides the primary are called secondary nodes and can replace the primary in case of its failure. It works in a partial synchronous model and guarantees the security in a presence of $m$ faulty nodes out of a total $3m+1$, thus it provides safety and liveness after the GST event. 
Replicas need to be deterministic -- when given the same input and being in the same state they should always produce the same result. All replicas must start with the same state.

The goal is that all non-faulty replicas agree on the request execution order.
The \Gls{pbft} consensus round roughly proceeds as follows:
\begin{enumerate}
    \item primary node receives a request from a client,
    \item the request is broadcasted by the leader to all secondary nodes,
    \item all replicas execute the request and send the replay to the client,
    \item the operation is terminated when a client receives $m+1$ the same replies.
\end{enumerate}

\Gls{pbft} does not scale well as it requires extensive communication overhead ($On^k$), therefore the number of nodes participating in the consensus shall be small. Also, it is susceptible to Sybil attacks (more in Section  \ref{sec:sybyil}) where a node can gain an advantage on the network by forging many identities.

\paragraph{Other \Gls{bft} algorithms}
many \Gls{bft} algorithms variants are used in blockchain projects. The Tendermint \Gls{bft} algorithm uses \Gls{pbft} in combination with \gls{dpos} \cite{buchman2018latest}.
In contradiction to \Gls{pbft}, Tendermint algorithm uses rotation for the leader role, therefore, the leader change is a normal part of the algorithm, and not as it was in case of \Gls{pbft}, the leader can stay the same and is changed only in case of a failure, keeping a stable leader results in a smaller message overhead. However, a malicious leader kept in place, poses a security threat. Therefore the \Gls{pbft} algorithm performs well in a permissioned setting, e.g. it is used in Hyperledger Fabric blockchain \cite{cachin2016architecture}.

Another example is SBFT algorithm that by using threshold signatures reduces the message overhead of the \Gls{pbft} algorithm to linear complexity $O(n)$ while providing about two times better throughput and scalability than traditional \Gls{pbft} \cite{gueta2019sbft}.




\section{Directed Acyclic Graph}
\label{sec:dag}
\gls{dag} is a data structure that is being widely used in distributed systems \Gls{dlt}. 
It consists of vertices and edges, that can not create cycles, thus the whole structure allows to move in one direction. \Gls{dag} structure may consist of multiple parallel chains, the special case of DAG is a single chain structure used in blockchain-based systems. Single chain structure allows for total ordering of the data stored in vertices, however, the new vertices in the blockchain structure, called blocks, cannot be added in parallel, which is known as a blockchain bottleneck. This problem is the main source of blockchain scalability issues. Therefore, many systems started to use multi-chain DAG, such as Hashgraph \cite{baird2016hashgraph}, Avalanche \cite{rocket2019scalable} or IOTA. 
A DAG-based system allows only for partial-order, as the new vertices can be added to the structure simultaneously and it is not possible to determine the total order without additional protocol rules. The DAG structure allows for efficient usage in the asynchronous setting where each network participant can have a different view of the current state of the ledger due to delays present in the network.

Another important characteristic of \Gls{dag} based system is the way value transfers are handled, the two most popular concepts are:
\begin{itemize}
    \item account-based -- fund transfers are tracked via balance on the user's accounts
    \item UTXO-based -- funds are represented by the unspent outputs, which are the outputs from the transactions that have not yet been spent. Funds are transferred by using unspent outputs as an input to the transactions, which in turn creates new unspent outputs.
\end{itemize}

The security and irreversibility of the ledger are provided by the cryptographic functions and protocol rules.

\section{Introduction to IOTA} 
\label{sec:iota}
The idea behind IOTA  was born in 2014 on one of the blockchain-related forums. The discussion arose around the topic: \textit{why does it need to be a single chain of blocks}? The Directed Acyclic Graph was proposed to be used instead, which could allow for the creation of multiple conflicting, individually valid branches. In principle, a \Gls{dag} allows for simultaneous writes to the ledger and the parallel processing of messages. After the conceptualization phase, the IOTA cryptocurrency was created, to remove scaling issues and transaction fees, utilizing a new type of distributed ledger structure. In 2015 S. Popov published "The Tangle"\cite{popov2018tangle}, which established the backbone of  IOTA. After an initial funding phase of the first version of the IOTA network went live in 2016 \cite{iotaWiki}.

Designing a system is not an easy task. Along with a new design idea, many new questions emerge. For example: who would validate transactions and data in the network, and what will be the incentive if there are no transaction fees? How should the protocol behave in case of conflicts, and how should consensus be achieved? Additionally, a new design opens the possibility of new unknown attack scenarios. Is the network secure enough and how to measure the health of the \Gls{dlt}? 
To answer all those questions in-depth studies must be carried out along with the system stress testing in practice. The IOTA protocol went through various stages of research and development. Many protocol components have been changed up until this point. In the next sections, we are going to explore this evolution and introduce core concepts of the protocol, that will be necessary to understand attacks and vulnerabilities discussed in Chapter \ref{ch:attacks}.
We will explore the Coordicide project which is the transition event for the update of the protocol to IOTA 2.0 that will be a major software update and a big step for the IOTA network towards full decentralization.

\subsection{The pre-Coordicide IOTA}
\label{sec:pre-coo}
Due to the novelty of this technology, the first version of the IOTA needed some temporary solution that could help with keeping the network safe during its infancy development period. In older versions of IOTA, the \Gls{pow} was used for limiting message throughput and for consensus finding. 
Therefore, without a sufficient amount of honest \Gls{pow} the network would not be able to sustain an attack from a powerful adversary that has a significant amount of hashing power. As a bootstrap mechanism, IOTA utilizes a centralized node, called the coordinator. The Coordinator runs slightly modified software. It is a source of trust and every message that is confirmed by the coordinator is also valid for the rest of the network. The Coordinator node periodically issues messages called milestones. These messages reference messages that are considered as voting winners. IOTA was not the only cryptocurrency that decided to use such an interim state, another example can be the Alert System in bitcoin \cite{alertSystem}.

Although it has been a couple of years from the introduction of the coordinator to the IOTA mainnet,  the Coordinator's removal was always a big part of the protocol development road map. There are many reasons why this centralized solution should not be included permanently, such as:
\begin{itemize}
    \item it is a single point of failure
    \item theoretically it gives the foundation that is running the Coordinator a possibility to confirm transactions with chosen priority or halting the confirmations of certain messages, which should not be possible in an open and honest distributed system,
    \item IOTA users need to follow exactly one official coordinator if someone starts to follow any malicious actor it might result in a fork,
    \item the coordinator might hinder problems with network scalability.
\end{itemize}
Indeed, there are real-world examples of the single point of failure threat, as it happened in the past that Coordinator went down. Note that, the ledger consistency was not broken, as the rest of the network kept processing transactions according to the protocol rules. However, transaction confirmations were halted until the Coordinator node was restarted.

\subsection{The reason for Coordicide}
Taking into consideration all Coordinator drawbacks explained in the previous Sec. \ref{sec:pre-coo} it comes with no surprise that removing the Coordinator will become one of the major objectives for the IOTA's further development. At the end of 2018 IOTA announced the Coordicide plans and test network to validate their ideas. 
The first proposal for a Coordicide solution was provided in Barcelona in 2019. The whole new protocol was imagined and outlined, changing the fundamental assumptions.  As the approach was a new design, testing was crucial, which created the necessity for a test network with a full node implementation. Thus, soon after the summit, the GoShimmer prototype implementation started. GoShimmer is the first prototype node software for the Coordinator-free IOTA network. Along with practical development, the team started to work on the topic from the theoretical point of view, so the next major step was to summarize all theoretical concepts in the Coordicide whitepaper \cite{popov2020coordicide} published in 2020, which was supposed to establish more organized, comprehensive explanation and provide more technical details needed for future real-world implementation.
    
Currently, the GoShimmer software, many versions later has much progressed, bringing it into a state where, together with a large array of analysis tools, scientific studies can be performed. With the current state of the node software, it is now possible, to validate the perceived theoretical ideas and test performance. At present, the prototype software is used not only by IOTA Foundation but also by the community members willing to help with testing the Coordicide progress. 
This enables a more rapid identification. The first full working prototype version of Coordicide was released in April 2021 on the DevNet and up to the time of writing this paper has resolved many conflicts and has processed many messages, proving that the Coordicide development came long impressive way.
In regards to the analytical discussion, the role of GoShimmer as a tool to validate the theoretical ideas has an important task: it allows for testing theoretical problems such as attack vectors in a more controllable and real network environment. The test network node software is provisioned with quite wide support for metrics and measurements collections. This allows measuring how the protocol behaves under different circumstances and can be used to fine-tune the best set of default parameters for the protocol. Or as in the case in this work, to perform the certain attack scenario and analyze the results of the experiment with a help of data collected by the GoShimmer software.


\subsection{The Tangle}
\begin{figure}
    \centering
    \includegraphics[width=\linewidth]{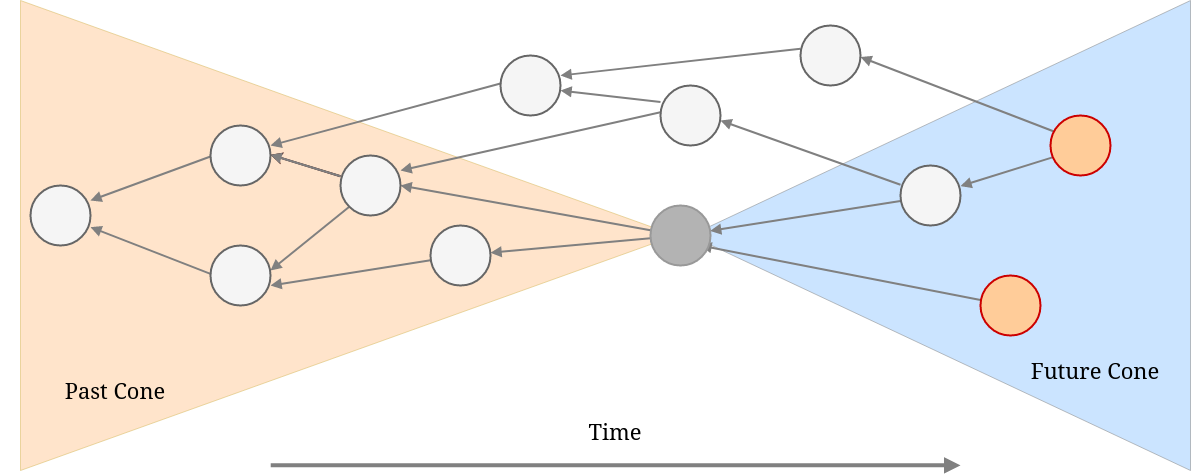}
    \caption{The Tangle. The past and future cone of the grey message are shown. Tips are marked in orange.}
    \label{diag:tangle}
\end{figure}

The Tangle is the core structure of the IOTA protocol. It is \Gls{dag} structure in which vertices are made up from messages and where messages reference each other. We call the referenced message a \textit{parent}. Each node has its local version of the Tangle and updates it based on the messages received from its neighbors. To add a new message to the Tangle, a node has to select and reference (approve) $k$ other \textit{parent} messages that are already part of the Tangle. Messages at the end of the Tangle data structure that has not yet been approved are called \textit{tips}. All messages that directly or indirectly reference a message are called its future cone. And respectively, all messages that are directly or indirectly referenced by a message are called its past cone.
The Tangle structure has its beginning in the message called the \textit{genesis}. A graphical representation of the Tangle is presented in Figure \ref{diag:tangle}

\subsubsection{Message}
\begin{figure}
\centering
\hspace{1cm}
    \begin{subfigure}{0.35\textwidth}
        \includegraphics[width=\textwidth]{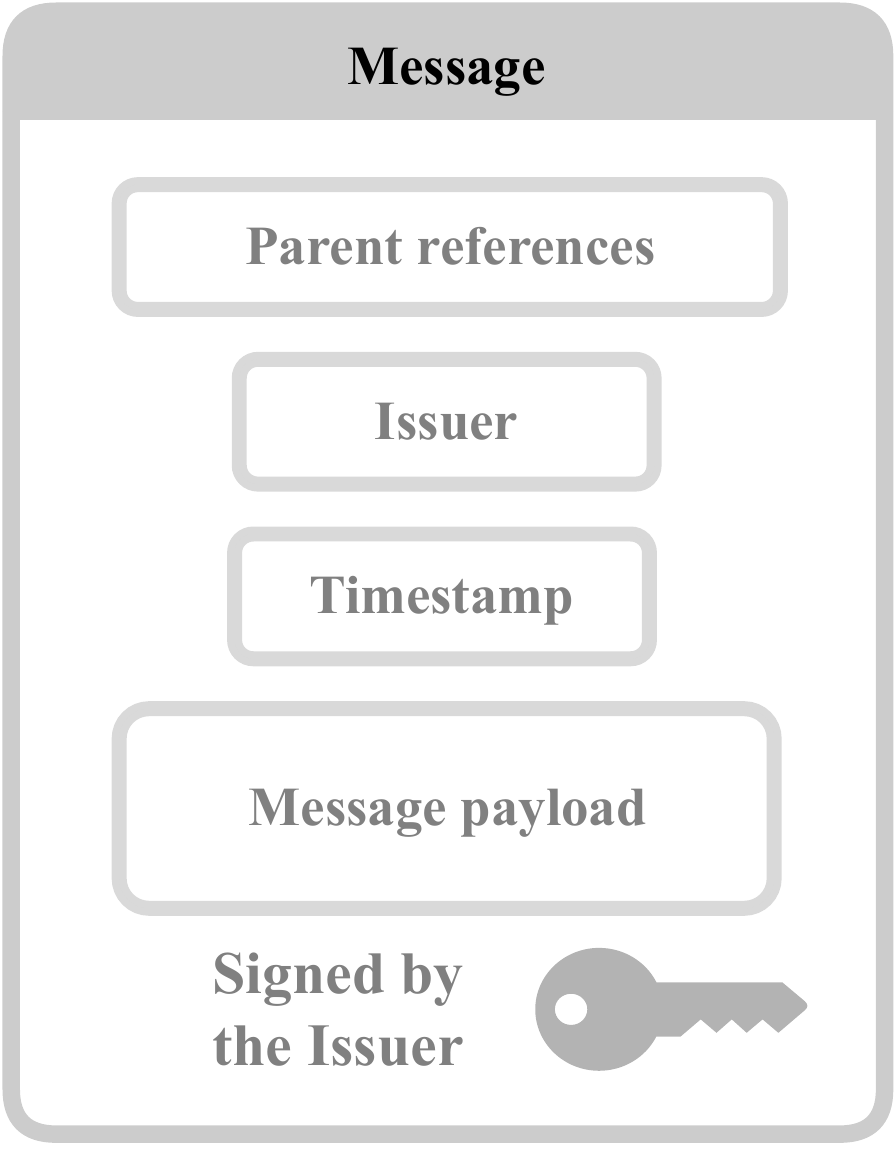}
        \caption{The simplified message layout.}
        \label{diag:message}
    \end{subfigure}
\hfill
        \begin{subfigure}{0.35\textwidth}
        \includegraphics[width=\textwidth]{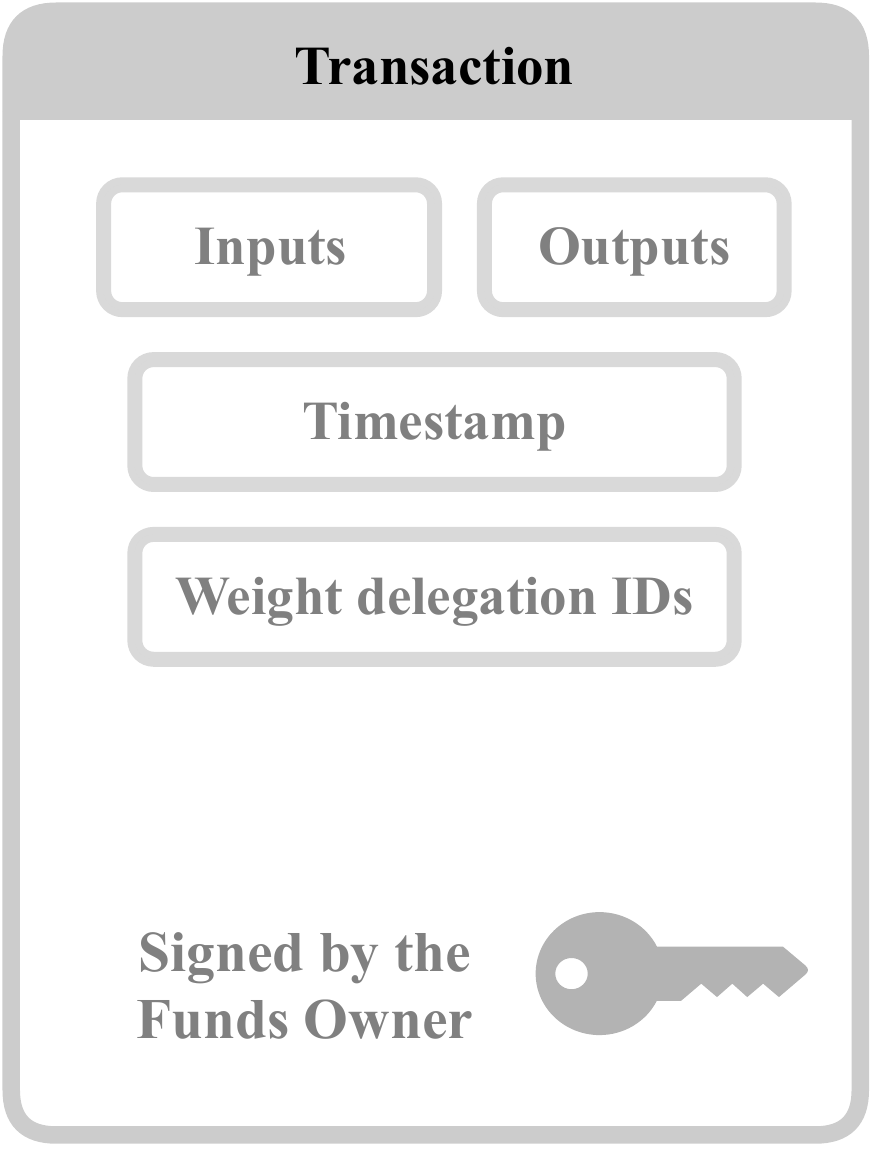}
        \caption{The simplified transaction layout.}
        \label{diag:transaction}
    \end{subfigure}
\hspace{1cm}
    
\end{figure}

The simplified layout of the message is presented in Figure \ref{diag:message}. Each message in the Tangle consists of parent references, the timestamp (the time when the message was issued, the issuer identifier, the payload that contains value or data transfers, and the signature, as each message is signed with the private key of the message issuer, through which all other nodes in the network can validate the owner of the message.

\subsubsection{Transaction} 
A transaction is a value transfer that is recorded in the ledger. A created transaction is contained as a payload in the message and is signed with private keys corresponding to addresses of the used unspent outputs. Each transaction consists of unspent outputs - transaction inputs, transaction outputs, timestamp, and signature. An output stores the information about the address connected to the funds, and the balance that will be transferred. The simplified layout of the transaction is presented in Figure \ref{diag:transaction}.

\subsubsection{UTXO and branch DAG}
The Tangle is not the only data structure used in the IOTA protocol. The transactions are also tracked with the UTXO DAG data structure that keeps track of all value transactions exchanged in the network, and allows for efficient conflicts recognition.
Whenever there are conflicting messages (transactions that send the same output) introduced into the network, a branch corresponding to each transaction that introduced a conflict is created. Each new conflicts introduce new possible future versions of the ledger depending on which transactions will be accepted. Spending from conflicting transactions creates a child branch of the origin conflict branch, thus forming a DAG.

All transactions that spend the same output belong to the same conflict set and only one transaction within this conflict set will be accepted through the consensus mechanism to the ledger.

\subsubsection{Coordice vision of the IOTA protocol}
\label{sec:iota-coordicide}
The protocol consists of multiple components that can be divided into three layers: network, communication, and application layer. 

The network layer manages connections with peers and the gossip protocol. It provides the underlying network structure and is mostly independent of the rest of the layers.

The communication layer contains the core data structure - the Tangle, which serves as a medium for messages exchange.  Components that belong to this layer are Rate Setter and Congestion Control.
Rate Setter is responsible for adjusting a node's issuing rate, in a way that fairly uses its bandwidth share. The Congestion Control, on the other hand, is controlling if other nodes are not abusing the network resources by exceeding allowed issuing rates.

The application layer is connected to the Consensus component and provides modules necessary for secure value transfers.

The next Sections will describe components of the protocol that are essential for this work. 

\paragraph{Network and nodes.}
The IOTA network has only one type of node, this is different from other DLTs, where multiple roles exist, e.g. miners in bitcoin. Those additional roles are not needed because there are no transaction fees and the work for the network maintenance is performed whenever a node sends a message. Each node is uniquely identified with a node ID and has weight assigned to its identifier (more on this topic in \ref{sec:mana}). Each node has up to eight neighbors which are selected by the autopeering mechanism \cite{muller2021salt} or manually. Nodes are using a gossip protocol to exchange messages. The gossip protocol is a communication protocol that aims to distribute any messages to all nodes in the network. It relays on the fact that each node should pass the gossiped message to its neighbors. 

\paragraph{Syncing.} The number of nodes in the network varies over time, as nodes can go offline and back online. Each node has its local version of the Tangle representing its view on the state of the ledger. Thus, if the node goes offline for some time, his ledger is not up to date after its return. Therefore, it needs to request missing data from its neighbors. 
A node that is not in sync should not issue messages as it will attach to the wrong part of the Tangle.  

\paragraph{Solidification.}
\label{sec:solidification}
The Tangle data structure is partially ordered because each message contains a reference to $k$ other messages that have been already included in the Tangle. Due to network asynchronicity nodes can receive messages in a different order. Also, messages can be lost during the broadcast. However, to attach a message to the Tangle, all past cone of the message has to be known. To continue with message processing and find out missing messages, the node is requesting missing parents through the process called solidification. Whenever a message is missing any of their parents, the node sends the solidification request to its neighbors and stops processing this message until all parents are known. Received parent messages go under the same process as the child message, thus the solidification process is a recursively repeated mechanism that allows retrieving the entire message history. It strengthens the network synchronicity and the assumption that each message will be eventually delivered to all nodes in the network.

Right after the message knows its past cone, the \textit{maximum parent age check} is performed. The node checks if the difference between timestamps of a processed message and each of its parents is not older than $\zeta$. If a message is not referencing too old messages the message is marked as \textit{solid}. If a message does not comply with the parent age requirement is discarded.

In the prototype software, the component that is responsible for the solidification process is called \textit{solidifier}.

\paragraph{Mana.}
\label{sec:mana}
Because each node in the network has a corresponding identity, there was a need for a Sybil protection system, which prevents the malicious actor from gaining a dishonest proportion of control over the network by forging many identities. Mana is a Sybil protection system that can be thought of as the reputation of the node or a scarce resource in form of a weight attached to each node. A node can gain reputation by contributing to the network, e.g. creating value transfers.

Mana is calculated locally, as a function that takes transactions as an input and returns the Base Mana Vector as an output, Which consists of reputation scores of all nodes. 
Mana values of a single node can be changed by pledging the corresponding funds transferred in a transaction to the node ID. Mana is derived from the token value, however, owner of the funds can delegate Mana to nodes it favors.  
There are two types of Mana, that serve different purposes:
\begin{itemize}
    \item Consensus Mana (cMana). It is used in the autopeering mechanism, to connect nodes with similar weights, and in the voting system to make votes of Mana-rich nodes more significant. cMana is constant over time if not interfered with value transfers.
    \item Access Mana (aMana). Is used as access control to the ledger. After a portion of aMana is gained during a value transfer it decays over time, thus it encourages active participation in the network.
\end{itemize}

\paragraph{Tip Selection Algorithm.}
\label{sec:tsa} 
\gls{tsa} is a process of selecting where to attach a newly created message. IOTA 2.0 uses \gls{rurts} algorithm that is an extension of \gls{urts} which has very simple principles. Each node maintains its own local Tip Set which is the collection of all messages that have not been referenced yet. When a new message is created nodes select tips from the Tip Set with a uniform probability. 
The difference between \Gls{rurts} and \Gls{urts} is that the first one puts a restriction on the message's parents' age. This means that if a tip was not selected during $\zeta$ time it is removed from the Tip Set.

\paragraph{Consensus and approval weight.}
\label{sec:aw}

In blockchain \gls{pow} based systems the longest chain rule is causing many blocks to be left behind, which leads to a waste of processing power. The DAG structure allows for the existence of multiple branches and forks is accepted as a part of the protocol, which allows for creating a mechanism that will allow for merging many of those outcast messages back into the main Tangle and significantly reducing the waste of resources.

\gls{aw} is one of the backbones of the \gls{otv} consensus mechanism that is used for conflict resolution. \Gls{aw} is a mechanism that allows a node to express its opinion on a conflict or a message, by attaching a newly created message to the part of the Tangle he likes and by using specific types of parent references. \Gls{aw} is used in a twofold way. Firstly, it is used directly on the Tangle as a weight accumulated on the message by its future cone. Whenever a message is referenced directly or indirectly by any other message attaching in its future cone, it gains \gls{aw} that corresponds to the referencing message issuer weight. If the weight collected by the message exceeds the known to all nodes threshold $\theta$ it is marked by the node as \textit{confirmed} (\textit{finalised}).

The second form of \Gls{aw} is approval weight collected on branches. It is the core concept for the \Gls{otv} mechanism, on which nodes express their opinions on conflicts by issuing messages. 
There is no additional communication overhead for exchanging votes, as support for conflicts is expressed by referencing them directly or indirectly on the Tangle structure. Nodes select a tip from the tip set and decide if it agrees with the ledger state it represents. If the selected message has a conflicting transaction in its past cone and the node does not agree that this conflicting transaction should be accepted, we can use a special type of reference to express our disagreement, which does not allow us to count our node weight. And the same is when the node likes the selected tip and the state of the ledger it approved, node can approve this message in a way that his vote will be assigned to the underlying branch. 

\Gls{otv} works in a way that nodes can shift their opinions if the majority of the network has a different opinion on which conflicting transactions should be accepted. This way, a node is adjusting its own opinion based on the information written to the Tangle by other network participants.

\paragraph{Rate setter and Congestion Control.}
\label{sec:cc}
To allow for the participation of smaller and less powerful devices, the total network throughput needs to be limited in a way that allows for the participation of devices with limited processing power. The congestion control algorithm should provide safety, fairness, and consistency.
Safety property guarantees that any malicious actor can not bypass the rules stated by the algorithm. The fairness property states that each node should get its share of the throughput. In IOTA 2.0 this share is calculated based on the node's access Mana.
Consistency means that if a message is accepted by one node and written to the Tangle, it must be received by all nodes in the network within some time-bound.

Therefore the Congestion Control component is regulating the flow of the messages and watches if other nodes meet their network share limits. When the network is fully utilized -- its whole throughput is used -- the congestion control mechanism does not allow for a node to exceed its bandwidth.

However, during under-utilization periods, when nodes are not using their full share of the access to the network, the protocol allows for using those unused resources. In this way, nodes with limited aMana supplies can issue messages at rates excessively exceeding their normal network share during full utilization periods. Usually, it is a desirable characteristic, however, it opens a possibility for an adversary with limited aMana resources to use those under-utilization periods to flood the network with malicious messages.

The rate-setting mechanism allows for a node to estimate its own allowed bandwidth and helps to choose a correct issuing rate.

\subsection{Protocol evolution over time}
The objective of this section is to review how the protocol changed over time. Knowing the legacy version will be useful in understanding the obsolete security flaws present in IOTA in the past. Only the first version of IOTA 1.0 will be discussed here and compared to the Coordicide version. The reason is that most of the previous attack analysis papers that will be investigated later were based on the legacy version. 
The first version of IOTA node software was Ivy, however, it was not fast nor stable enough, therefore the new two major projects were started together by IOTA Foundation and the IOTA Community: Hornet (implemented in Go) and Bee (Rust implementation). Both stable versions are currently used in the IOTA mainnet, the official primary network on IOTA version 1.5 released during Chrysalis update. Also, IOTA mainnet version is still not fully decentralized, due to the need for Coordinator existence. 

It is worth mentioning that IOTA protocol is just a set of rules that network users are advised to follow. Based on rules provided in the protocol specification there can be many software node implementations. Thus, IOTA can be considered as truly free protocol, if most of the network users follow those rules and official implementation, the network is able to sustain some percentage of malicious users, behaving in a way that does not contribute to the common interest. The goal of protocol improvements is to find a set of rules that will be most beneficial to the honest participants and will punish or neglect any negative impact of adversarial behaviors. 

The first idea on IOTA given in \cite{popov2018tangle} was combining the consensus mechanism and the tip selection in one method called \gls{mcmc}. Each transaction had its weight corresponding to the amount of \Gls{pow} done during a transaction creation and the sum of weights of all transactions directly or indirectly approving this transaction. 
In the case of two transactions spending the same funds, the transaction that had a higher weight was considered valid. Firstly, the security of the Tangle was supposed to be guaranteed because the hashing power put into the Tangle by honest nodes should exceed the adversarial one, especially when the network will be fully grown. 
However as mentioned earlier, due to security concerns the Coordinator node was introduced, which follows the protocol and is the deciding entity in terms of what is accepted and what would be rejected by the network. Therefore, the transaction was considered as confirmed and accepted by all honest nodes in the network only when it was referenced by the milestone message.


\chapter{Attack vectors review}
    
\label{ch:attacks}
Blockchain-based systems are being developed and used for over a decade now. According to \cite{slowmist}, at the moment of writing, there were more than 600 hack events related to the blockchain technology recorded. The losses are estimated at over 25 billion dollars. The report \cite{charoenwong2021decade}  summarizes the largest cryptocurrency thefts within 2011-2021 and indicates 30 thefts with value lost between \$7 to \$88 billion. Attacks are related to many areas of the crypto industry, such as exchange hacks, wallets security issues, exploits of the \gls{dapp}, \gls{defi} platforms, \gls{nft}, public blockchain security incidents, and many more. 
As shown in \cite{blockchainReview2021} attacks targeting applications and platforms built upon the base blockchain layer account for over 80\% of all incidents. One of the most important characteristics of blockchain technology is its security. The security of an ecosystem built around a given \Gls{dlt} project depends heavily on the security of its base layer. Security breaches that happen on this base layer are much rarer, however, as the report indicated in 2021 there were at least 8 (1,3\%) attacks on public blockchains.
In this thesis, we will focus on the security incidents and vulnerabilities related closely to those public blockchain technologies and their protocol flaws.

\section{Review of known attack vectors}
People's creativity has no limits. One of the greatest examples are clever security violations represented in numerous threats.  Carefully designed attacks by malicious actors target the crypto space daily. In addition to adversarial behavior, failures can also occur due to random incidents caused by protocol flaws. All of the above can be divided into different subgroups, depending on what kind of vulnerability is exploited or occurs,  what the goal of a potential attack is and which part of the protocol is targeted. 

By considering the attacker's objective and the type of the caused damage, two general groups can be distinguished. In the first case, individuals or specific user subgroups that use protocol utilities through client libraries and applications are targeted. Examples for such an attack are eclipse attack (more in Sec. \ref{sec:eclipse}), double-spending attacks, and other attempts to break through a cryptographic shield that is protecting a user. Additionally, an attacker could try to create the funds "out of thin air" or revert confirmed transactions. The second type is targeted at the network and the protocol itself. There, a malicious actor attempts to cause damage to the wider range of users, or even paralyze the activity of the whole network, which could lead to both financial and trust losses. Examples of these kinds of attacks are spam attacks, denial of service attacks, or attempts to keep the network undecided, i.e. attacks on the liveness of the protocol.

For the purpose of this thesis attacks will be discussed in the following groups:
\begin{itemize}
    \item Network layer -- denial-of-service (DOS), spam attacks, solidification attack, eclipse attack
    \item Sybil attacks -- forging multiple identities, attempts to obtain unfair network resources proportion.
    \item Agreement failure -  attacks on the consensus, keeping network undecided, double spending, berserk attacks.
    \item Structure attacks -- spamming or issuing purposely designed message structures, and not following the general protocol guidance, such as ignoring the standard \gls{tsa}. e.g. Orphanage, Blow Ball, parasite chain.
    \item Security exploits -- exploits on the used cryptography methods, replay attack (reuse of addresses plus reattachments).
\end{itemize}

\subsection{Threats in the network layer}
    
    In this section attacks on the network layer will be discussed. These kinds of attacks usually are not meant to steal the funds or bring to the adversary any tangible benefits. However, the consequences of such an attack can be fatal to the network and thus could bring about irreparable damage to the reputation of the DLT. Additionally, any interruptions in the continuity of network operations could lead to severe damage, and financial losses to individuals or enterprise projects may be incurred. Some of the undesirable outcomes of the violation of network layer robustness are halting of the network, connections issues, prevention of the ledger inclusion for honest transactions, and crash of nodes in the network. Certainly, if a protocol cannot defend itself against such an attack its credibility is at stake together with reputation loss to the community and the wider ecosystem.
    
    \subsubsection{Spam attacks}
    The \gls{dos} attack is a primary attack leading to disruptions in the network layer. It is not specific only to the DLT space, for example, quite often it is targeted at centralized hosting servers or services, resulting in denying access to the web-service resources to users\cite{schuba1997analysis}. During the DoS attack, a bad actor is flooding the network or a specific target server with excessive internet traffic. The purpose is to overload servers or network nodes, thus causing problems for nodes to process requests and respond efficiently.  Eventually, this can make the network unusable for its honest participants. A DoS attack can be performed with a single machine. 
    A more powerful attempt to destabilize the network is \gls{ddos} \cite{nazario2008ddos}. Where several machines, usually distributed around the world, are used in coordination. Those machines are often part of a botnet - a collection of infected computers, that can be controlled remotely \cite{hoque2015botnet,liu2009botnet}.
    In the cryptocurrency realm, an attacker could create thousands of transactions to increase the congestion in the network, overload the nodes, and even halt the confirmation of honest messages, which can result in an unusable network. For most of the existing cryptocurrencies, the defense mechanism against this kind of bad behavior is to incur a  cost for performing such an attack. Typically, creating a transaction requires to include a transaction fee, so issuing thousands of transactions can lead to expenses that outweigh any supposed benefits. This is especially true when a spamming attack does not result in direct profits.
    Nevertheless, when the project is still in its infancy, the price of issuing transactions is low or even feeless. Thus, the project is particularly vulnerable to spamming attacks. Therefore, it is important to have working defense mechanisms in place, e.g. a Congestion Control component, described in \ref{sec:cc}.
    
    There are many real-world examples of DoS and DDoS attacks. 
    \paragraph{Nano.}
    In March 2021 a spamming attack on the Nano network was performed, during which about 4.6 million transactions were issued within the first 24 hours \cite{nano1}. At first, it seemed that the network will sustain the attack, however spamming continued for a couple of weeks. It was possible with limited expenses, as Nano has no transaction fees.  The attacker was able to send transactions at almost zero cost, using only a little computing power to solve insignificant  PoW, which was supposed to work as spam prevention.  Users reported problems with halted transactions and long confirmation times through the social media channels. Many of the nodes went out of sync, as they could not keep up with processing a higher network throughput \cite{nano2}. To tackle the problem, the node's operators were asked to intentionally lower the performance of nodes, to allow other nodes that went out of sync to recover \cite{nano3}. These kinds of intervention, needed on the broader scale, showed that not only was the network not sufficiently prepared but also that it is not truly decentralized, as this kind of external intervention would not be effective in a fully distributed system. Another issue was a significant increase in storage due to the flood of very small transactions, which is related to the dust protection problem. 
    
    \paragraph{Solana.}
    In September 2021 the Solana network was entirely shut down, causing a 17-hour outage, after an enormously high increase in the transactions overload. As the result, the network started to fork, which on the other hand, caused increased memory consumption, making some nodes go offline. 
    The attack was compared to an attack on Ethereum from 2016 \cite{sol2}. 
    However, in the case of Ethereum, the production of blocks was not stopped, in contrast to Solana, where it did.
    To recover after the attack,  Solana's network had to be restarted, and the fact that this process was very well coordinated might indicate that the network was strongly centralized at the time of the attack \cite{sol1}. The problems caused by the attacker have triggered discussions around the centralization of the cryptocurrency.  Many critical voices indicated that Solana's consensus mechanism \gls{poh} requires further research to uncover potential vulnerabilities \cite{solgrayscale}. For example,  Solana's protocol requires the selection of a leader among the validators nodes, thus for a short period of time, the selected node becomes a single point of failure, especially since the list of the next validators is known upfront \cite{yakovenko2018solana}.
    In December, the Solana network suffered another attack from high congestion. At first, a DDoS attack was suspected. However this time, the issue was not initiated by any malicious actor. Those assumptions were clarified by  Solana's co-founder explaining that the congestion was caused by an unusually high network activity due to the game SolChicks, which is built on Solana \cite{solDec}.

\paragraph{ChainLink.}
    ChainLink is a decentralized Oracle network that is allowing for the interaction of different chains and external data sources. It is built on Ethereum and provides a bridge for smart contracts and real-world use cases. At the end of August 2020, ChainLink became the target of a spam attack \cite{chainLinkAttack}. The attack lasted for a few hours and did not affect the performance of the network, which successfully resisted the attack. However, as  ChainLink is an Ethereum smart contract, during the spam nodes had to pay higher gas fees (Ethereum's fees for the smart contract execution). \cite{chainLinkCompensation} indicates that node operators lost in total around 335,000 USD due to increased fees during the attack and ChainLink needed to compensate those losses.
   
   \paragraph{EOS.}
  
   EOS blockchain is built to support large-scale applications and it was designed to work both in permissionless and permissioned setting. EOS offers transactions without fees, as it pays node operators from the tokens minted by the inflation. Also, the available resources for transaction processing are connected to the amount of staked funds. In \cite{xu2018eos} authors performed an in-depth study regarding the EOS architecture, security, and performance. 
   Results indicate that EOS does not achieve claimed goals as a blockchain system and it may rather be considered as a distributed database system. 
   Due to its design EOS can be susceptible to spam attacks targeting a certain type of user. In January 2019 the network was under a transaction congestion attack. EOS allows for deferred transactions. In this process transactions  are prepared before they are issued at a specified time. Deferred transactions are prioritized over user-signed transactions. 
   If the attacker issues large amounts of transactions in dead loops, he can cause over-congestion in the network and use up all node's CPUs resources resulting in paralyzing targeted EOS application. Attacks exploiting transaction congestion were repeated many times and caused the loss of thousands of EOS \cite{EOSattacks}.

    \subsubsection{Eclipse attack}
    \label{sec:eclipse}
    An eclipse attack is a well-known attack vector present in almost all distributed systems. In an Eclipse attack the adversary that acquired a number of nodes, attempts to convince an honest node to accept the adversary's nodes as its only neighbors, thus covering the access of a node to the honest part of the network. This allows the adversary to control the in- and out-message flow and provide it with an incorrect state, allowing for manipulation of the targeted node, cause denial-of-service, or even to double-spend confirmed transactions. \cite{singh2006eclipse}.
    The details about the attack depend on the system setting. 
    In \cite{wust2016ethereum} authors describe an Eclipse attack on an Ethereum network, that can lead to \Gls{dos} situations or even allow for double-spend. A vulnerability is presented, where a victim  accepts a longest chain that has a lower total difficulty than the main chain. Additionally authors found a bug in Ethereum's \Gls{pow} difficulty calculation. 
    
    Another example of an eclipse attack designed for a specific network is \cite{yves2018total}. An attack  on a bitcoin network is presented, which can cause a total eclipse of a bitcoin node. The attack was successfully performed in practice by the authors. During the attack, the victim was completely isolated, thus could be provided with an incorrect ledger state.
    
    The Coordicide version of IOTA uses a salt-based autopeering mechanism that serves as a defense mechanism for the eclipse attacks \cite{muller2021salt}. Nodes are periodically generating a public and private salt. After a salt update of a node, the node starts to look for new outbound connections, and at the same time accepts the incoming connections that have smaller scores (calculated based on the generated private salt value). This way the network topology will be constantly reorganized whenever any node's salt expires. The autopeering mechanism provides some level of protection against  eclipse attacks, as long as the salt updates will not happen synchronously across the network.
    
    \subsubsection{Routing attack}
    In \cite{perazzo2020analysis} authors indicate that routing attacks are not a well-researched topic for the IOTA protocol, as the existing studies usually focused on consensus or cryptography issues. They distinguish three types of attacks: address freeze, general denial of consensus, and targeted denial of consensus. The attack focuses on BGP hijacking. BGP is an internet routing protocol, and a malicious node can advertise the route that it does not have. It can cause the denial of service by dropping the intercepted traffic. Authors reconstructed an IOTA network topology and indicated that the attack can be harmful to the network due to the existence of a single point of failure, the Coordinator node. If the adversary intercepts the Coordinator node it could stop the confirmation of transactions in the whole network. Authors also referred to the post-Coordinator version of IOTA, which would be much more secure when fully decentralized. However, they indicate that there still could be some vulnerabilities, and propose to use secure channels for the communication between the nodes.

\subsection{Sybil attacks}
\label{sec:sybyil}
    Adversarial strategies known as Sybil attacks are commonly acknowledged in \Gls{dlt} systems. They are an inseparable part of all distributed systems, as shown in \cite{douceur2002sybil}. The network parties, with no initial knowledge about each other, are not able to trustfully establish distinct identities where one identity corresponds to only one party. In a centralized system, each party can be verified by a trusted authority that is a bridge between a real-live entity and an identity in the system. In such a system it is not possible for an entity to present more than one identity. This kind of approach is not possible in a distributed system. Thus, it is necessary to implement an effective Sybil protection mechanism that will stop any malicious actor from gaining more power in the network only by frauding false identities to gain an advantage over honest network users.

    Sybil attacks can be divided into subcategories for better understanding and to be able to respond with a correct defense mechanism for different types of an attack\cite{newsome2004sybil}. The subcategories most relevant in this work are:
    
    \begin{itemize}
        \item Routing --  a malicious actor can appear in multiple places in the network layer and peer honest network users while pretending to be multiple separate identities.
        \item Voting -- if there is no protection from Sybil attacks, then an adversary could take control over the voting mechanisms and gain an unfair advantage on the voting outcome simply by creating false identities. 
        \item Fair Resource Allocation -- whenever there are limited resources it is a common approach to design rules on how they will be allocated. If the resources are distributed per identity it leaves the open window for an adversary to obtain an unfair proportion of the network resources.
        \item Misbehaviour detection -- if the network uses any mechanism for misbehavior detection, any malicious actor could exonerate itself, or even force accusation onto a legitimate user.
    \end{itemize}

\subsubsection{Unsuccessful Sybil attack on Monero}
In November 2020 the Monero project leader confirmed an unsuccessful Sybil attack on Monero \cite{moneroSybilAttack}. An attacker created many node instances in an attempt to connect transactions to the IP address of the broadcasting node. The breached data was revealed publicly on a website\cite{moneroHackerWebsite}. However as explained by R. Spagni the attacker can not prove a link between the revealed user Monero transaction and corresponding IP address, as the attack was not large enough to be effective against Monero's defense mechanism Dandelion++. It is a broadcast system in which transactions are passed through several nodes before it is broadcasted to the network, in a way that intermediate nodes do not know if the node that provided them with the transaction is the sender.

    
\subsection{Byzantine fault and agreement failure}
The consensus mechanism is usually the core of any blockchain project. The ability to reach an agreement on the state of the network is indispensable for any system that preserves valuable data. Attacks that aims to disrupt the system's ability to come to an agreement are usually divided into two groups: 
\begin{enumerate}
    \item attack on liveness of the system to halt the consensus in a metastable state
    \item attack on the security of a system, spending the same funds twice or reverting previously considered final transactions
\end{enumerate}

\subsubsection{51\% attack}
The 51\% attack is applicable to  \Gls{pow} based blockchains in which an adversary that controls over half of the total hashing power of a system intends to fork a blockchain and perform a double spend~\cite{sayeed2019assessing}. This kind of attack usually is considered unlikely to happen as the cost to perform the attack typically   outweighs the gain. In this attack the adversary starts to secretly build a malicious version of a chain, and one the chain is long enough it is presented to the network. Honest nodes apply the longest chain rule and follow the chain proposed by the attacker.
The 51\% attack is also possible for \Gls{pos} systems, however, the cost may be even greater, as an attacker would need to own more than half of the targeted cryptocurrency. Therefore, the threat is potentially lower compared to \Gls{pow} systems.

In \cite{aponte202151,sayeed2019assessing} the problem of centralization of \Gls{pow} system is raised, where mining pools (groups of miners that combine their hashing power and split the reward proportionally) are in the possession of a large amount of hashing power in the network. E.g., in Bitcoin seven out of the ten biggest mining pools are based in China.

A large theft happened on the Bitcoin Gold network in May 2018. The double-spend amount was estimated to exceed \$18 million \cite{sayeed2019assessing,bitcoinGold}. Another double-spend attempt documented in \cite{bitcoinGold2} happened in January 2020. The deposit of stolen BTG coins was received on the Binance exchange after six-block confirmations and could be withdrawn after 12 confirmations. Since this incident, the number of required confirmations was increased up to 20.



\subsubsection{Stellar protocol security}
In April 2019 \cite{kim2019stellar} brought  attention  to the centralization of the Stellar network. Stellar is  a blockchain-based platform that uses \gls{fba} consensus mechanism, which differs from \Gls{pbft} because it allows for open membership as long as nodes are trusted by others.
The study demonstrates that the system is highly centralized, and that three nodes belong to the Stellar Foundation. Additionally, the system is susceptible to cascading failure, in which only two nodes need to crush to take down the entire network.

After the study has been revealed, Stellar researchers confirmed the centralization issue and assured that they are currently working on a solution.

Afterward, in May 2019 the Stellar network went down as too many new nodes were added to the system over a short period time. The network became unstable and two nodes crashed, causing the failure of the whole system and violating the liveness \cite{stellar}.

To bring the network online, the number of validators was increased from two to three, nevertheless, as the authors of \cite{kim2019stellar} state,  this does not solve the fundamental problem.


\subsection{Cryptography vulnerabilities}
Cryptographic structures are essential in every blockchain project. They often guarantee the security of the system. With the growing demand for improvements in privacy, scalability, and functionalities for the systems the used solutions introduce complexity. Along with system complexity, the chances of vulnerability being unnoticed also increases. In this Section, we introduce some examples of such security flaws in different projects.

\subsubsection{Ethereum direct piracy attack}
    SlowMint, the company focused on blockchain security, published a report describing the flaw in an Ethereum node's authentication mechanism that was exploited for two years (2016-2017) and led to \$20 million lost \cite{EthereumAPI}. 
    An attacker scanned the network to find nodes with RPC API ports enabled, then repeatedly tried to transfer the balance to the attacker's wallet. If the user happens to execute \texttt{unlockAccount} on his wallet, there is a time when he does not need to reenter the password. The attack will succeed if an attacker will try to transfer the balance during this duration period.
    The team performed an extensive network scan and discovered more than 10,000 Ethereum nodes exposing the RPC API, thus being at high risk of a direct piracy attack. 

\subsubsection{Monero and Zcash}

Most blockchain-based systems, despite hiding the user identity behind digital identities or addresses, do not guarantee privacy. Usually revealing only one of the used addresses might allow interested parties to find out other addresses owned by the user and track its activity over the network. To tackle this issue many privacy coins projects have been initialized, among them are Monero\cite{van2013cryptonote} and Zcash\cite{hopwood2016zcash}. Those projects aim to hide the users' activity and hide any information that could lead to revealing user identity.
There exist examples of attacks aiming for the deanonymization of the user. 

The ITM linkability attack explained in \cite{leto2020attacking} on the example of Zcash protocols is also applicable for Monero. Zcash Protocol is using shielded transactions that do not reveal the addresses of sender and receiver, nor the transferred amount. However, as the author indicates a large amount of metadata is produced for each transaction at the protocol level.
The ITM attack takes the transaction id and \textit{zaddrs}, which happened to be revealed publicly by users on social media websites. It allows for deanonymization of the \textit{zaddrs} and connecting them to social media accounts, IP addresses, and more data. The attack is costly as it needs a specific infrastructure, so it would not be performed by a standard user, but rather big companies or organizations.
Additionally to the attack description, the mitigation strategy \emph{Sietch} is proposed, which is already implemented in the Hush crypto coin.

Another deanonymizing attack was revealed by Stanford University in \cite{tramer2020remote}. The vulnerability was first exposed to the Monero's and Zcash's team to allow for fixing the issue before publishing the finding to the public\cite{remoteSideChannels}. The remote side-channel attack allowed to identify the node of the payee in an anonymous transaction. Which enabled the adversary to find out all transactions sent by a user, their IP address, and finally a user's diversified address. Additionally, the attack allowed to remotely crush any Zcash node for which the attacker knows the payment address. 

\subsubsection{Replaying attack on the legacy IOTA}

    In 2018 when the \gls{mcmc} algorithm was still used as a tip selection mechanism in the IOTA protocol, there was a need for transaction reattachments, which is the possibility to resend the same transaction again by selecting different parents. Due to the walk, there is a positive probability, that an honest non-conflicting transaction would never be chosen as a parent by other transactions, and thus would never be added to the ledger. If the transaction was still pending after several minutes, the user needed to repeat the attempt to include the transaction into the Tangle, by issuing a reattachment. 
    At that time IOTA was using base-3 numbers (ternary) and the quantum immune hash function Curl-P with a one-time Winternitz signature. In the Winternitz signature scheme transactions should not be issued twice, since with each repeated signature a part of the private key is revealed. Therefore, the official IOTA guidance was to never use the same address more than once. 
    Although this rule was officially supported in all officially available applications, there was still a possibility that a user could not follow this rule if he decided to interact with the network through the API client library. In  \cite{de2018break}, the authors show that it is possible to reuse an address as a user and that it is possible to exploit this by performing a replay attack.
   
\subsection{Exploits on the protocol structures}

\paragraph{Selfish mining in Tezos}
Tezos is a \Gls{pos} blockchain with a self-amending governance system that went live in June 2018. In 2020 the group of Harvard researchers discovered a selfish mining strategy that could allow for an attacker to gain rewards for a longer period if an attack will be undetected \cite{neuder2019selfish}. The reward gained from an attack is low compared to the staked funds that an attacker might lose if he will be caught. 
Also, the issue can be fixed, by introducing changes to the protocol without the need for a fork, through a governance structure. Therefore, the discovered strategy is not a significant threat for the Tezos network, however, it rose a discussion about the security of the \Gls{pos} systems and showed that research about the \Gls{pos} systems is still in the early stages and many vulnerabilities can still be exposed.

\paragraph{Parasite chain}

The \gls{pc} attack is a well-known and researched topic that was applicable in the old versions of the IOTA protocol. The \Gls{mcmc} \Gls{tsa} was introduced to counter the parasite chain attack. The PC attack falls into a category of double-spend attacks, when an attacker is aiming on spending funds more than once, by reverting the ledger state to some point in the past after certain funds have been confirmed and accepted. The ledger state forks and the previously confirmed transaction is considered invalid in favor of the issued later double-spending transaction. In the initial white paper version of IOTA this attack was theoretically possible in the older version of IOTA when the attacker had access to reasonable power resources. In the attack, he could secretly create a chain of transactions starting from the double-spending transaction and reveal it whenever the collected weights of transactions that reference the double-spend (the starting point of a parasite chain) outweighed the honest part of the Tangle. Practically this attack was not feasible, since a centralized node called the Coordinator is required to approve transactions before they get accepted. 
 The attack was first mentioned in the IOTA whitepaper. The practicability of the PC attack (without a Coordinator) was shown in \cite{cai2019parasite}. The IOTA development network was used for the experiment. The attacker was able to outpace the cumulative weight of the previously confirmed transaction and keep it in this state through the whole duration of the experiment, which was 35 minutes. However, as mentioned above, even if the double-spend transaction had a higher weight it remained unconfirmed as the network was guarded by the Coordinator, that would not approve this transaction.

\cite{penzkofer2020parasite} introduced a method to detect the PC attack based on the differences in the structure created by the attacker and the main part of the Tangle. The proposed solution was to measure the distance between distribution retrieved from the Tangle data and the expected distribution describing the number of direct approvers. 

\paragraph{Large weight and split attack on IOTA}

The Large weight attack, similarly to the \Gls{pc} attack, attempts  to build a special structure in the DAG consisting of transactions to perform a double-spending attempt. The attack is applicable for the first versions of IOTA as the attacker attempts to create support for the double-spending transaction by trying to out-weigh the main part of the Tangle containing the first transaction, which has been already confirmed. The double-spending transaction is  attached to the Tangle such that it is outside of the future cone of the conflicting transaction, to increase the weight of conflicting transactions separately.
In \cite{vries2019iota} the large weight attack was executed on one of IOTA test networks. The double-spending transaction gained more weight than the first conflicting transaction, however, the Coordinator did not approve the double sending transaction, therefore the attack did not succeed.

A split attack,  performed in \cite{brady2021dosing}, attempts to keep the network undecided as long as possible (until the time the next milestone will be issued and only one conflicting transaction will be selected). The goal is to make as many honest messages to be left behind when they attach to the losing message, thus increasing the orphanage in the network. The results indicated that the attacker was able to slow down the confirmations as the number of confirmed honest messages dropped to about 60\%.

Attacks involving collecting the weight by issuing messages on top of the conflicting transactions are no longer applicable in the IOTA network, as in the update to the Chrysalis version of the node software  the concept of the white flag has been introduced \cite{whiteFlagRFC}. White flag utilizes the advantage of having a single source of truth in form of a Coordinator node which allows for total ordering of the transactions. With a total order, all nodes in the network can select the same conflicting transaction which will be applied to the ledger state, and ignore all the others.

As the Coordicide version of IOTA introduced the concept of Mana, flooding the network with a large number of transactions will not be possible as the access to the network is guarded by the access Mana, except during  under-utilization periods (more on this in Section \ref{sec:orphanage}).

\section{Orphanage problem}

\label{sec:orphanage}
    In this section, the main topic of this thesis will be introduced: the orphanage attack. Additionally, the next Chapter \ref{ch:experiment} will cover the practical experiment and carry out the orphanage attack scenario.
    A specific definition of the orphanage problem might vary depending on the underlying ledger and the protocol rules. The name derives from 'orphan' that is referring to a completely valid message, or a block that is left behind in the ledger structure. In the DLT setting, this means that the message most probably will never become accepted by the network, i.e. added to the ledger. This is not a desirable phenomenon, especially from a user perspective, as it forces the user to resend a transaction that was rejected even though it was completely valid. A further concern is a waste of resources consumed during the preparation and attachment of a message to the ledger, such as the carried out \Gls{pow}.

    The orphanage problem is not IOTA specific only. Many other DLTs are facing or even accepting this problem as an inevitable necessity. For example in Bitcoin, each time when more than one block is created simultaneously and all of them attach to the same parent blocks a fork of the chain is created. From that event on, the miners will decide, by building upon the preferred chain, which chain will be considered valid. All forked chains that did lose the competition for the title of the longest chain, are no longer considered as a part of the blockchain and become orphaned. Hence, the energy used to solve the cryptographic puzzle is wasted.

    \subsection{IOTA and orphanage in the past}
    \label{sec:old-orphanage}
    In the first version of the IOTA protocol presented in \cite{popov2018tangle} the consensus algorithm responsible for conflict resolution was a random walk \Gls{mcmc}. Despite problems with the performance, in terms of time taken to resolve a conflict, there were other issues impacting user satisfaction and usability of the protocol. Whenever an honest user of the IOTA network wanted to send funds, he needed to create a transaction, send it, and wait for funds to be confirmed. However, it could happen that after 30 minutes, his transaction was still pending. The problem that caused this, from the user's point of view, inconvenient suspension, was how the tip selection algorithm and consensus mechanism worked. The \Gls{mcmc} selection algorithm was not only serving the purpose of selection which tips would be approved but also was reducing the probability for a successful double-spend, i.e., it served as a consensus mechanism. 
    New messages would commit their approval weight that was related to the invested PoW to the validity of all transactions in its past cone in the Tangle. 
    
    The idea is then that to overcome the main network a double-spending transaction would have to be placed on a part of the Tangle that has lower cumulative weight and is unlikely to be selected by the tip selection. This was protected through a combination of a random walk through the Tangle and protection of the message issuance by PoW. As a consequence, the attacker would require a significant hashing power and accumulate more weight than the rest part of the honest Tangle. As an additional initial protection mechanism, a central node was introduced that should protect the network in its infant steps. Transactions approved by the Coordinator were considered confirmed / final. The \Gls{mcmc} benefit of reducing the chances of the double-spend attack success came, however, with a disadvantage. 
    
    Since messages were made interdependent through their approval there was a probability that a transaction would never be picked up by a new message since it is attached to the Tangle in place that is disfavoured by the network. The behavior of the random walk used in \Gls{mcmc} algorithm was controlled by the $\alpha$ parameter. Increasing the $\alpha$ value increases the probability that the transaction with a large cumulative weight will be selected. Thus, it ensures that only the most reliable tips will be selected. This reduces the chances for a double-spend attack success.
    However, increasing  $\alpha$, increases the chance of an honest transaction being not selected, especially if it was placed on a part of the Tangle structure that is unfortunate enough to be avoided by the Random Walk. If during the creation of a new transaction, the algorithm will follow a path with a low probability, the chances of this new transaction being approved, and eventually confirmed will be equally low. Additionally, the chance of being selected decreases over time, as the further the transaction is from the end of the Tangle the less accumulated weight it has compared to the main part of the Tangle that has been already supported by newly added transactions. The further from the top of the DAG the transaction was attached, the less probable it was that it will ever be approved. Thus, some of the transactions that selected old or unfavoured tips could be left behind and never be approved. We call this pending transaction an orphan.
    Even though this kind of situation is unlikely to happen, it still can be the source of a major inconvenience for a normal user. 
    
    To determine the frequency of these events we need to determine with which probability valid and honest transactions are being orphaned. We call the frequency of these events the \textit{orphanage rate}.

    To solve this problem, many solutions were proposed. The simplest one was to find the best compromise by choosing the lowest $\alpha$ parameter that will provide enough security for given network throughput. However, this is challenging and a given parameter might not be optimal in all circumstances. Other solutions were presented in \cite{ferraro2019stability}, where modifications to the \Gls{tsa} were introduced.
    
    \subsection{Orphanage problem now}
    The \Gls{mcmc} tip selection algorithm, mentioned in the section above is no longer used in the IOTA protocol. The newest version of the protocol separates the \Gls{tsa} from the consensus mechanism. IOTA 2.0 is using the \gls{rurts} algorithm (Sec. \ref{sec:tsa}) that selects the tips uniformly from the tip set that is the collection of not yet approved messages, which is based on a node's local perception of the Tangle. 
    This has some positive consequences. The chances of the valid message for being orphaned in an idle network situation should be extremely low. Nevertheless, there is no guarantee that all the network participants follow the rules, and there exist certain strategies (for adversaries)  that can increase the orphanage rate, e.g. the orphanage attack that will be covered in the next section.
    
    In the IOTA 2.0 the separation of the consensus mechanism from the tip selection is achieved by the introduction of new references. More specifically, e.g., weak references are introduced that allow approving the transaction within a message without approving the past cone of the message. This way, even if a message selected the losing side of a conflict it can be still referenced by other messages.
    The \Gls{rurts} algorithm implied the time restriction on the allowed age of a message's parents. All messages that are referencing messages older than a maximum parent age $\zeta$ are considered invalid, so if the message will not be selected before the elapse of the time $\zeta$ since its issuance it will become an orphan. 
    
    By taking into consideration all the protocol changes introduced for  IOTA 2.0, we can state the orphan definition as follows:
    
    If a message age or age of all tips in the future cone of this message has surpassed the $\zeta$ time and has not been selected by any other message, this message and all messages in its future cone are \textit{orphaned}.
    
    Where \textit{age} is the time difference between the issuance time and the current time.
    
    \paragraph{The cause of the orphanage.}
    If the network is not overloaded, this situation should rarely happen, as each tip has a similar chance for being selected by \Gls{rurts} \Gls{tsa}. One possibility to increase the probability of an orphanage event is to significantly increase the size of the tip pool, as the probability of a tip being selected with \Gls{rurts} is inversely proportional to the tip pool size. 
    Usually, in a healthy environment and the presence of high network congestion, the increased width of the Tangle is a manifestation of the Tangle scalability characteristic, as the existence of multiple interconnected chains allows for parallelization of operations in the network.
    However, if the flow of transactions exceeds the processing capabilities of nodes, it can cause an increase in the tip pool size and creates the possibility of orphanage occurrence. This problem is controlled with the Congestion Control module, introduced in \cite{cullen2021access} that governs the allowed issuing rates for the network participants based on its aMana supplies. When the network allowed throughput is fully utilized the issuance rates are fairly distributed among the network participants, and without violation of the Sybil protection mechanism, flooding the network with too many messages is not possible.
    However, forbidding low Mana nodes to issue transactions even if a network is underutilized would be wasteful. Therefore, IOTA 2.0 allows the utilization of unused network resources and omits Sybil protection requirements, when there is not enough demand from the allowed parties. The long periods of a network under-utilization open up the possibility for an attacker to gain a huge proportion of the network throughput and by applying malicious strategies inflate the size of the tip pool set.
    
    Therefore, in addition to a bad user experience, orphanage can induce additional problems. An attacker can, by not contributing to the reduction of the tip pool set,  cause tip pool size inflation and affect the finalization time of messages. When the Tangle is wide, the whole data structure grows slower and more time is needed to collect enough \Gls{aw}.
        
    \subsection{Orphanage attack}
    \label{sec:orphanage-attack}
        The orphanage attack strategy takes advantage of the Tangle structure and the fact that the protocol allows to use of underutilized network throughput even if nodes do not possess much Mana, see the previous section. The goal of the attack is to damage the network and increase the orphanage rate for all messages. As discussed in the previous section, it can be achieved by caused by tip pool inflation. An attacker is not simply spamming, he is aiming to decrease the number of tips removed by his messages as much as possible while following the protocol rules. 
        As each node can have a different Tangle perception due to delays present in the network, it is not possible to validate if a given node is following the default \Gls{tsa}, which is \Gls{rurts} in the case of IOTA 2.0. Attaching to older messages could be caused by clock or synchronization issues and not necessarily means malicious intentions. The protocol imposes that any new message can not reference a message in the past that is older than the time duration specified by the protocol - the $\zeta$ parameter.  As long as the attacker chooses messages within this time range, his behavior will be accepted by other nodes, and messages issued by the adversary will be processed. 
            
        The orphanage strategy is as follows:
        \begin{enumerate}
            \item Select the oldest possible messages.
            \item If possible select only own tips.
            \item Use a minimum number of allowed message references.
        \end{enumerate}
        
        The first point is responsible for most of the damage caused during an attack, as by selecting the messages closest to the maximum parent age an attacker is maximizing the possibility of referencing a message that has been already selected by others, this is not contributing to the reduction of the tip pool size.
        The second point reduces the chances of an honest message being selected, and the third one minimizes the tip pool size reduction. Nevertheless, the situations, when the adversary benefits by applying points 2) and 3), will not happen too often during an attack, because the attacker attaches to old messages that would be removed from the tip pool a moment after.

\begin{figure}
            \centering
            \includegraphics[width=\linewidth]{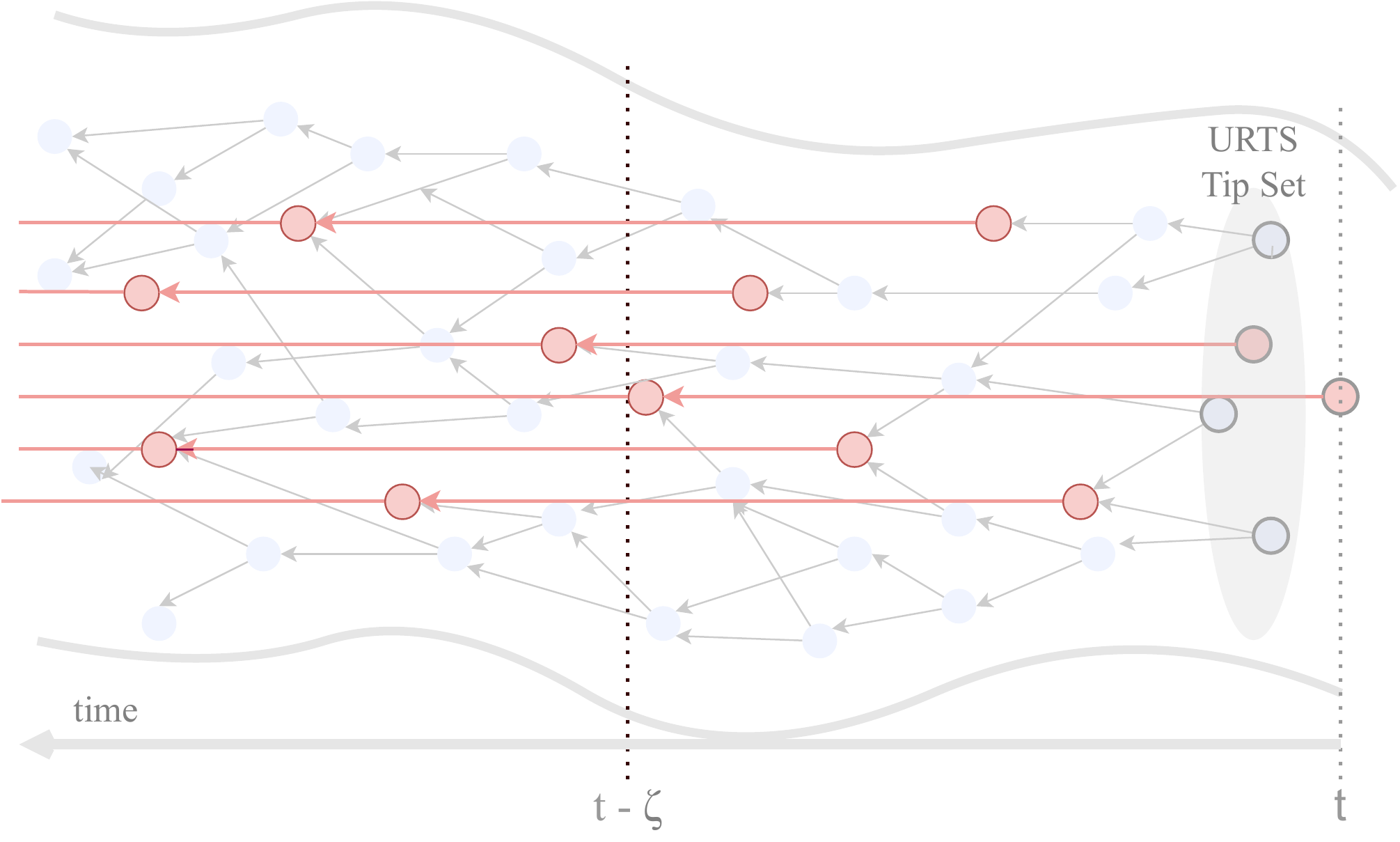}
            \caption{Illustration of the orphanage attack. Adversary messages marked in red, honest messages in blue. At the time $t$ an attacker is adding a new message to the Tangle and references his own, oldest possible message, despite being completely aware of the URTS tip pool set he should use.}
            \label{img:orphanage-attack}
        \end{figure}

        When the attack starts, an adversary can simply attach to his own, same message until the message's age will get close to maximum parent age $\zeta$. After this time, the attacker keeps attaching to his own oldest, yet valid message. Attack is presented in Figure \ref{img:orphanage-attack}. The structure of the Tangle changed, as the attacker references create long connections. It is visible as the whole structure is supported only with honest messages. The visualized network is not overwhelmed by the adversary, as honest nodes keep approving all messages, despite additional adversarial tips.
        However, if the attacker increases the spamming rate, the honest nodes might no longer be able to keep the tip pool size under control.

        \subsubsection{Attack consequences}
        The majority of nodes, that follow the default \Gls{tsa} will select the newest messages, that from the given node's local view have not been referenced yet. As the local perception of the network state differs among the nodes, it may happen that a  certain message will be selected more than once. 
        
        As proven in \cite{penzkofer2020parasite} the average number of references that message received is equal to $k$ for high network throughput confirming the intuition that each honest message will remove $k-1$ messages.
        
        The ability of honest nodes to reduce the size of the tip pool depends on the allowed $k$ parameter, which represents how many tips can be removed from the tips set while adding a new message, which in turn contributes to the speed of the tip set reduction.
        When the number of messages exchanged in the network does not change in time and the network is attacked by an adversary aiming to inflate the tip pool, the $k$ parameter dictates the ability of the honest part of the network to counter the spam. However, when the proportion of adversary messages within the total network throughput $q$ increases, honest nodes at a certain point would be no longer able to keep up with selecting messages and the attack will succeed.
        
        \subsubsection{Critical point}
        \label{sec:crit-point}
        Assume the total network throughput does not change in time and all nodes in the network keep sending messages at rate $\lambda$. The adversary spam rate $\lambda_A$ and honest nodes spamming rate $\lambda_H$ are distinguished. The proportion of an adversary spamming rate $\lambda_A$ to the total rate is called $q=\lambda_A/\lambda$. The number of messages referenced by honest nodes is denoted by $k$. 
        The critical point for when the above becomes an issue depends on the number of parents $k$ that each honest message selects. 
        To provide an estimate for the critical point $q_{crit}$ we apply the following simplifications:
        \begin{enumerate} 
            \item The current state of the tip pool set is globally known to all nodes in the network.
            \item Each honest node reduces the tip pool size by $k-1$.
            \item Honest nodes always select $k$  number of unique parents.
        \end{enumerate}
        
        The first assumption ignores the fact that information flow in a real network is delayed by the network connection and processing delays, and that messages are added to the Tangle in parallel. In a real network environment it is common that due to network delays, one message might be selected by more than one node. As a result of assumption 1) each node immediately knows that a message was referenced by other nodes, thus from 1) we obtain 2), i.e., every honest reference removes exactly one tip. Furthermore, as we will see in \ref{sec:experiment-resuts} the obtained equation for the critical value agrees well with experimental results.

        The critical point $q$ happens when the number of tips introduced by both honest and adversary nodes is equal to a number of tips removed from the tip set. Due to the orphanage attack strategy,  an adversary does not contribute to the reduction of the tip pool set.
        Additionally, by applying the second assumption that each honest node decreases the tip pool size by $k-1$ it can be denoted:
        $$
        \lambda_A = (k-1)\lambda_H
        $$
        From $q=\lambda_A/\lambda$ we can substitute $\lambda_A$ with $q\lambda_H/(1-q)$. From this we obtain a   critical value  for which the tip pool becomes inflatory, thereafter:
        \begin{equation}
            q_{crit}=\frac{k-1}{k}
        \end{equation}
        
        \subsubsection{Increase after the critical point}
        \label{sec:above-crit}
        To predict the behavior of the tip pool size above the critical point we can extend the assumptions introduced in the previous section by adding: the time is divided into intervals with length $t_{i+1} - t_i = h$ which represents the delay, caused by the time needed to do \Gls{pow} to process message and the network delay. 
        As proven in the \cite{penzkofer2020parasite} the average number of approvers is equal $k$ and the tip pool size at time $t_i$ is equal $L_i = k\lambda + 1$.
        By following the same line of reasoning, to derive the formula that will describe the tip pool size increase during an orphanage attack, the tip pool size increase during one interval $h$ can be written as:
        \begin{equation*}
            L_{\Delta} = \lambda_A - \lambda_H(k-1)
        \end{equation*}
        where $\lambda_A$ and $\lambda_H$ are respectively the adversary and honest spamming rates. By assuming the initial condition $L_0=1$ it can be deduced that
        \begin{equation*}
            L_i = i(\lambda_A - \lambda_h(k-1)) + 1
        \end{equation*}
        After replacing spamming rates with the proportion of adversary spam $q=\frac{\lambda_A}{\lambda}$ the tip pool size at time $t$ can be described with
        \begin{equation}
            L_i = i(\lambda q - \lambda(1-q)(k-1)) + 1
        \end{equation}
        
        \subsubsection{Confirmed orphans problem}
        \label{sec:confirmed-orphans}
        Three major consequences of the orphanage attack introduced in the previous sections were connected to the tip pool size inflation that causes both confirmation time increase and increase of the orphanage rate. The orphaned messages created during an orphanage attack usually will have no children or the orphaned chain will be rather short, because the increase in the orphanage rate is caused by the tip pool size inflation and what usually will be orphaned are tips. Since these messages will not be confirmed,  their rejection will not cause any problems within the ledger state. 
        
\begin{figure}
            \centering
            \includegraphics[width=\linewidth]{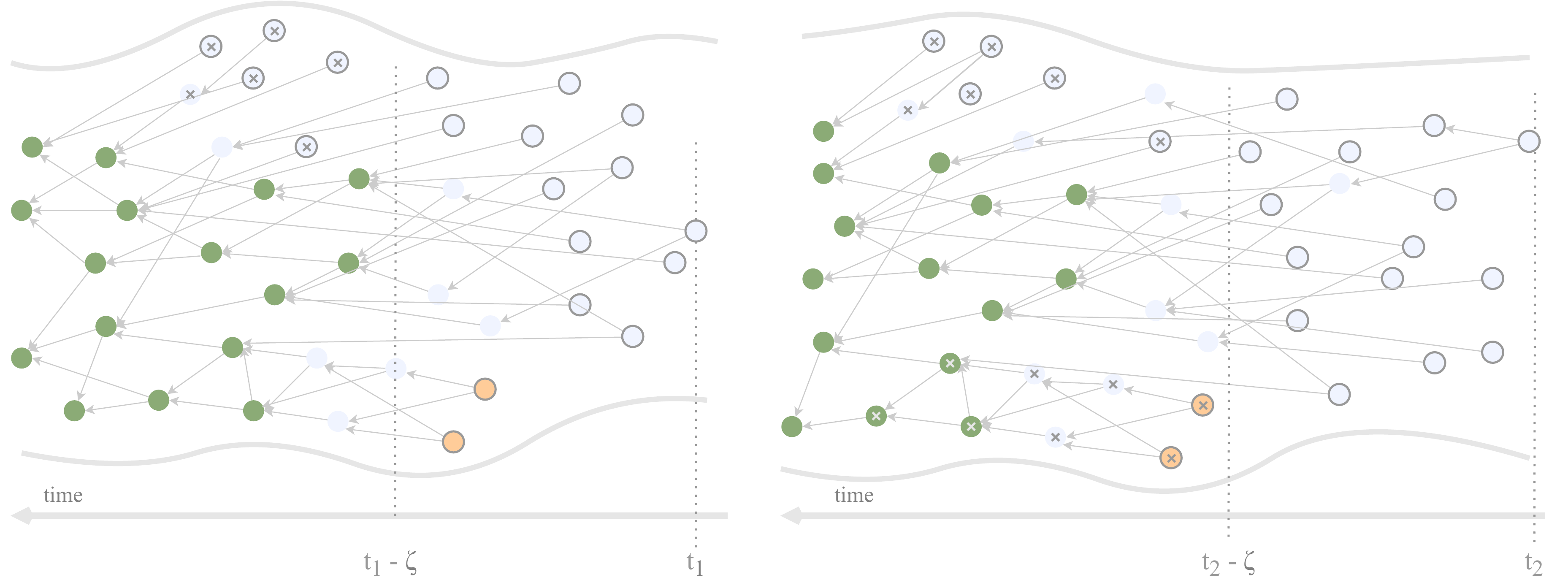}
            \caption{The Tangle during an orphanage attack. Confirmed messages are in green, the gray border indicates the tip. Cross means that the message is orphaned. On the left, two orange messages have confirmed messages in their past cone. However, they are only two tips in the future cone of those confirmed messages. For the situation on the left, there is still a chance for the side Tangle that ends in orange messages to be not rejected by the network, as the age of the orange messages is $<\zeta$. In the left Figure, the sub-tangle with orange messages is orphaned, along with confirmed messages.}
            \label{img:confirmed-orphans}
        \end{figure}

        However, one can consider situations in which the network is rejecting confirmed messages. This situation is visualized in Figure \ref{img:confirmed-orphans}. It can happen when certain messages will collect enough \Gls{aw} and get confirmed, however, their future cone will consist of a sub-tangle of which the tips become orphaned. Whenever those tips will cross the $\zeta$ age check, the whole past cone of those orphaned messages is also rejected by the network, as it is not possible to collect more AW.
        
        The orphanage of the entire sub-tangle can introduce inconsistencies in the ledger, which can lead to consensus problems.
        Specifically, ledger state inconsistencies could appear. The confirmed and orphaned messages are no longer reachable during the walk from the growing end of the Tangle. The only way to find them would be to walk forward starting from Genesis or some old checkpoint. However, at the same time, the unspent output within those transactions will be considered valid, as the transactions are considered confirmed, and there are no contradictions from the UTXO DAG perspective to use them in the new transaction. At this point, different states on the Tangle DAG structure and the UTXO DAG structure can be obtained. There is a requirement for the unspent output that it has to be in the past cone of the transaction that tries to spend it. When the confirmed and orphaned output will be used, the past cone check could never be fulfilled, as the output is no longer reachable while performing a backward search algorithm in the  Tangle.
        Note that in the current prototype implementation, the past cone check is not performed due to performance reasons. Thus the implementation has very strong assumptions about the ledger consistency and requires that the divergence between ledger of different nodes will not happen. As each node should be able to retrieve all messages (sync) by walking the Tangle starting from the tips and requesting missing messages through the solidification process. However, when considering the above situation with confirmed orphans, the syncing node would never know about those transactions as they are not reachable by a simple backward search algorithm. Consequently, we end up in a situation where two nodes have different ledger state perceptions. 
        
        A ledger inconsistency has also impacted other components apart from the balance states. In particular, its derivative -- Mana -- can be used as a Sybil protection mechanism, as is the case in the prototype software.
        Many components rely on Mana as a Sybil protection. For example, the consensus mechanism and the rate control. 
        Nodes that have an incorrect view on the Mana vector have an outdated view on \Gls{aw} collected by each conflict branch, which can result in voting for a losing branch. The Congestion Control component which also uses Mana is responsible for controlling the message flow in the network. A Mana inconsistency can lead to punishing honest nodes by not gossiping their messages at the rate that would available if the node knew about those confirmed and orphaned transactions. 
All of this would be impossible to notice by the node that has an incorrect ledger state, as from his point of view he solidified all messages. 
        
    \subsection{Blow Ball attack}
    \label{sec:blowball}
    The blowball attack is a variation of the orphanage attack. Its implications are similar to the ones from an orphanage attack, however, those results are achieved by different spamming strategies and by building different structures with spamming messages. An attacker instead of issuing messages at a constant, maximum possible rate is creating the 'ball', which is a collection of messages that reference only messages that are part of this blowball and usually are attached to the Tangle via a single reference from the message that ties all messages in the blowball. This kind of structure is prepared upfront, and all messages from the blowball are issued at once. Therefore, the adversary messages join the network with bursts, causing sudden tip pool size increase, and are followed by an idle period of only honest activity, up to the time when an attacker issues the next blowball.
    
    The blowball can be attached to the last message of the previous blowball, to avoid selecting any honest message. In this case, the time difference between two consecutive blow balls should not exceed the maximum parent age $\zeta$.
    
        \begin{figure}
            \centering
            \includegraphics[width=\linewidth]{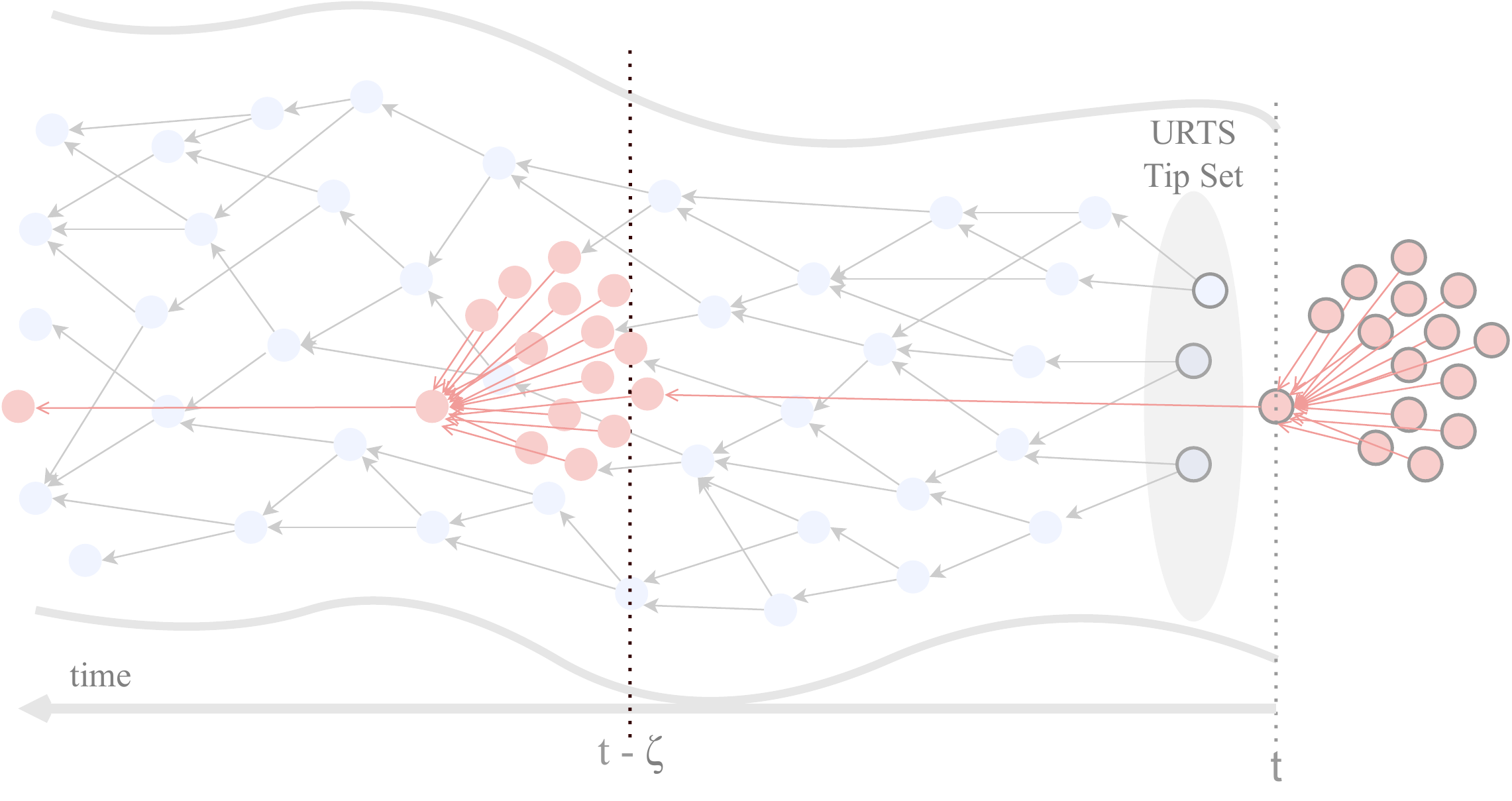}
            \caption{Depiction of the blow ball attack. The adversary messages are marked in red, honest messages in blue. At time $t$ an attacker is adding a new blow ball to the Tangle and references a message from the previous blow ball. Now he starts to prepare the next blow ball, which will be issued as soon as the last one start passing the $\zeta$ age.}
            \label{img:blowball-attack}
        \end{figure}

\chapter{Experimental validation of the orphanage attack}
    
    \label{ch:experiment}
    \section{Experiment setup}
     In the previous Chapter \ref{ch:attacks} many adversarial scenarios were introduced, starting from real-world examples to the theoretical discussion about the security of the protocols, and possible ways to damage the distributed system. One important threat for the IOTA DLT is the orphanage attack, see Section \ref{sec:orphanage-attack}. Although relevant for all iterations of the IOTA protocol, this is the first thorough investigation of this topic.  In particular, for IOTA 2.0, poses a threat to the data structure during periods of under-utilization, when the adversary is more likely to be able to gain a high portion of the network throughput. More precisely, since vacant throughput can be occupied at no cost an adversary can cause a real threat to the \Gls{dlt} with, potentially, very little cost and effort \cite{cullen2021access}.
    
    It is common to investigate possible security flaws from a theoretical side, to cover edge cases for attacks, even though practically these attacks may be hard to achieve. Therefore many proposed attack scenarios require from an adversary:
    \begin{itemize}
    \item an exceptional knowledge and skills,
    \item access to hardware resources greatly exceeding one of an average user,
    \item gaining a significant amount of control over the communication layer.
    \end{itemize}
    
    This is not the case for the orphanage attack. Its simplicity imposes a major threat for the system since it can be imposed without an adversary needing to ramp up its capabilities. There are no significant costs related to the transaction fees or any other additional on-ledger payments. More specifically the resource limitation  \Gls{pow} is negligible in IOTA 2.0 and unused bandwidth is free. Furthermore, even in the Chrysalis network, where PoW would be noticeably increased, compared to IOTA 2.0, an adversary may obtain the necessary specialized hardware to overcome a given \Gls{pow} spam protection. 
    To support those assumptions, the orphanage attack is performed in a form of experiments aiming to measure the impact and damage it can impose on the network. To perform the malicious scenario the prototype node software is used. As such the experiment is reflecting well how the \Gls{dlt} would be affected by the orphanage attack. The experiment is performed under different conditions and with a different setup of network prototype parameters.
    The following sections introduce the technical details of the experiment, tested parameters, and measured network characteristics. Finally, the results of the experiment are presented and discussed.

    \subsection{Technical details}
    
    The experiment is set up in a virtual environment, with  Docker -- an open-source project allowing for automatic deployments of self-sufficient virtual environments within containers. The container abstracts the program dependencies, the underlying system, and access to storage, processing power, and other resources thus allow to run applications fast and easily, regardless of the platform we use. Additionally, the Docker Compose application is used to create a running multi-container network with multiple GoShimmer node instances. It allows for performing an experiment on a single machine, and at the same time using a full version of a node software to create an artificial network with close to real-world network behavior. Performing an experiment within such a controlled environment brings benefits, such as being in control of spamming rates and behavior of all network instances, and the experiment environment can be set up quickly with support for tasks automation. Also, changing any of the network parameters is not problematic, as all necessary parameters are exposed as program flags. Most importantly, `docker-compose` allows for defining the initial setup of the network instances. 
        
        The local network created for the experiment is based on the `docker-network` tool available in the GoShimmer repository\footnote{https://github.com/iotaledger/goshimmer/tree/develop/tools/docker-network}. The tool provides a `docker-compose.yml` file, that describes the base setup of the network. All network nodes are running with the 8.3 version of GoShimmer. 
        The created network consists of seven honest node instances, where their behavior does not deviate from the default prototype implementation. The eighth node is representing an attacker instance and follows the modified version of the prototype software. The introduced changes follow the orphanage attack strategy. The adversary node's codebase has been modified to accommodate a mode in which nodes perform an alternative operation, during which malicious actions are executed. The node's software modifications and all necessary measure tools are implemented in a forked Gohimmer GitHub repository \footnote{https://github.com/daria305/goshimmer-orphanage}.

        \paragraph{Network setup.}
        For this experiment the following default properties of the prototype software are altered:
        \begin{itemize}
            \item Disabled activity messages. The message flow in the network can be controlled with a message spammer via \gls{api} calls.
            \item Disabled faucet plugin. The experiment is performed only with data messages.
            \item Enabled metrics collection for all nodes. All nodes are connected to the data collection dashboard.
            \item Identity and private seed for each node. It enables the allocation of aMana and cMana.
        \end{itemize}
         
        \paragraph{Snapshot.}
        The default Mana distribution recorded in the snapshot file is divided only between two nodes, thus is heavily centralized. After assigning identities to all honest nodes the new custom snapshot is created. For the purpose of this experiment the aMana, see Section \ref{sec:mana}, is distributed equally among all nodes to allow for undisturbed spamming. The cMana is pledged equally to honest nodes with an exception for the faucet node that received all cMana from the adversary node. In this way, only honest nodes are contributing to the message confirmations, as messages issued by the adversary do not have any impact on the approval weight collection. This represents the worst-case scenario, as slowed down \Gls{aw} collection in combination with wide Tangle results in increased confirmation times. Also, all nodes have an equally large portion of aMana. Thus the message exchange during an attack is not interrupted by the Congestion Control module.
        
        \paragraph{Data collection.}
        For the collection of data, we use a Grafana dashboard that is configured for the prototype software. Grafana is an application that provides a graphical interface and facilities the presentation of the collected metrics and data. The dashboard is fed by the Prometheus server set up along with the Docker network.  Prometheus is a database that collects data from each configured node that is connected to it and then sends the data to Grafana. For the experiment, the metric collections were expanded to gather the data from all nodes, and some of the data views were modified to allow the collection of:
            \begin{itemize}
                \item tip pool size changes over time for each node instance,
                \item confirmation events count for each node,
                \item issuing rates for each node.
            \end{itemize}
         The first two metrics are used for the analysis of the experiment results. The third is collected for data processing purposes only, thus will not be visualized.
         
         Finally, we measure also the orphanage rate. To calculate this measure, an approximation has been adopted to identify which messages are orphaned: we walk the Tangle starting from the Genesis, and check each message older than maximum parent age $\zeta$. The implemented method counts orphans based on their confirmation status, which is a sufficient approximation for this research. Nevertheless, as stated in Section \ref{sec:confirmed-orphans}  exceptions can occur and there might be some messages that are both orphaned and confirmed. However, for the sake of simplicity, we omit this special case. The walk and orphanage metrics collection can be triggered for each node via an API call after a part of an experiment has finished.

        \subsection{Plan of the experiment.}
        
 The experiment is divided into three major parts:
        \begin{enumerate}
            \item Time restriction on tips ($\zeta$=60s), variation of $q$ and $k$.
            \item No time restriction on tips ($\zeta$ is greater than experiment duration), variation of $q$ and $k$.
            \item Variation of time restriction length $\zeta$ for fixed $q_crit$ and $k$.
        \end{enumerate}
    
    \subsubsection{Orphanage with maximum parent age restricted}
   The first part of the experiment is fairly aligned with the Coordicide specification\footnote{https://github.com/iotaledger/IOTA-2.0-Research-Specifications}.  As mentioned in \ref{sec:tsa} IOTA 2.0 uses the \Gls{rurts} tip selection algorithm, which is also implemented in the prototype software. In this algorithm, tips are selected at random from the available tips, with a restriction on the maximum time since issuance. The default value for the maximum allowed age $\zeta$ is 30 minutes. For the purpose of this experiment, all data is collected for $\zeta$=1 minute. The maximum parent age period is shortened to allow for measuring the network characteristics in a reasonable time when an attack lasts for more than one $\zeta$ period. This allows the stabilization of the experiment. 
        
    In an initial phase of the experiment the  docker network is set up with  $$k=\{2, 4, 8, 16\}$$ and $\zeta$=1 minute. Next, there are spamming periods when an orphanage attack is performed for $12\cdot \zeta$ duration. Each attack period has a different proportion $q$ of adversarial messages within the total network throughput that is set to 50 \gls{mps}:
    \begin{equation}
    \begin{split}
        q_{k=2} &= [0.35\quad 0.4\quad 0.45\quad \textbf{0.5}\quad 0.55] \\
        q_{k=4} &= [0.6\quad 0.65\quad 0.7\quad \textbf{0.75}\quad 0.80] \\
        q_{k=8} &= [0.7\quad 0.75\quad 0.8\quad \textbf{0.88}\quad 0.93] \\
        q_{k=16} &= [0.8\quad 0.85\quad 0.9\quad \textbf{0.94}\quad 0.99] \\
    \end{split}
    \end{equation}
    
\begin{figure}
        \centering
        \includegraphics[width=\linewidth]{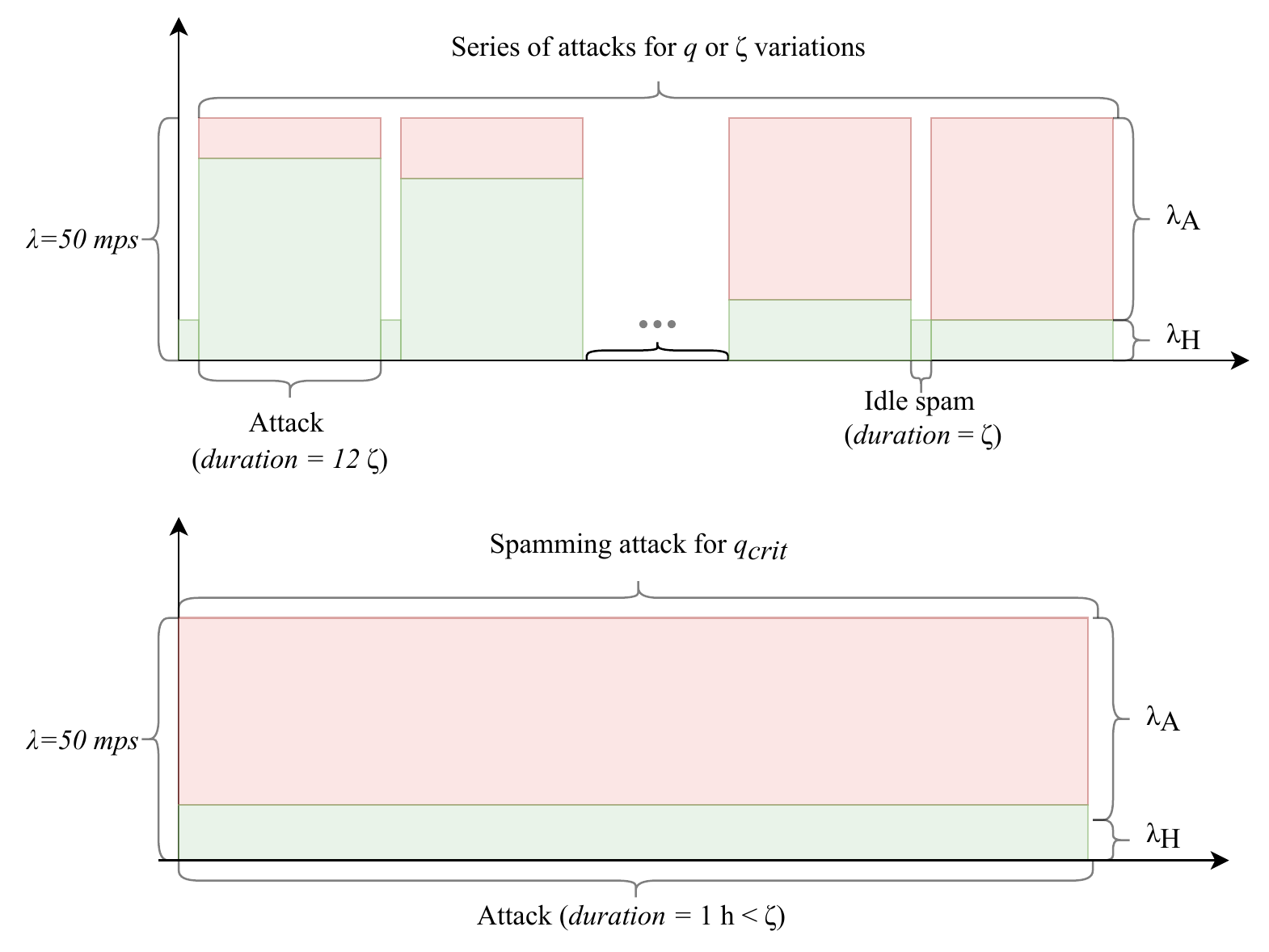}
        \caption{Experiment setup. The diagram illustrates the experiment methodology for a chosen $k$. The top diagram reflects scenario 1) (fixed max parent age, and $q$ variations). With each attack period, $q$ is increased. All attack periods are followed by an idle spam period with only honest messages. In scenario 2) (no time restriction) spam duration is increased and only $q=q_{crit}$ is investigated.}
        \label{img:exp-setup}
    \end{figure}

    All attacks are separated with idle spamming periods when only honest nodes are issuing activity messages. It is necessary to allow the network to recover and each node to reduce the tip pool size. The total throughput is set to be approximately constant during each attack spamming period, i.e. $\lambda_H+\lambda_A=const$. The experiment starts with the smallest proportion $q$, which is increased with each next attack period. We investigate the $q$ rates up until a value slightly above the critical point. 
The flow of the experiment for a particular $k$ is presented in the top diagram of Figure \ref{img:exp-setup}.     The red and green volumes show the issuance rate of the adversary and honest nodes, respectively. 
    
\subsubsection{No restriction on parent age}
    The second part of the experiment is investigating the tip pool size and confirmation times increase if there is no  $\zeta$ restriction on the maximum parents' age. The experiment is run for 1 hour with no restriction on the parents' age. This setup is used only for $q_{crit}$. The experiment is shown in the bottom diagram of Figure \ref{img:exp-setup}. 
    
    \subsubsection{Impact of the maximum parent age restriction on the orphanage}
    In the last part, we investigate how the $\zeta$ parameter changes the orphanage results. We keep  $k$ and $q$ fixed during the experiment.
     
    \subsection{Adversarial behavior}
    To implement the orphanage attack strategy, the Tip Manager component of the  GoShimmer software has been modified. As described in Section \ref{sec:tsa}, honest nodes keep track of the current set of available tips. Whenever there is a new fully processed and correct message it is added to the tip pool. At the same time, there is a timeout set to remove the tip from the tip pool if it gets too old according to the maximum parent age parameter $\zeta$. Additionally, the node is removing tips from the tip pool whenever any other message referencing it is added to the tip pool. 
        The adversary aims to:
        \begin{itemize}
            \item reference only its messages 
            \item reference messages that are closest to the $\zeta$ limit
            \item  reference the least possible amount of parents allowed by the protocol to minimize the tip reduction
        \end{itemize}
        There are many ways in which the malicious version of the software can be implemented, to achieve the goals stated above. The strategies used for this experiment can be summarized accordingly: 
        \begin{itemize}
            \item adversary is no longer removing referenced tips from the tip pools even if he knows that a message has been already referenced,
            \item tips are tracked in two sets -- adversary and honest -- of which one is only for adversary's messages,
            \item the adversary is using the honest set only when tips are missing in his own tip set
            \item because tips are removed only when they are too old, the number of tips the adversary keeps in its memory is much higher than for honest node, the malicious tip pool size is limited by deleting more recent messages with a higher probability. 
        \end{itemize}
        
        Most of the codebase changes for the purpose of an orphanage attack affect the Tip Manager component. The default honest behavior of the component has been mocked, so the base implementation can be replaced whenever the adversary mode is enabled. The mode can be enabled through a node's configuration file by providing flag parameter \texttt{adversary.orphanageEnabled=true} Thus, all the network nodes within the docker network can be created from the same GoShimmer image, and only the change of one parameter is needed to introduce a malicious actor to the network. 

\begin{figure}
            \centering
            \includegraphics[width=\linewidth]{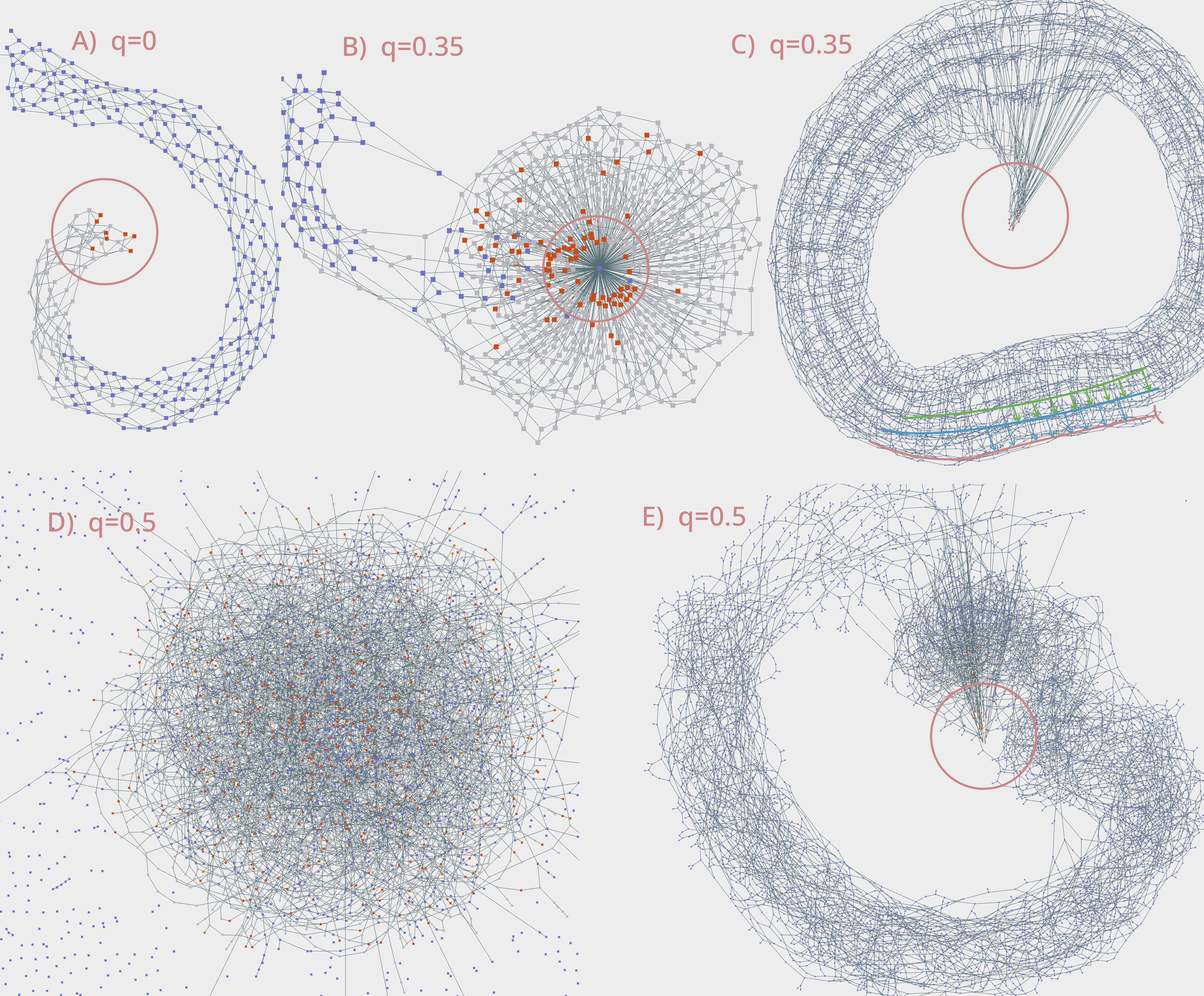}
            \caption{Collections of screenshots of the Tangle visualisation tool for different stages of an orphanage attack, for $k=2$. The situations in A) represents an honest scenario with $q=0$, B) and C) are taken for $q=0.35$, while in  D) and E)  $q=q_{crit}=0.5$.}
            \label{fig:screenshot-attack}
        \end{figure}

        The impact of the described above adversarial strategy on the graph structure of the Tangle depends on the attack duration and the adversary spam proportion. In an initial phase, when the attack duration is less than $\zeta$, the adversary keeps attaching to the same (his oldest) message as none of his messages are outdated yet. We show an example for such a case in Figure \ref{fig:screenshot-attack}, which shows several screenshots from the Tangle visualizer tool. 
        
        Situation A) shows how the Tangle looks in an honest environment. 
        B) shows the beginning of an attack up until the $\zeta$ time is reached. 
        C) presents the Tangle in the middle of an attack, where the adversary messages (new messages are marked with the red circle) are attached to old messages, thus creating connections to the older parts of the Tangle. 
        As  $q$ is less than the critical value for A), B) and C) all messages are timely confirmed and the width of the Tangle is small. The visualizer used to present the Tangle structures created during an experiment has a vertex limit. When the limit is reached the oldest vertices are removed. The visible end of the tangle was marked with a red line in C). Additionally added arrows point to how the Tangle structure is connected to the Tangle part about $\zeta$ time in the past, which are the attacker connections to the oldest possible messages. The whole structure is a circle due to the physic of the visualization tool.
        
        D) and E) show how the situation changes when the adversary throughput proportion achieves the critical value $q=q_{crit}=0.5$. Case D)  shows many red (tips) and grey messages (pending), far outnumbering the blue ones (confirmed), the structure is not visible as at the beginning of an attack the adversary attaches to only one message. In E) the maximum parent age is contributing to the reduction of the tip pool size. Thus a growing structure of the DAG can be observed. Note that the  Tangle in E) is visibly wider than in C).
  
    \section{Results}
\label{sec:experiment-resuts}
    In this section, the outcome of an orphanage attack adversarial strategy is described. Many theoretical assumptions introduced in \ref{sec:orphanage} have been confirmed in practice. Additionally, some of the performed attacks have to lead to the crashing of nodes, thus making the network unusable. This, however, is mostly connected to implementation details, and the robustness of the software might increase with future versions. In the following sections analysis of the results is presented. For the complete visualization of all gathered data, we refer to the Appendix \ref{appendix}.
    
    \subsection{Restriction on the parents' age.}
    The orphanage attack exploits the structure of the Tangle and the fact that each node tracks tips locally. With that, an attacker can cause serious damage to the network, while not breaking any protocol rules. In the first series of experiments, we check how the network will behave under an ongoing attack. 
    
\begin{figure}{h}
        \centering
        \includegraphics[width=\linewidth]{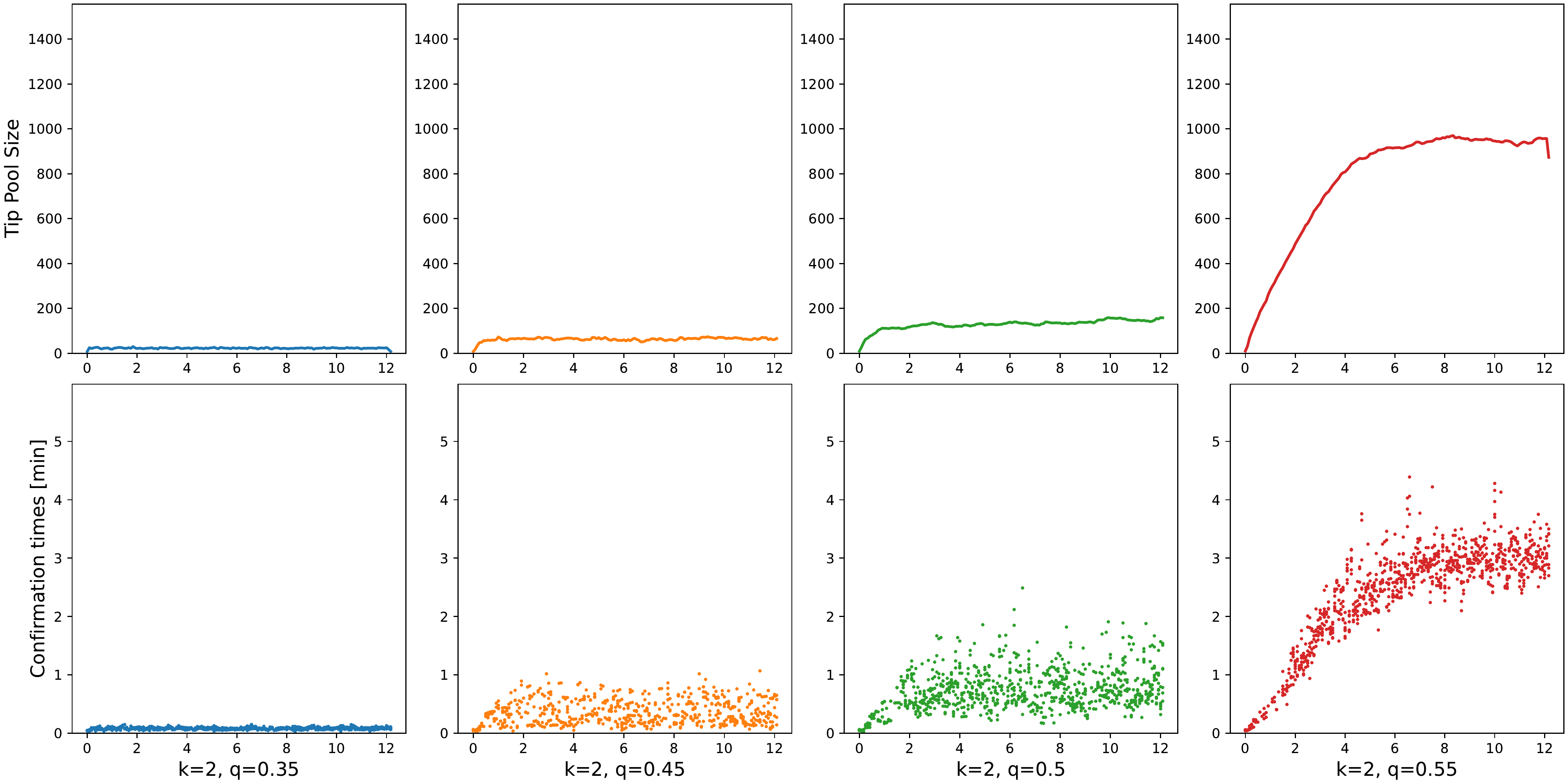}
        \caption{The average tip pool sizes (top subplots) and finalization times (bottom subplots) for $k=2$, measured for all honest nodes during separate experiments with different adversarial spam proportions $q$. The third column shows the critical value $q_{crit}=0.5$. The network setup with parent age restriction $\zeta$=1[min]. The total network throughput $\lambda=50$mps. The x-axis represents the duration of the experiment in minutes.}
        \label{fig:graf-vary-q-summary}
    \end{figure}

    Figure \ref{fig:graf-vary-q-summary} shows the attack impact on the network for $k=2$ and how it is changing when the adversary message proportion increases. The parent age restriction is set to $\zeta = 1[\text{min}]$ and the attack continued over the $12\zeta$ time period. 
    The critical point should be the value that represents when the attack starts to be harmful to the network. Indeed the tip pool noticeably grows significant around and even before $q_{crit}$. The increase of the tip pool is not a problem by itself, however along with tip pool inflation, the finalization times increase too. For $q=0.45$, below the theoretical critical point, the time needed for message confirmation might take up to minutes, instead of seconds as is the case without an attack.
    The tip pool size starts growing fast, flattens as the attack progresses, and finally, it stabilizes.
    All tests, performed for other $k$ values behave in a very similar way, as for the $k=2$ example. The slight increase of the tip pool size and confirmation times is visible even before the critical value $q_{crit}$ is achieved. 
        
\begin{figure}[h]
        \centering
        \includegraphics[width=\linewidth]{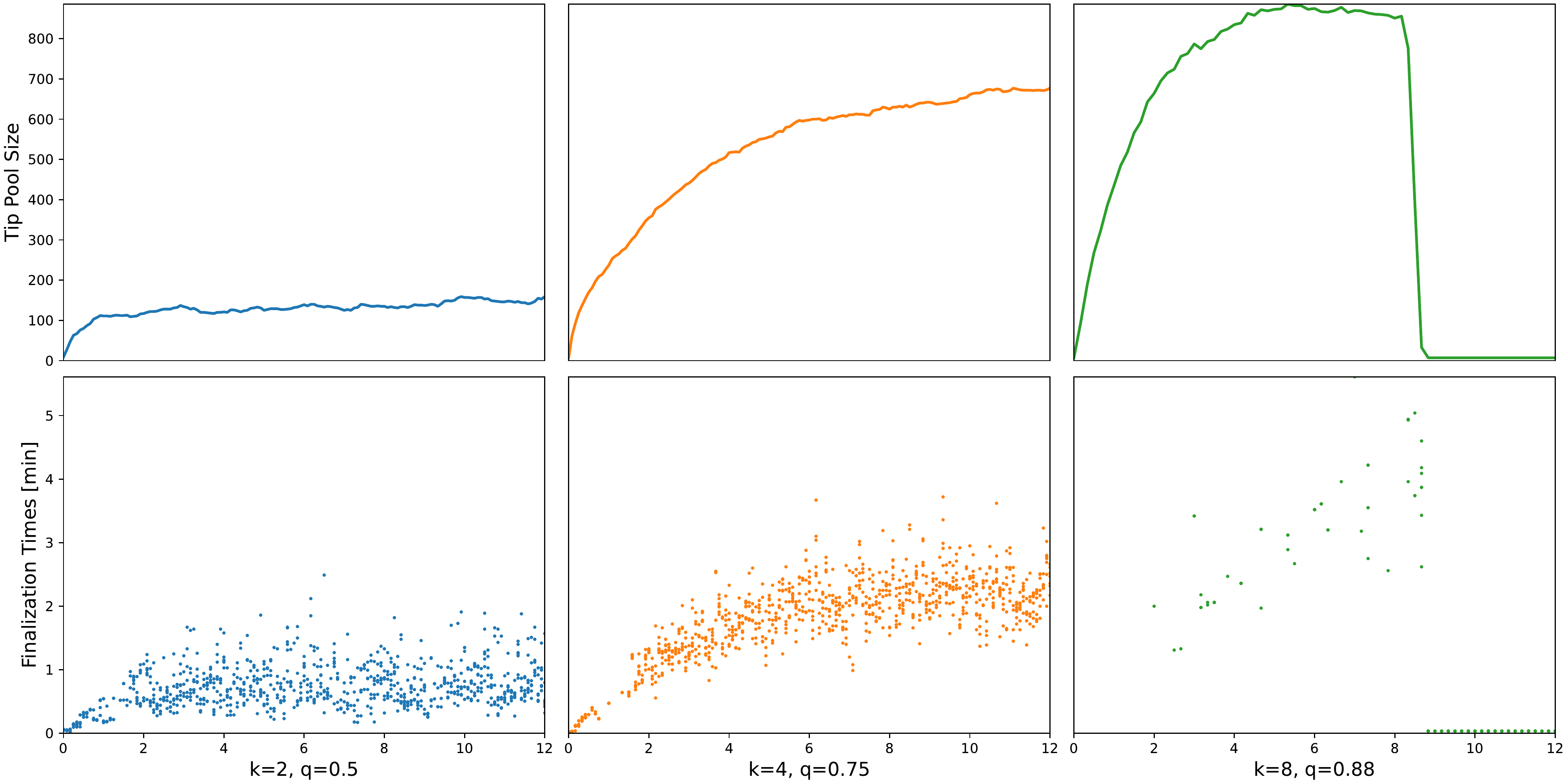}
        \caption{The average tip pool sizes (top subplots) and finalisation times (bottom subplots) for different $k$ values and their respective critical value $q_{crit}$, measured for all honest nodes. The network setup with parent age restriction $\zeta$=1[min]. The total network throughput $\lambda=50$mps. The x-axis represents the duration of the experiment in minutes.}
        \label{fig:graf-like-summary}
    \end{figure}

    Figure \ref{fig:graf-like-summary} presents a comparison for the critical point of different allowed parent numbers $k$.  Most importantly,  along with increased $k$ the network security against the attack increases. For $k=2$ the major effects on the network are visible already at $q\approx q_{crit}=0.5$, however, for  $k=4$ or more, an attack with this value for $q$ does not affect the Tangle (see \ref{appendix}). Moreover, unsurprisingly tip pool sizes for critical points of higher $k$ keep increasing, e.g. for $k=8$ the Tangle is over 800 tips wide, which is correspondingly reflected in higher finalization times that can reach up to 4 minutes for $k=4$, while for  $k=8$ it even exceeds the experiment duration. The reason for the bigger impact that an attack has at the critical point for higher $k$ values is connected to the percentage of network throughput controlled by the adversary. 
    
    \subsection{Tip pool size inflation}
\label{sec:tip-inflation}

\begin{figure}[h]
        \centering
        \includegraphics[width=\linewidth]{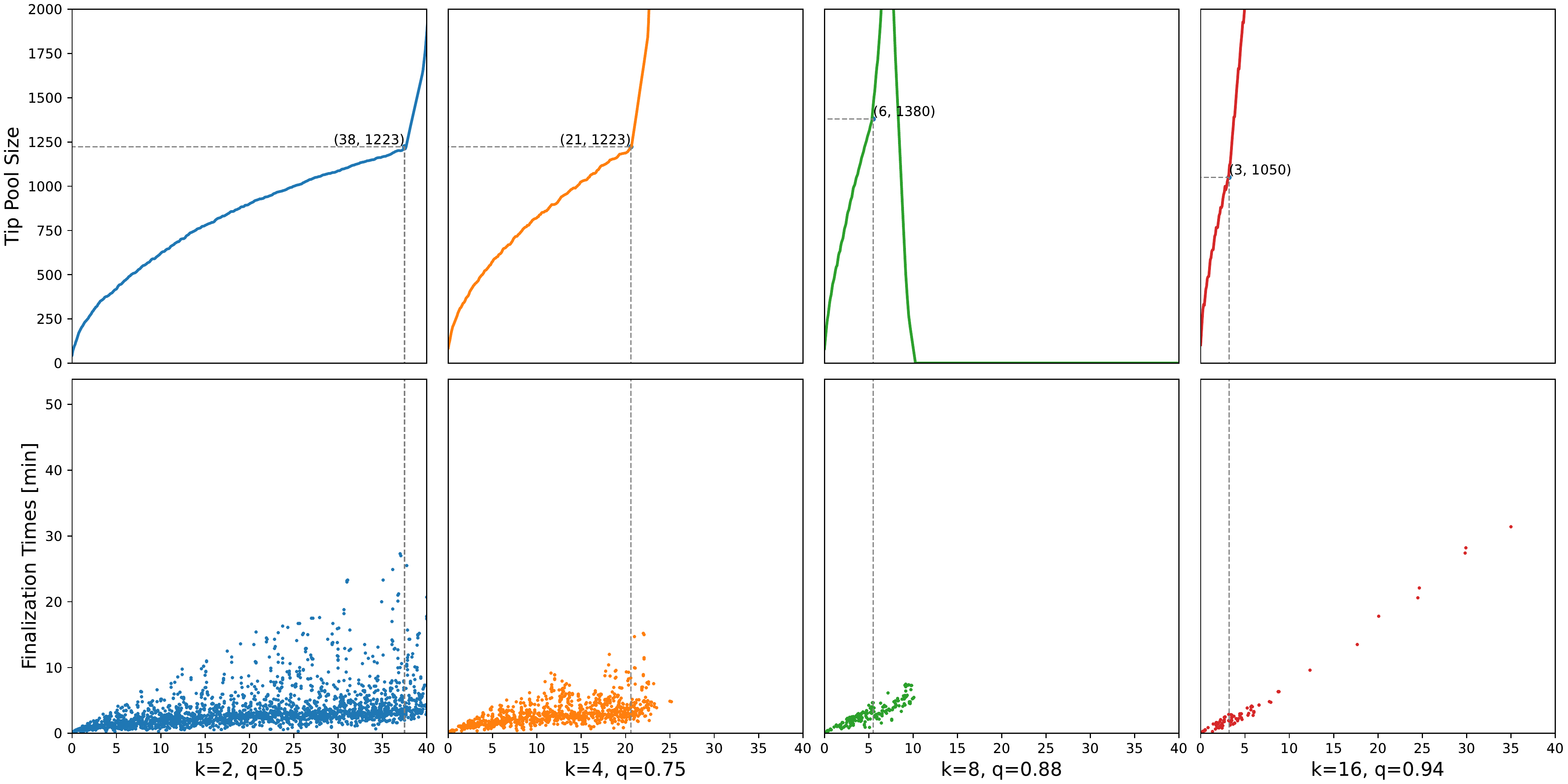}
        \caption{The average tip pool sizes (top subplots) and finalisation times (bottom subplots) for different $k$ values and their respective critical value $q_{crit}$. The network setup is without restrictions on the parent age. The total network throughput $\lambda=50$mps. Each subplot presents the points at which the network broke, and their corresponding time and tip pool size values. The x-axis represents the duration of the experiment in minutes.}
        \label{fig:inf-summary}
    \end{figure}

    The experiment is performed for different values of $q$ including the critical value $q_{crit}$. Results for the restricted parent age show that the tip pool size is under control for  $q<q_{crit}$.  As soon as the adversary spamming rate achieves the critical proportion, a small tip pool size inflation can be observed. After a few max parent age $\zeta$ periods pass, the tip pool size stabilizes. This is expected since whenever the first message issued by the adversary reaches age $\zeta$ it is removed from the tip pool set of all the nodes. Then the adversary is attaching to the next oldest message, which soon will be removed. If the adversary spams with the fixed rate for a longer time the number of tips introduced is equal to the number of tips removed to the parent age check. Thus, as  Figure \ref{fig:graf-vary-q-summary} shows, the number of tips slowly stops to increase and stabilizes.
    By looking at the tip pool sizes in Figure \ref{fig:inf-summary} it can be seen how the tip pool would be increased without the max parent age restriction, i.e. without the removal of old tips from the tip pool. Assuming that each message removes $k-1$ one may expect a linear increase in the tip pool size. However, to understand this effect we have to reassess the assumption in Section \ref{sec:crit-point} 1) about zero network delay that is no longer valid in a real network scenario of an experiment: a tip may receive more than one reference and the number of removed tips by an honest message will thus be less than $k-1$. This effect will decrease as the tip pool size inflates, due to the used \Gls{tsa}. Since the probability for each tip to be selected is equal, as the tip pool size becomes large, the chances for each honest message to be selected (more than once) decreases. The larger the tip pools size is, the closer it is to the assumption that each honest node will remove the $k-1$ tips. This assumption will limit the size to which the tip pool might grow for  $q=q_{crit}$, as the honest nodes will be able to remove exactly the amount of tips being added to the Tangle.

    During a longer attack period tested with no restriction on the parents' age, the experiment was interrupted by the crushing of some of the honest nodes. As indicated in Figure \ref{fig:inf-summary} the point in which the node software starts to crash matches the time when the tip pool size exceeds $\approx 1200$ tips. The fact that this breaking point is similar for different $k$ indicates that the cause could be in the used node's software implementation. Right after the moment when some of the honest machines disconnected from the network, the proportion of adversary messages $q$ became higher, which results in an enormous tip pool size increase and the stop of message finalization soon after that moment.
    
     \subsection{Similarities to blowball attack}
   An interesting point is to consider if similar results might be visible if the attacker use blowball strategy, see Section \ref{sec:blowball}, instead of the orphanage attack. In an Orphanage attack, with adversary network throughput proportion $q>q_{crit}$, the tip pool growth exceeds the honest nodes' capabilities of reducing tips number. Tips are selected randomly, therefore the bigger tip set is the higher is chance that a tip will not be selected during the $\zeta$ aging period. When the maximum parent age $\zeta$ is reached by the oldest tips, they start to expire with an increasing rate limited by $q-q_{crit}\lambda$. Which counterweights the tip pool inflatory rate caused by an adversary and results in limiting the tip pool size, as presented in Figure \ref{fig:graf-vary-q-summary}. 
    The blowball strategy might, therefore, have a different impact on the network, as it temporarily eliminates the factor by which old messages start to be removed due to parent age restrictions. A blowball attack causes a burst in the tip pool size, however, for the limited adversary spamming rate, the tip pool size increase will be still limited, similar to the case of an orphanage attack. Therefore, on average, the damage should be similar. However, by taking into practical considerations and software limitations, it could be possible that a burst of transactions will make it harder to process the data structure.
    
\subsection{Tip pool size linear growth}
     \begin{figure}
        \centering
        \includegraphics[width=\linewidth]{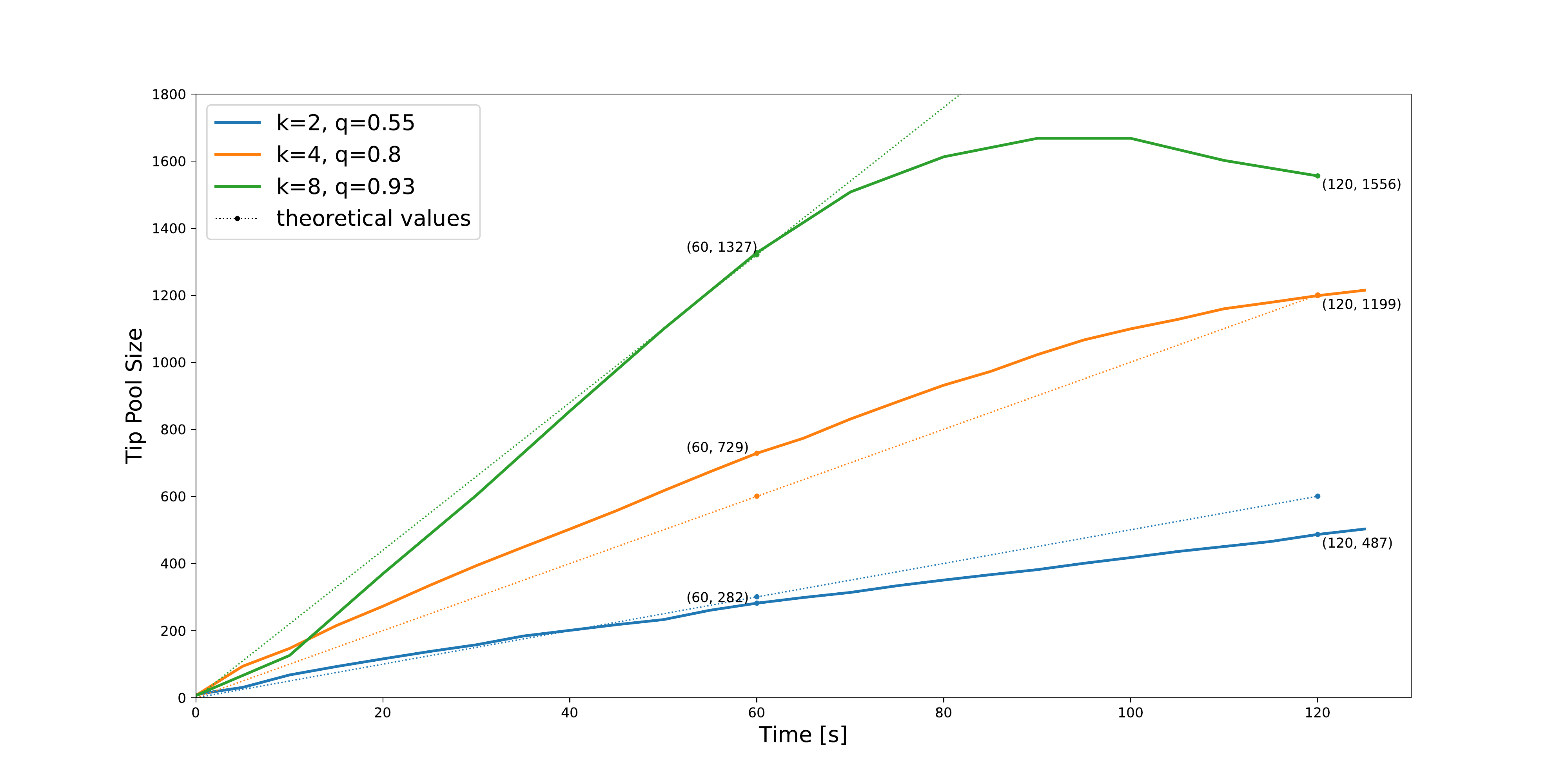}
        \caption{The first two minutes of the experiment performed for $\zeta=$1[min] shows the average tip pool sizes for different $k$ values and $q > q_{crit}$. The total network throughput $\lambda=50$mps. The plot presents the theoretical calculations of the expected tip pool sizes, marked with dashed lines.}
        \label{fig:tips-closer-look}
    \end{figure}
    In Section \ref{sec:tip-inflation} the growth of the tip pool was described for the critical point. There, in case of a continuous attack, the speed of tip pool growth would go down to zero for the network with no restriction on the parent age.
    With assumptions used for the critical point, as described in  Section \ref{sec:above-crit}, we expect that the increase of the tip pool is described by a linear function. This reflects the best-case scenario in which each honest node removes exactly $k-1$ messages from the tip set. Figure \ref{fig:tips-closer-look} shows the first two minutes of an attack with $k=2$ and adversary spamming rates exceeding the critical point. Additionally, the figure presents also the theoretical values for the tip pool size calculated with formula in Section \ref{sec:above-crit}. The results are close to the predicted values despite the major simplification of the model. Up to  60 seconds, the slope of the lines is matching the theoretical curve well. However, after one minute, which is $\zeta$ time, is exceeded, the increase is slowed down. This is because the oldest messages, that reached the maximum parent age are being removed.
    
    \subsection{Orphanage}
    
\begin{figure}
        \centering
        \includegraphics[width=\linewidth]{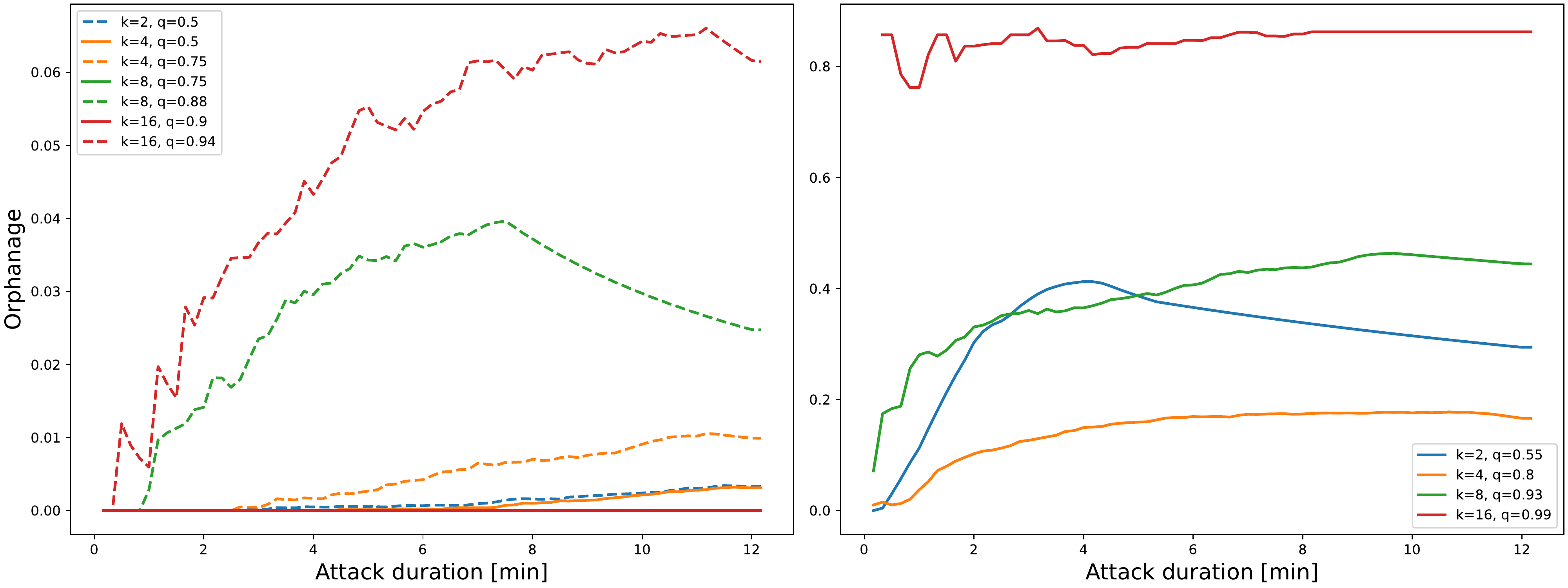}
        \caption{Orphanage results for the experiment wit $\zeta$=1[min] restriction. The left subplot shows the results for  $q=q_{crit}$ marked with dashed lines, and $q<q_{crit}$ marked with solid lines. The right subplot presents the results for $q>q_{crit}$. The total network throughput is $\lambda=50$mps.}
        \label{fig:orphanange-summary}
    \end{figure}

To investigate orphanage caused during an attack, the number of unconfirmed and older than $\zeta$ tips was recorded, which together with a total number of issued messages gives an approximation of an orphanage rate present during an attack. 
Results for the  experiment with $\zeta=1$[min] are showed in Figure \ref{fig:orphanange-summary}. The left subplot shows results recorded for critical values for each $k$ and the comparison on how higher $k$ value stands out for $q_{crit}$ of previous $k$.
For the critical point, the attack duration was too short for the orphanage rate to stabilize, as shown in the first subplot. However, for the $q>q_{crit}$ the rates stabilize in the second half of the attack. For $q_{crit}$ the numbers of orphans for each $k$ except $k=8$ keep increasing. The decrease in the orphanage rate for the $k=8$ parameter is related to node crashes that happened after 8 minutes of the spam, as visible in Figure \ref{fig:graf-like-summary}, and therefore should be ignored.
The lowest orphanage results correspond to the $q$ < $q_{crit}$ for each $k$. In such a case, there is no orphanage for $k=8$ and $k=16$. Surprisingly, for low $k$  some messages were orphaned even below the critical level. Low orphanage rates (below 1\%) for the critical point are achieved for  $k={2, 4}$. For the higher values, the orphanage present during an attack with critical adversary message proportion increases up to 5\%. The higher number of orphaned messages is related to higher tip pool sizes inflated for $k={8, 16}$, caused by higher spam rates of an attacker.
    
    In the case where an adversary controls a proportion of the network throughput exceeding the critical point, the orphanage rates increase significantly. For example, in the extreme scenario where 99\% of messages are malicious,   four out of five messages are orphaned. In a more realistic scenario with $q=0.55$ the orphanage rate reaches 40\% of all messages.
    Orphanage rates measured for $q_{crit}$ are monotonic in regards to $k$ decreasing. Nevertheless, it is not observed for the values above the critical point, where $k=2$ is more problematic than $k=4$. This indicates that the distance of $q$ to $q_{crit}$ has a greater impact for smaller $k$ values. In other words, an additional 5\% of the total network throughput for the attacker gives him much more power to damage the system in case of lower $k$ values.

    \subsection{Impact of changing the $\zeta$ parameter. }

\begin{figure}[ht!]
        \centering
        \includegraphics[width=\linewidth]{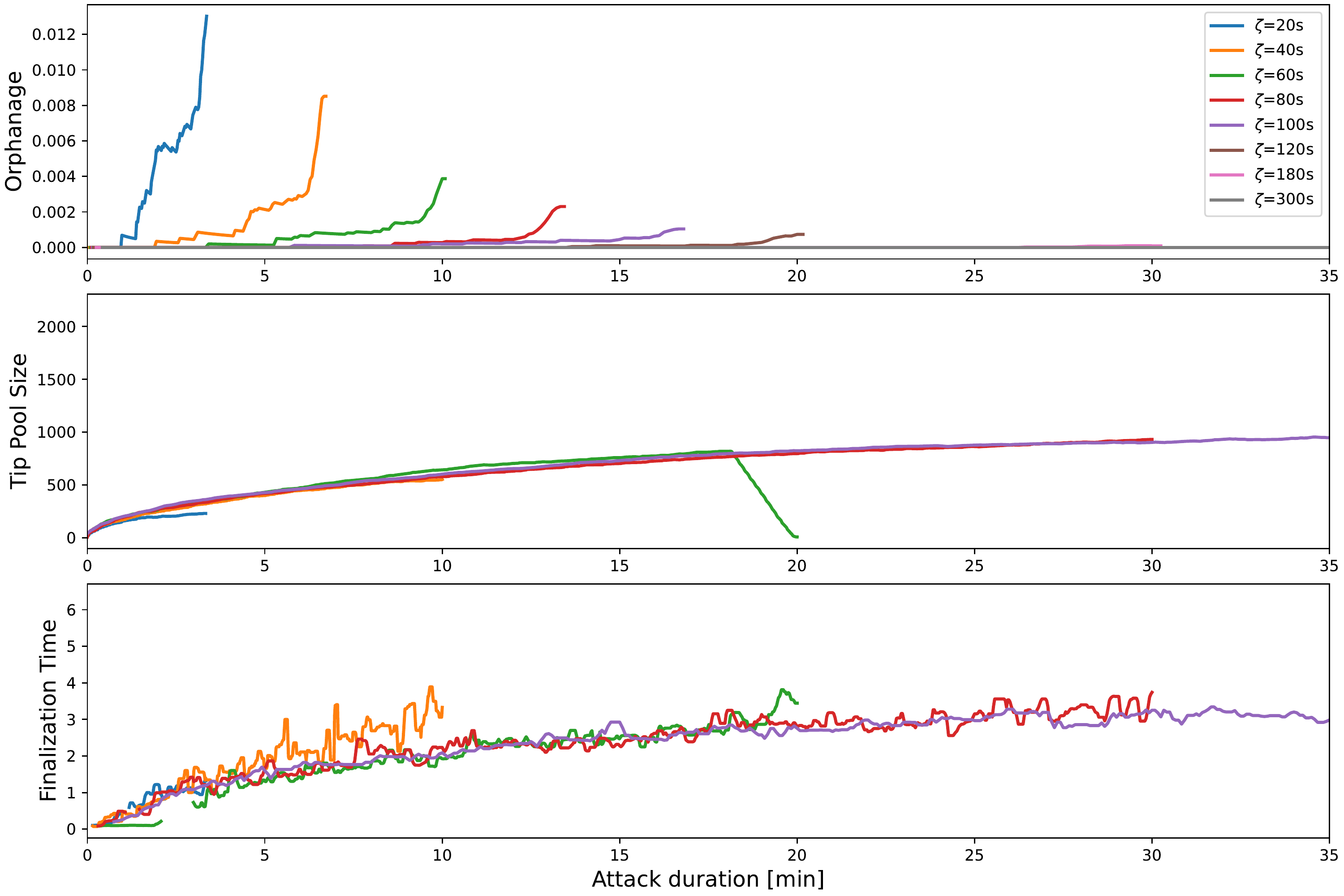}
        \caption{The experiment results for $q=q_{crit}=0.5$ and $k=2$: the orphanage results, the average honest node's tip pool sizes, and the moving average of nodes' confirmation times are presented for different $\zeta$ parameters. The attack duration of each experiment is $10\cdot\zeta$ and the total network throughput is 50 \Gls{mps}.}
        \label{fig:maxage-summary}
    \end{figure}

The previously discussed experiments are performed with the maximum parents' age parameter set to $\zeta=1$[min]. However, in a real network environment, this time range will be higher, e.g. currently the prototype software uses $\zeta=30$[min]. Figure \ref{fig:maxage-summary} presents the characteristics measured for different $\zeta$ parameters. 

 The outcome of the experiment indicates that the change of the maximum allowed age for the parents has no impact on both the tip pool size and finalization times for the critical point $q=0.5$. The results agree well with the measurements demonstrated above for the attack on the network with no parent age restriction. 
 
 Values reached up to $\zeta$ point are a reflection of data presented in Figure \ref{fig:inf-summary} and the experiment with no-restriction on the max parent age can be treated as an indicator of what would be the values of tip pool size and finalization times for different $\zeta$. This is because the behavior of the network before this point is reached is exactly the same in both situations.

\begin{figure}[h!]
        \centering
        \includegraphics[width=\linewidth]{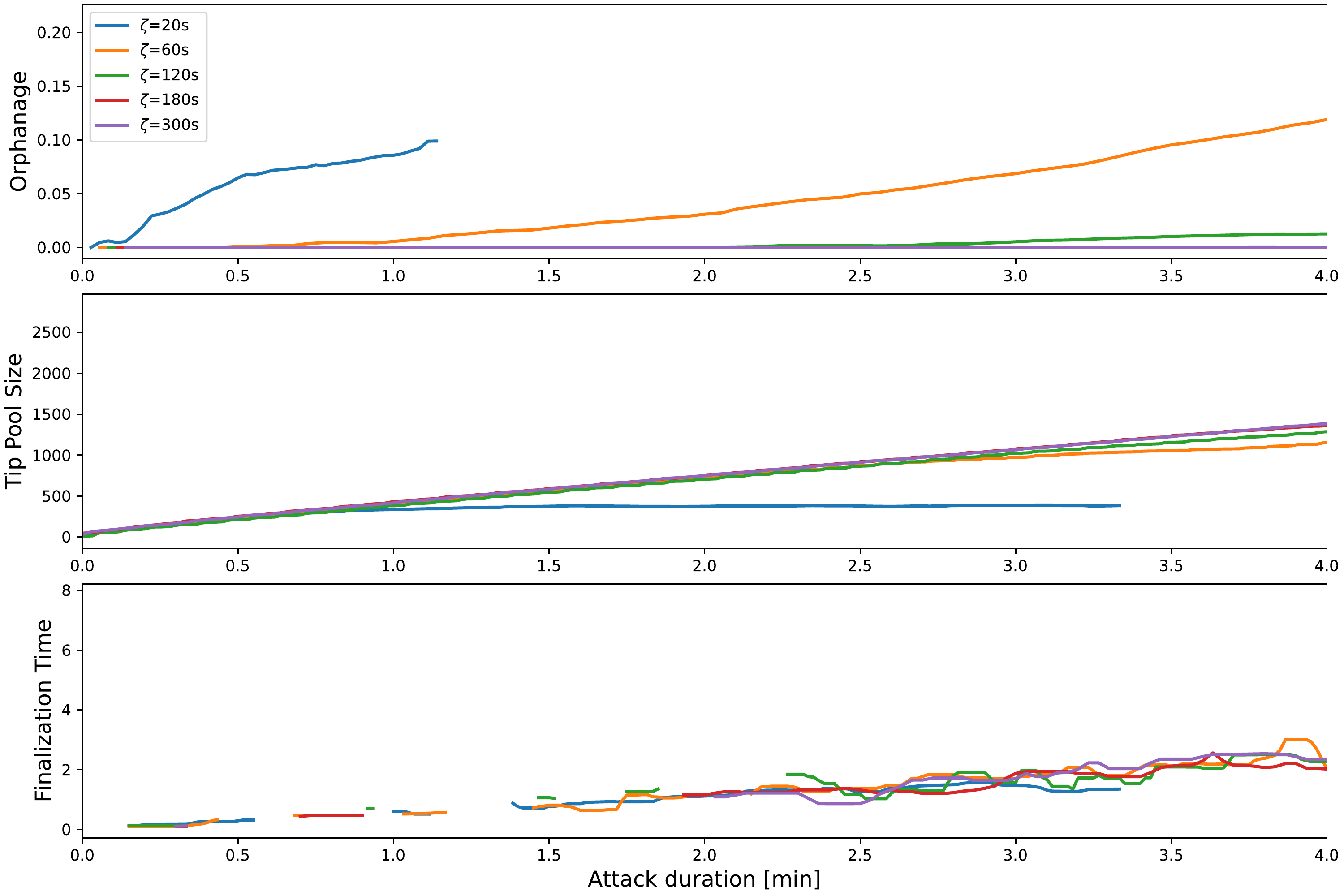}
        \caption{Experiment results for $q=0.55$ and $k=2$: the orphanage results, the average honest node's tip pool sizes, and the moving average of nodes' confirmation times are presented for different $\zeta$ parameters. The attack duration of each experiment is $10\cdot\zeta$ and the total network throughput is 50 mps.}
        \label{fig:maxage-summary-above}
    \end{figure}

However, the differences between tip pool sizes and the confirmation rates become visible when the adversarial spamming rate $q$ exceeds the critical point. As shown in Section \ref{sec:tip-inflation} the tip pool should increase linearly up to the point of reaching $\zeta$ time, as after this point adversary's tips start to be removed at a constant rate $\lambda_A$.
Following this reasoning,  the tip pool sizes for different $\zeta$ should not differ until they reach maximum parent age for the first time. Therefore, for the different $\zeta$ parameters, the tip pool size increase starts to slow down after different time periods. Thus the lines representing subsequent $\zeta$ values diverge monotonically, one by one, starting from the smallest $\zeta$. Experimental results confirm the applicability of this simplistic assumption, which can be seen in Figure \ref{fig:maxage-summary-above}. We show only the first 10 minutes for optimized visualization. For the presentation of the whole experiment duration, we refer to the appendix \ref{appendix}.  Since in the experiment the $q$ was above the critical point, the tip pool inflation was significant, which caused some nodes in the network to crash. Therefore,  only at the beginning of the experiment, we can observe that the lines start to diverge after the $\zeta$ time is reached for the first time. The finalization times stayed the same as due to network problems the experiment stopped too soon to notice the effect that a slowing down of the tip pool size increase has on the finalization of messages.

{\backmatter \chapter{Conclusion}}

\label{ch:conclusion}
After the summary of the most important concepts in the DLT space, the IOTA protocol has been introduced and portrayed in terms of  security flaws along with other blockchain-based projects. The evolution of the protocol and the reasoning for the Coordicide transition and components of the protocol were described. Many public blockchain attacks and existing vulnerabilities have been collected from the existing literature to give context to the studied attack scenario in this work. Also, a variety of online sources has been searched, for example real-world thefts and security violations, to increase the awareness of the importance of studies concerning the security of the systems, and where a small undiscovered vulnerability might lead to a loss of millions of dollars.

In the next part of the thesis  the orphanage problem was described and analyzed. Two attack vectors were introduced: the orphanage attack and the blowball attack. Both are exploiting the orphanage problem and causing harm to the network through the tip pool inflation and increase of the confirmation times. The orphanage attack was then implemented in a real network setting within a docker network environment. An extensive study was performed to measure the effects of the attack on the network for a variety of parameters, such as the part of the network throughput controlled by an adversary or the number of parent references for a message allowed by the protocol.
The experimental results have confirmed the theoretical assumptions about the tip pool growth and confirmed that the change of the number of allowed parent references can greatly increase network robustness towards the attack, by increasing the proportion of the network throughput an adversary needs to gain for the attack to succeed.

\bibliographystyle{ieeetr}
\bibliography{main}

\appendix

{\backmatter \chapter{Appendix A - complete experimental results}}
\label{appendix}

\begin{figure}

\includegraphics[width=0.8\linewidth]{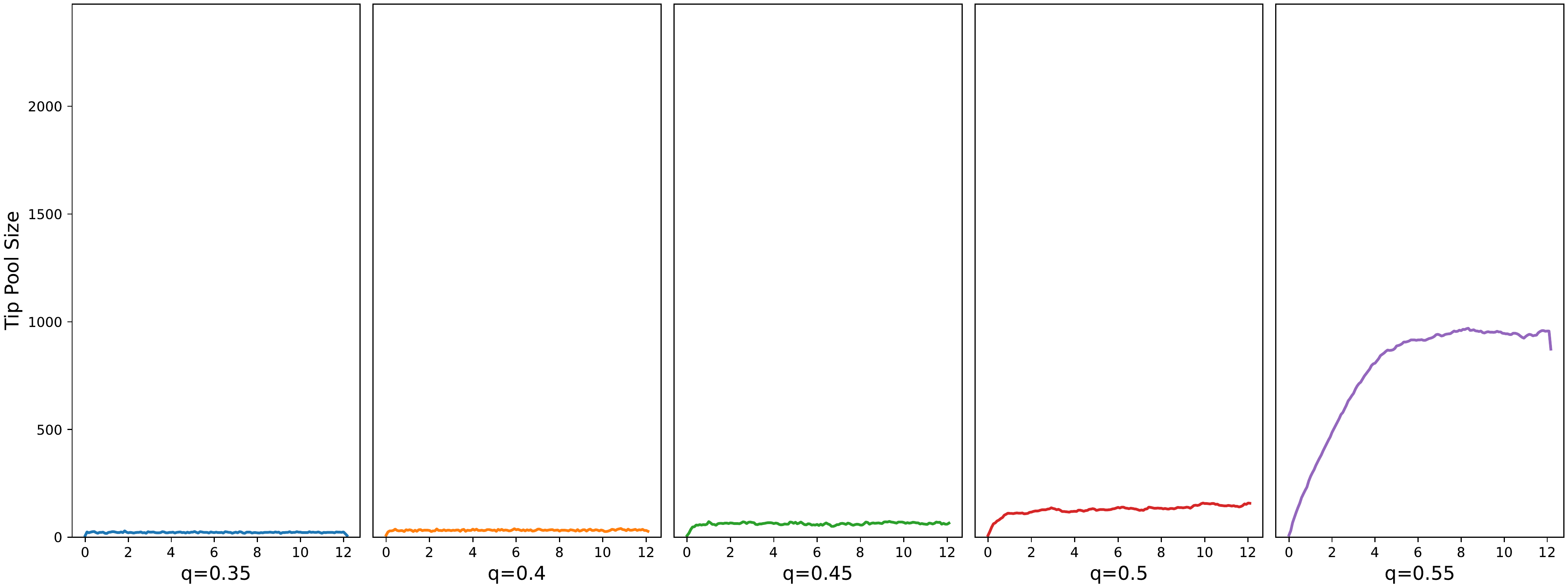}
\newline
a) $k = 2$

\includegraphics[width=0.8\linewidth]{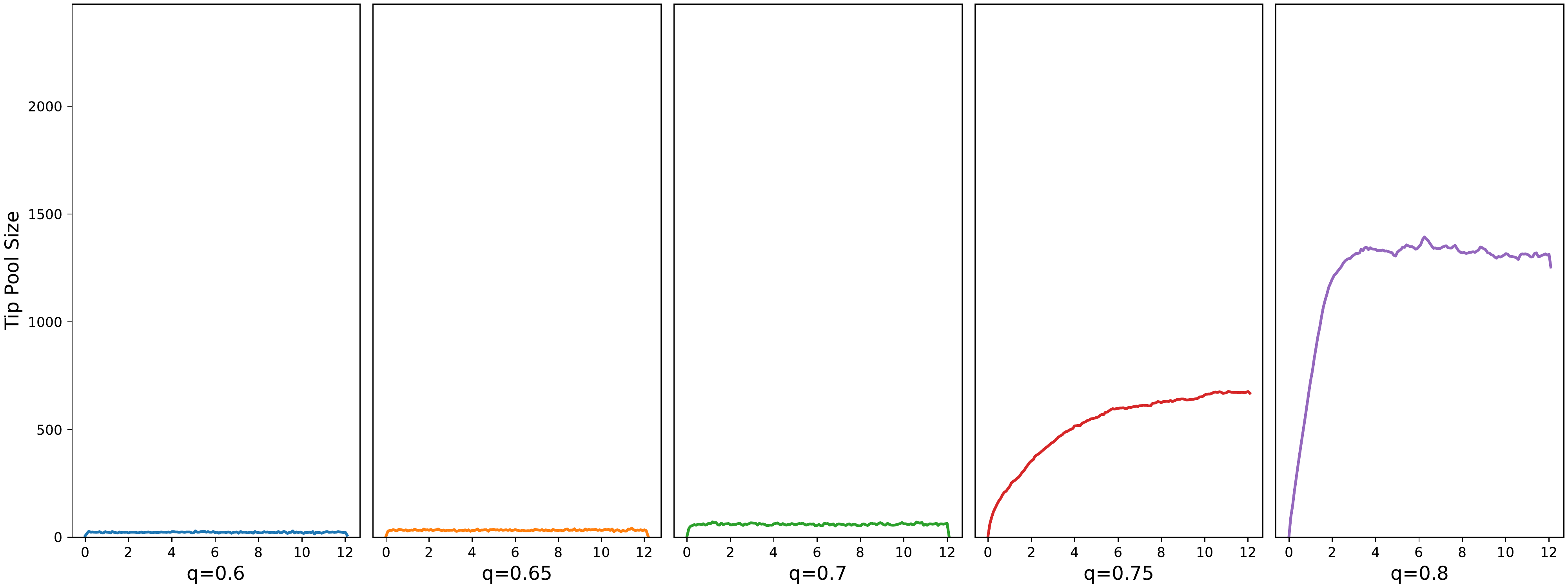}
\newline
b) $k = 4$

\includegraphics[width=0.8\linewidth]{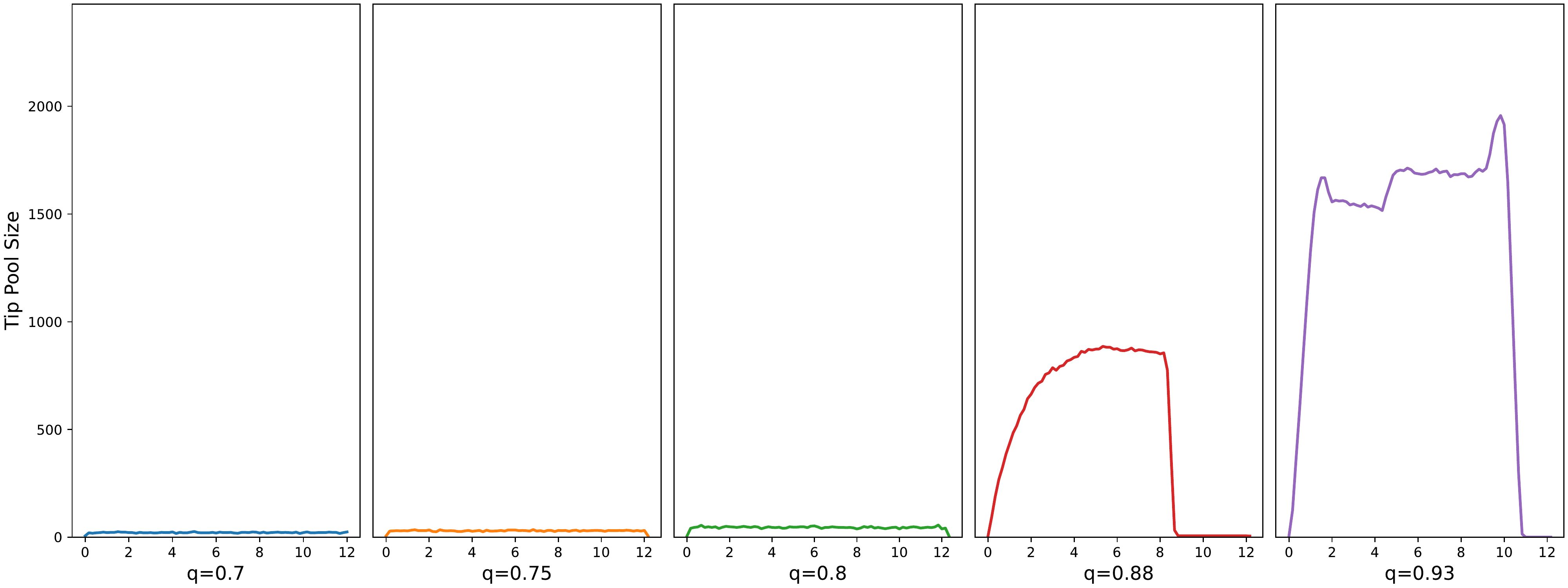}
\newline
c) $k = 8$

\includegraphics[width=0.8\linewidth]{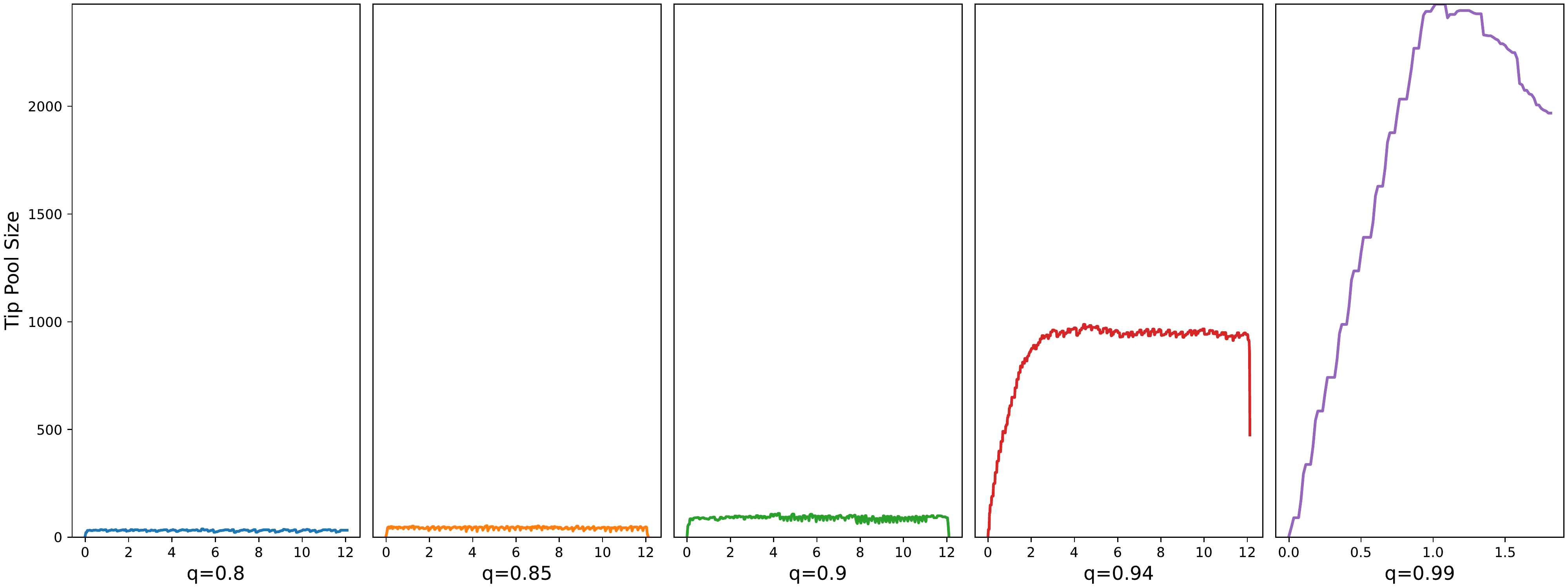}
\newline
d) $k = 16$

\caption{Tip pool sizes for different $k$ during an attack. }
\label{fig:apdx-grafana-tips}
\end{figure}

\begin{figure}

\includegraphics[width=0.8\linewidth]{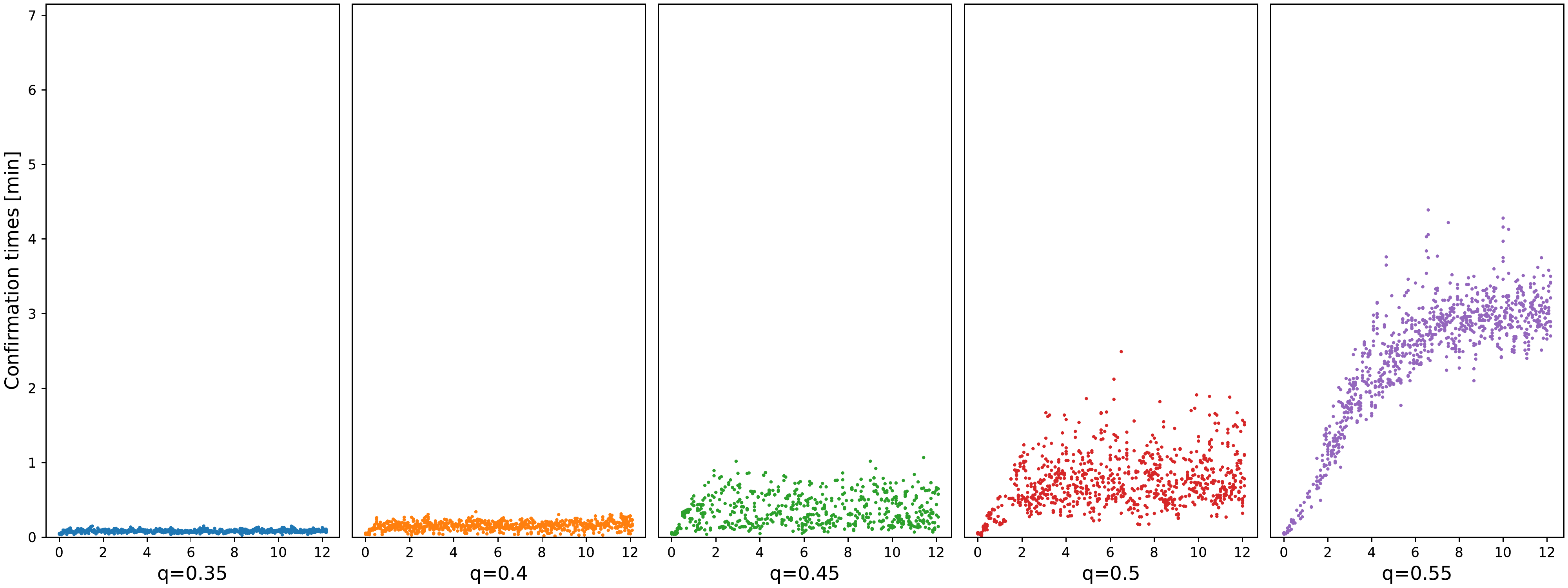}
\newline
a) $k = 2$

\includegraphics[width=0.8\linewidth]{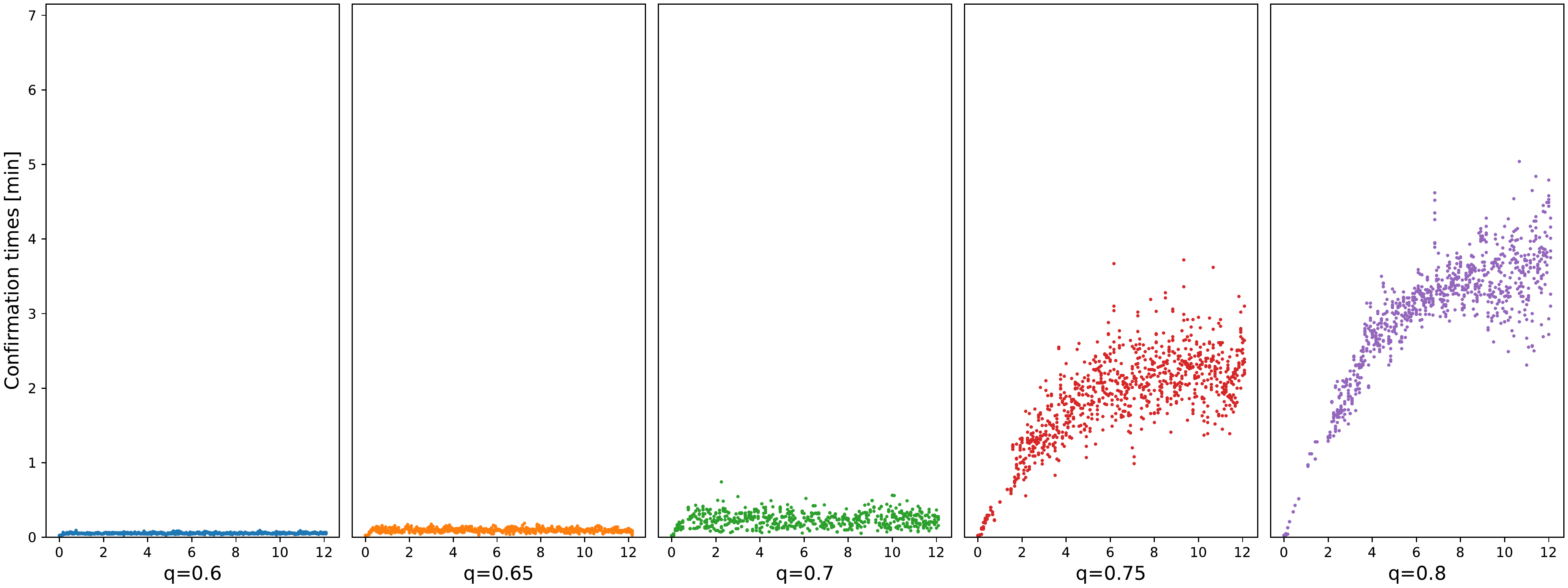}
\newline
b) $k = 4$

\includegraphics[width=0.8\linewidth]{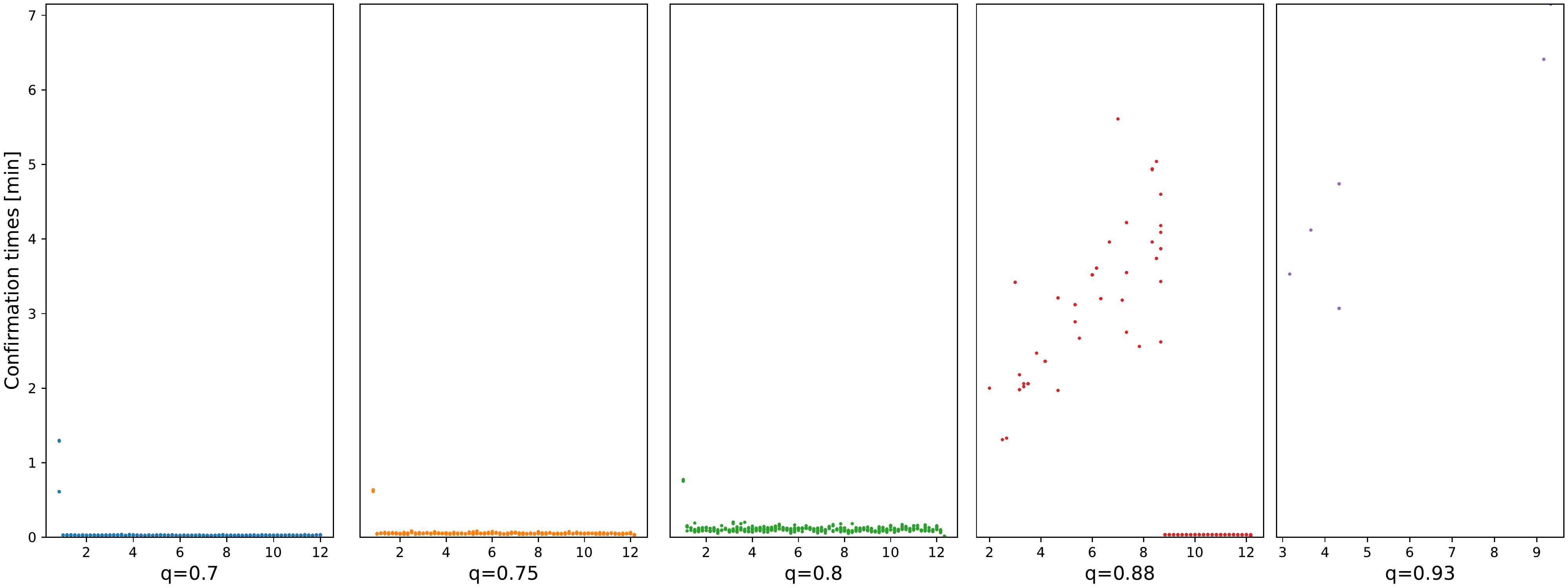}
\newline
c) $k = 8$

\includegraphics[width=0.8\linewidth]{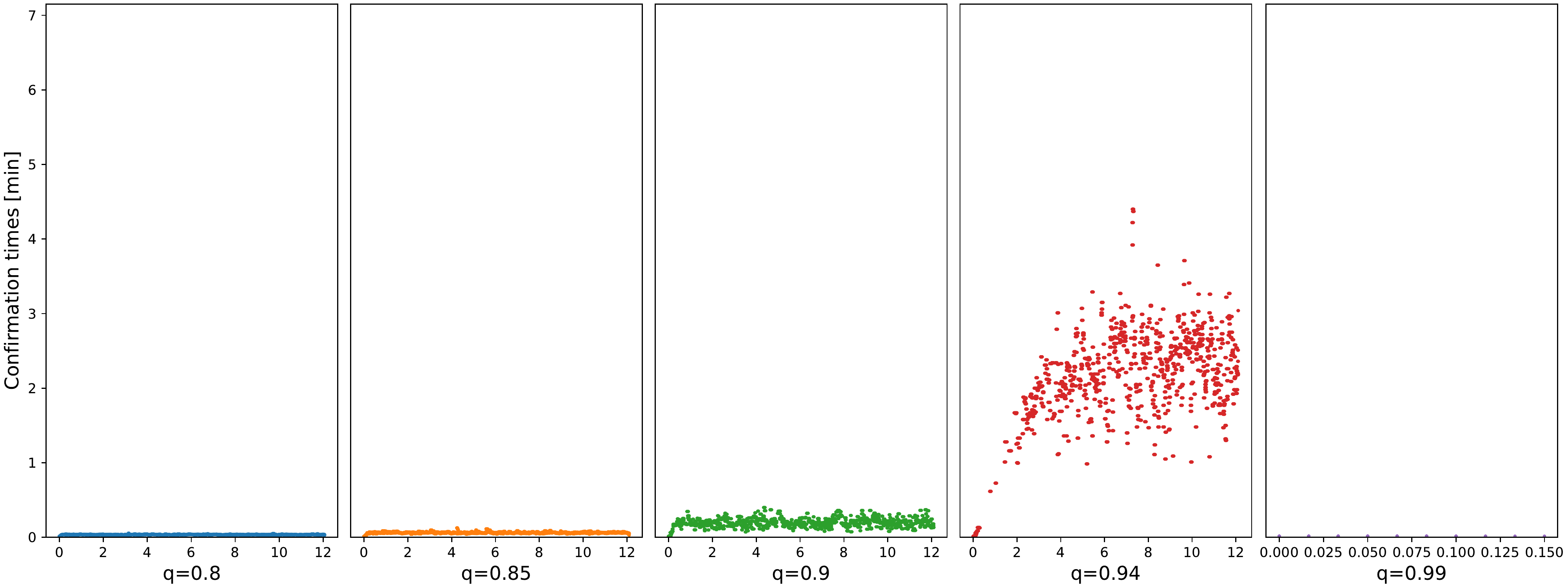}
\newline
d) $k = 16$

\caption{Confirmation events and corresponding times [min] for different $k$ during an attack. }
\label{fig:apdx-grafana-conf-times}
\end{figure}

\begin{figure}
\includegraphics[width=0.8\linewidth]{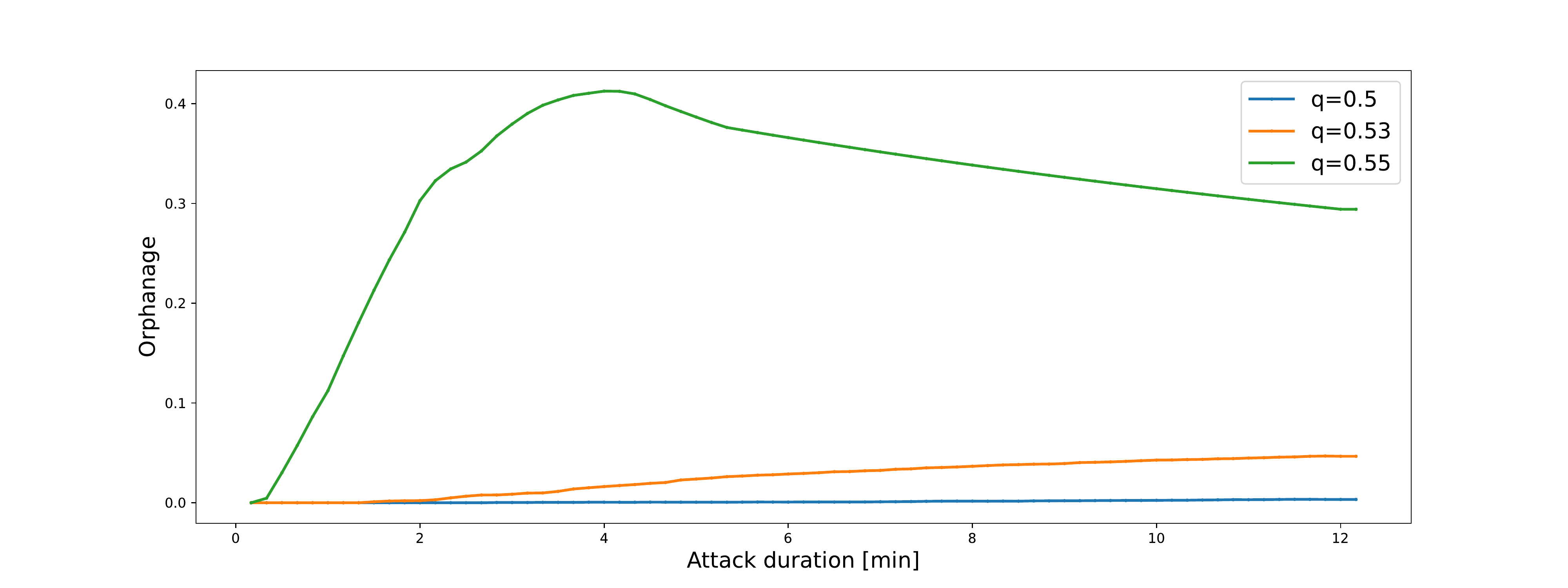}
\newline
a) $k = 2$

\includegraphics[width=0.8\linewidth]{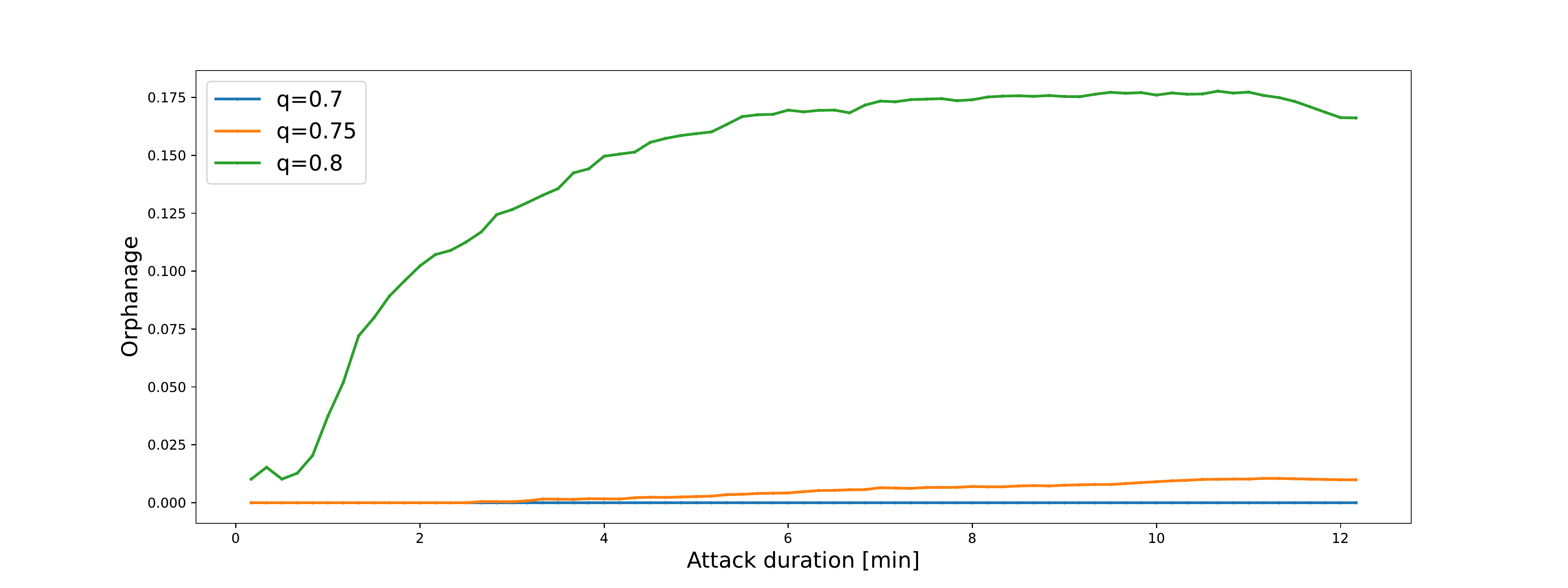}
\newline
b) $k = 4$

\includegraphics[width=0.8\linewidth]{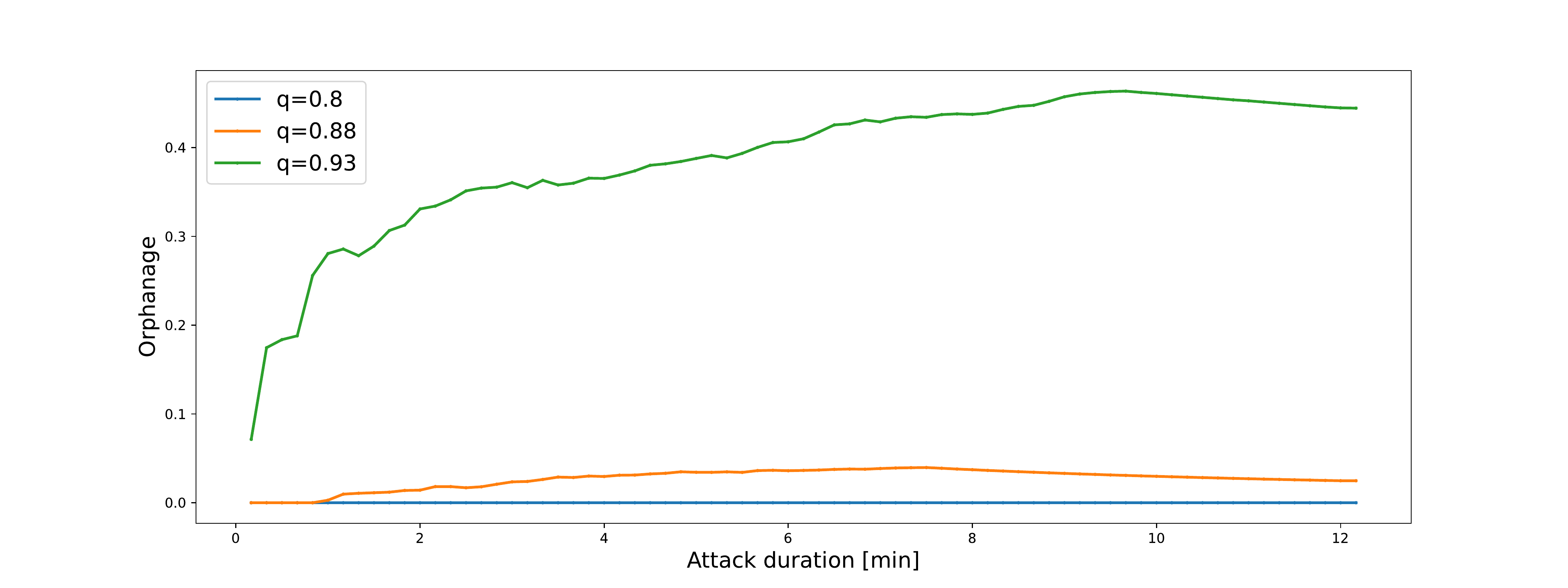}
\newline
c) $k = 8$

\includegraphics[width=0.8\linewidth]{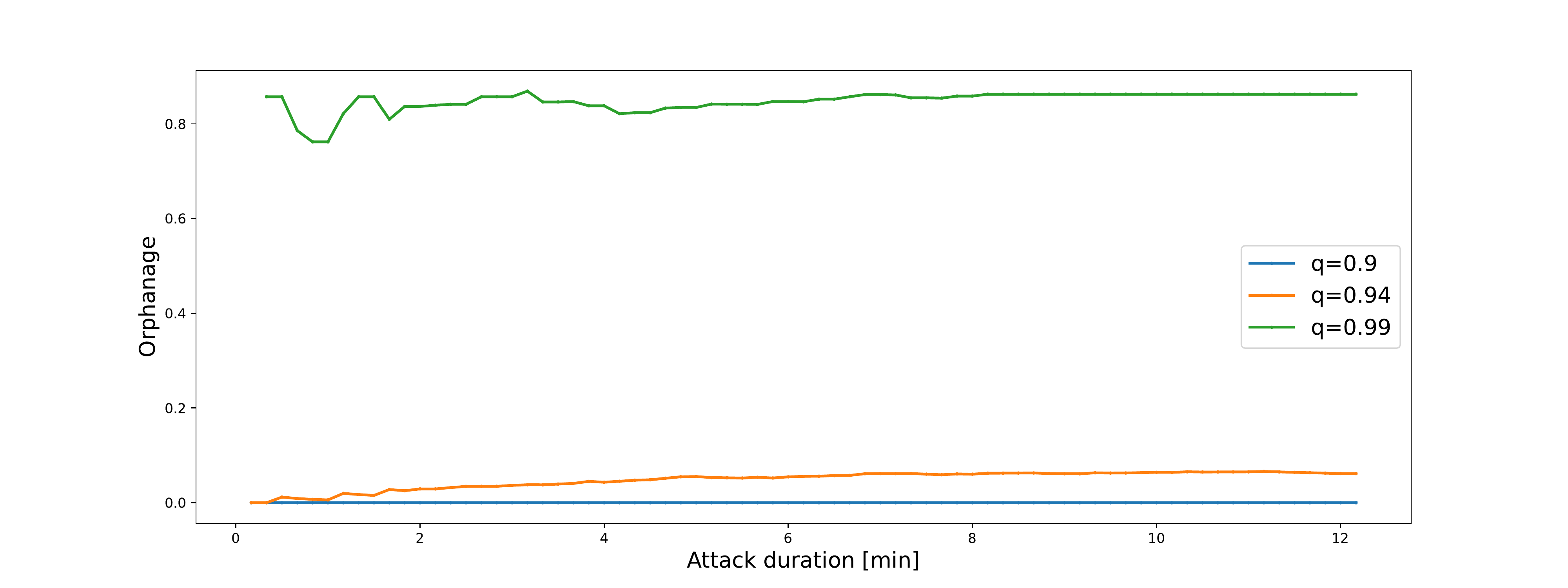}
\newline
d) $k = 16$

\caption{Orphanage rates calculated from cumulative orphan counts and total issued messages for different $k$. }
\label{fig:apdx-orphanage}
\end{figure}

\begin{figure}

\includegraphics[width=0.8\linewidth]{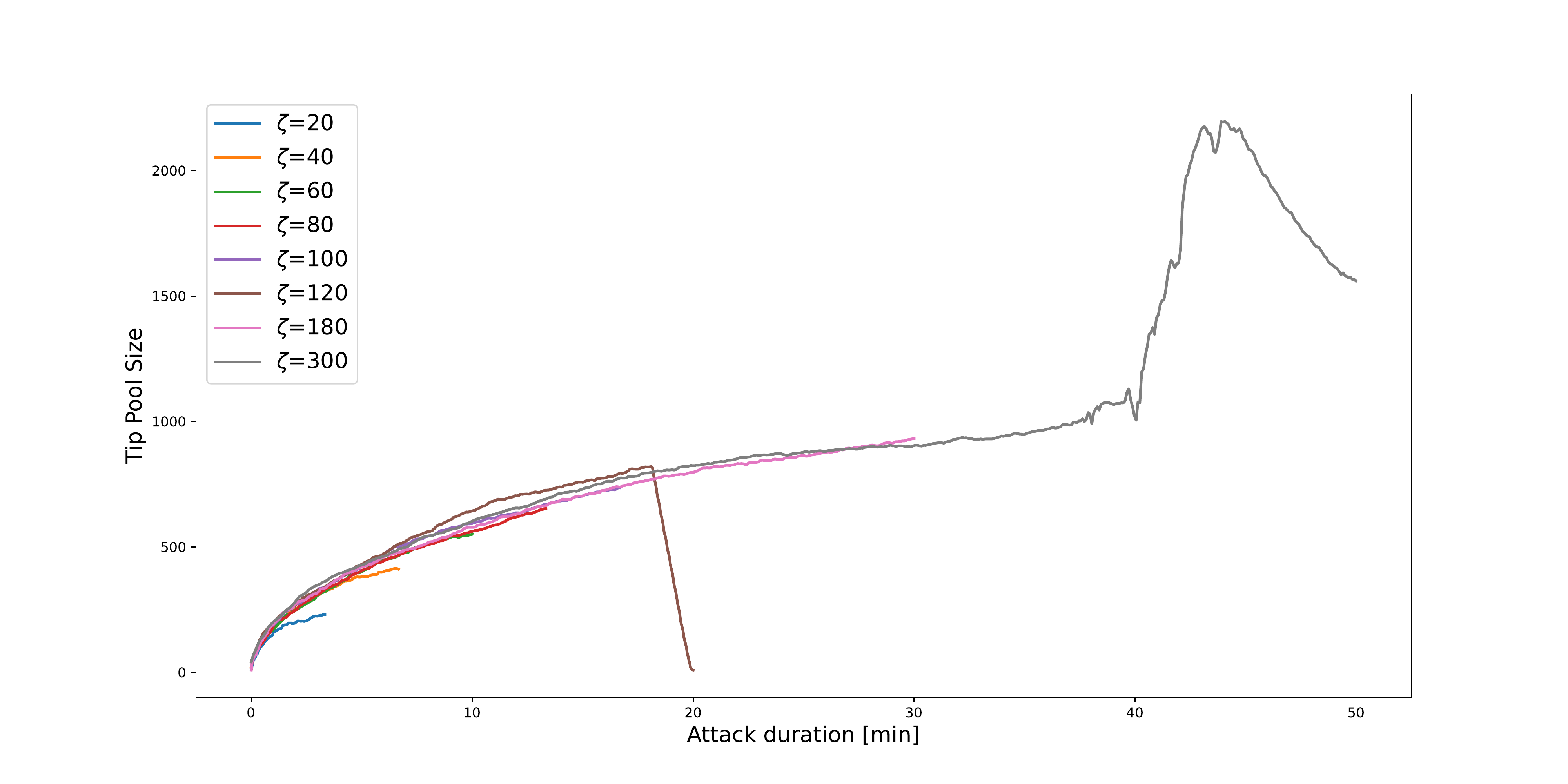}
\newline
a) Finalization times for different $\zeta$ and $k=2$.

\includegraphics[width=0.8\linewidth]{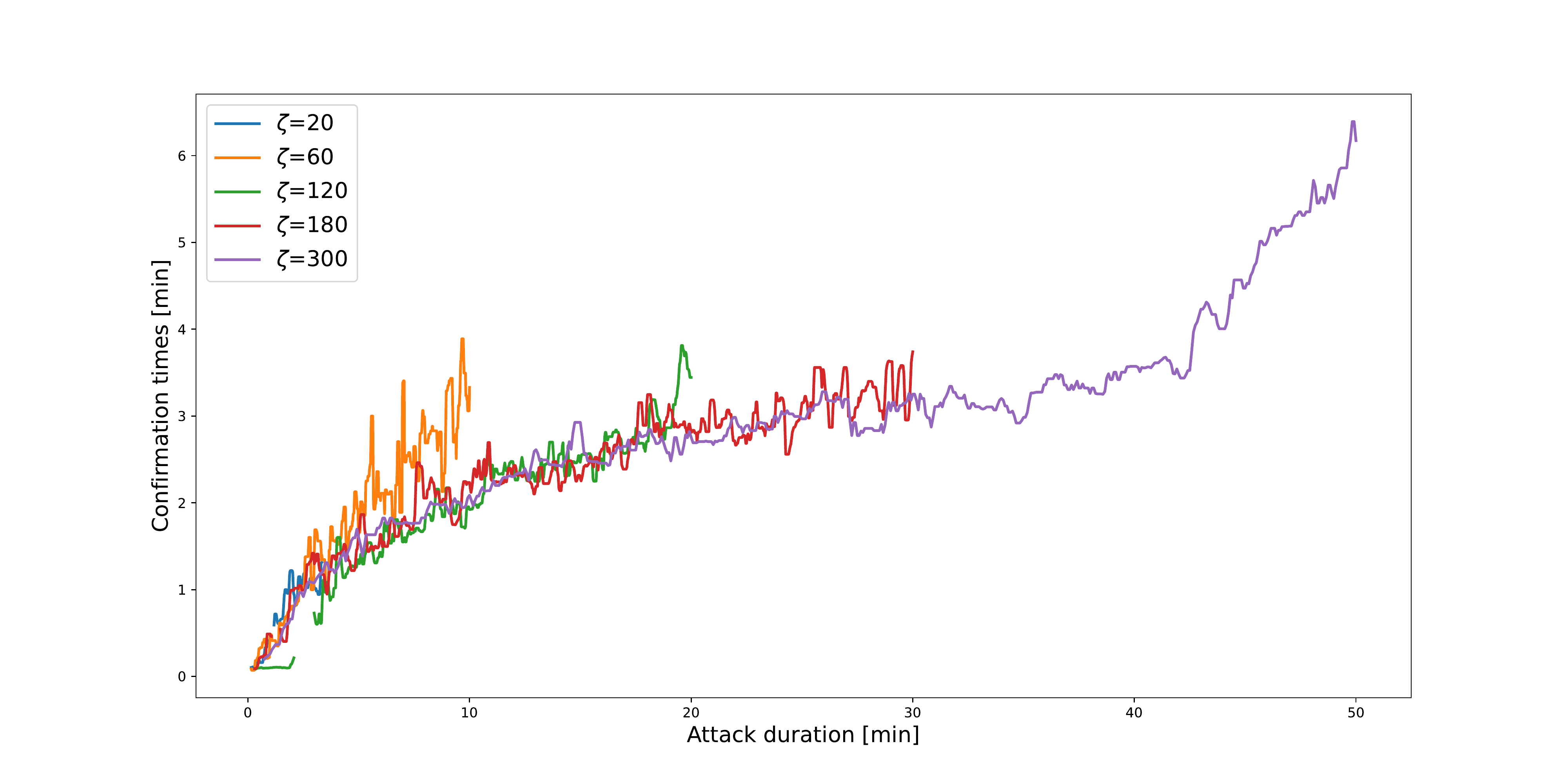}
\newline
b) Finalization times for different $\zeta$ and $k=2$. 

\includegraphics[width=0.8\linewidth]{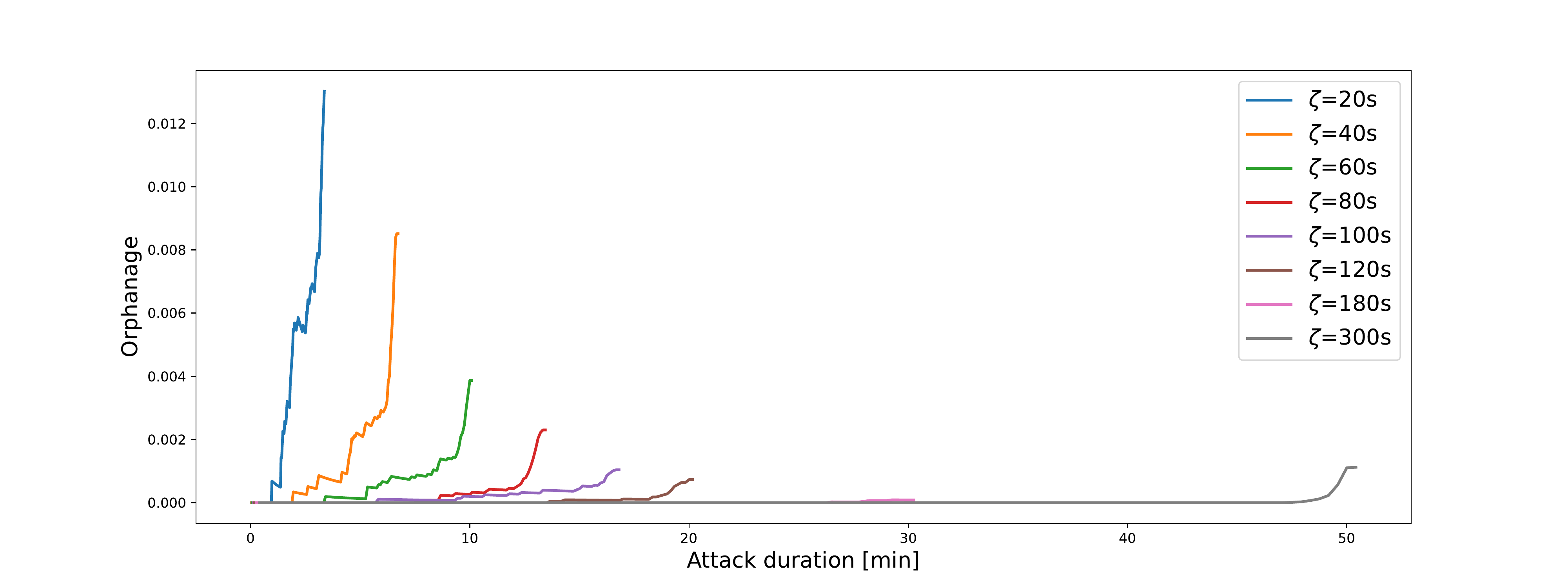}
\newline
c) Orphanage rates calculated from cumulative orphan counts and total issued messages for different~$\zeta$.

\caption{Attack performed for different max parent age $\zeta$ values. Each attack sustained for $10\cdot\zeta$ time period. }
\label{fig:apdx-maxage}
\end{figure}

\begin{figure}

\includegraphics[width=.8\linewidth]{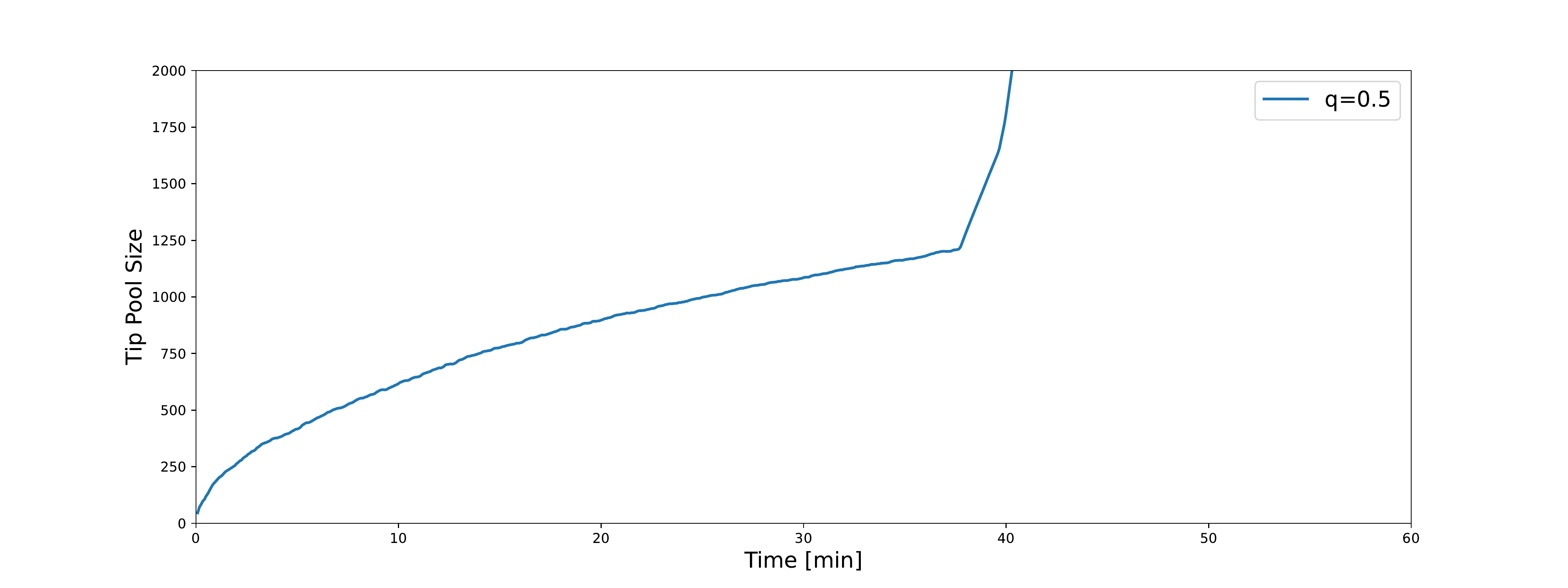}
\newline  
a) $k = 2$

\includegraphics[width=.8\linewidth]{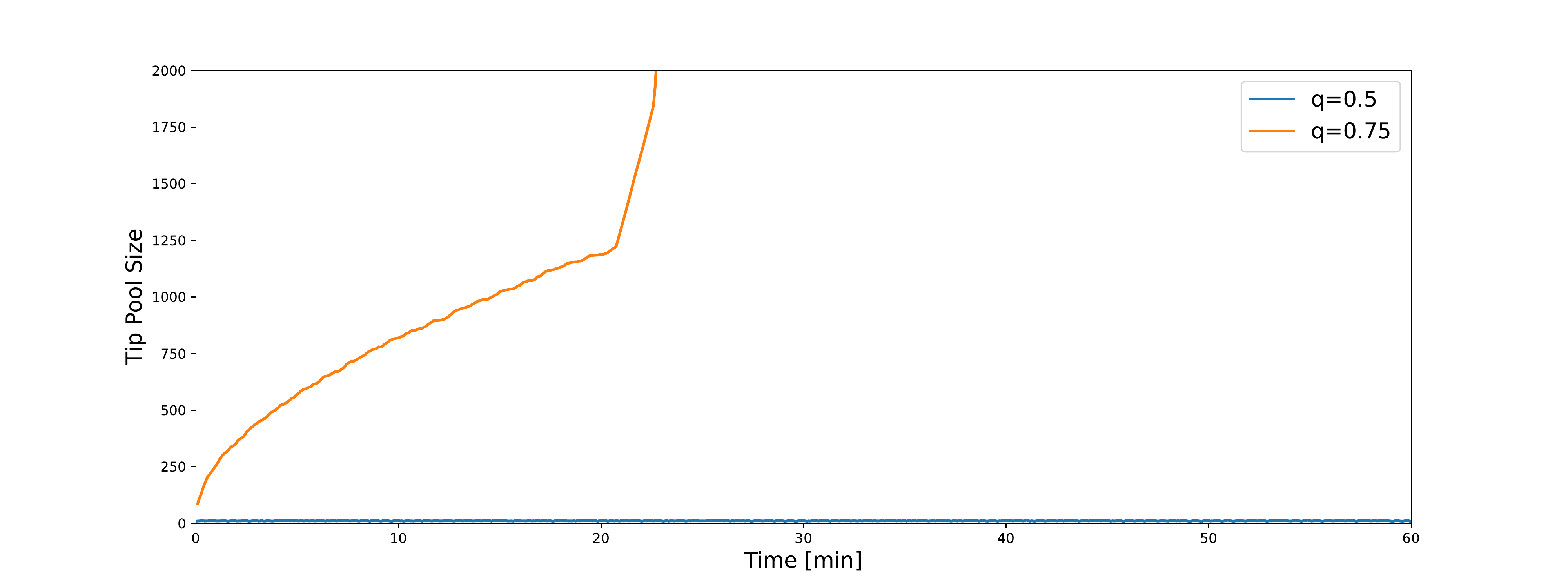}
\newline 
b) $k = 4$

\includegraphics[width=.8\linewidth]{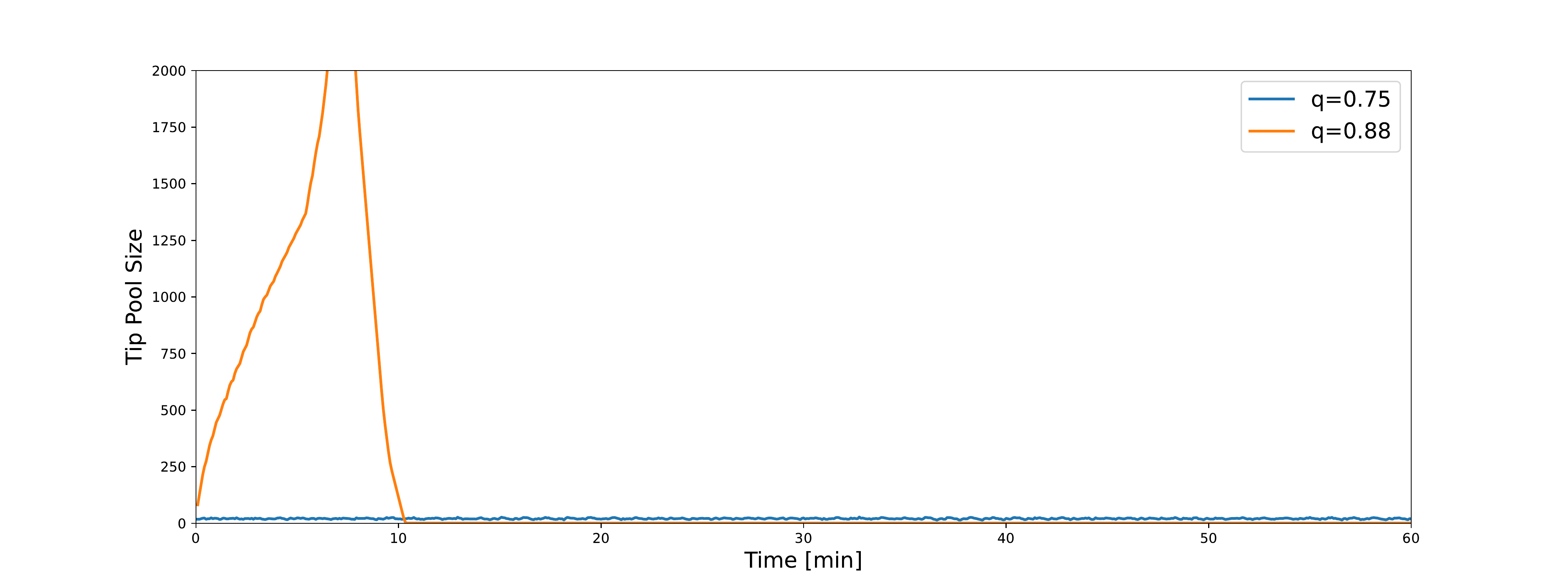}
\newline
c) $k = 8$

\includegraphics[width=.8\linewidth]{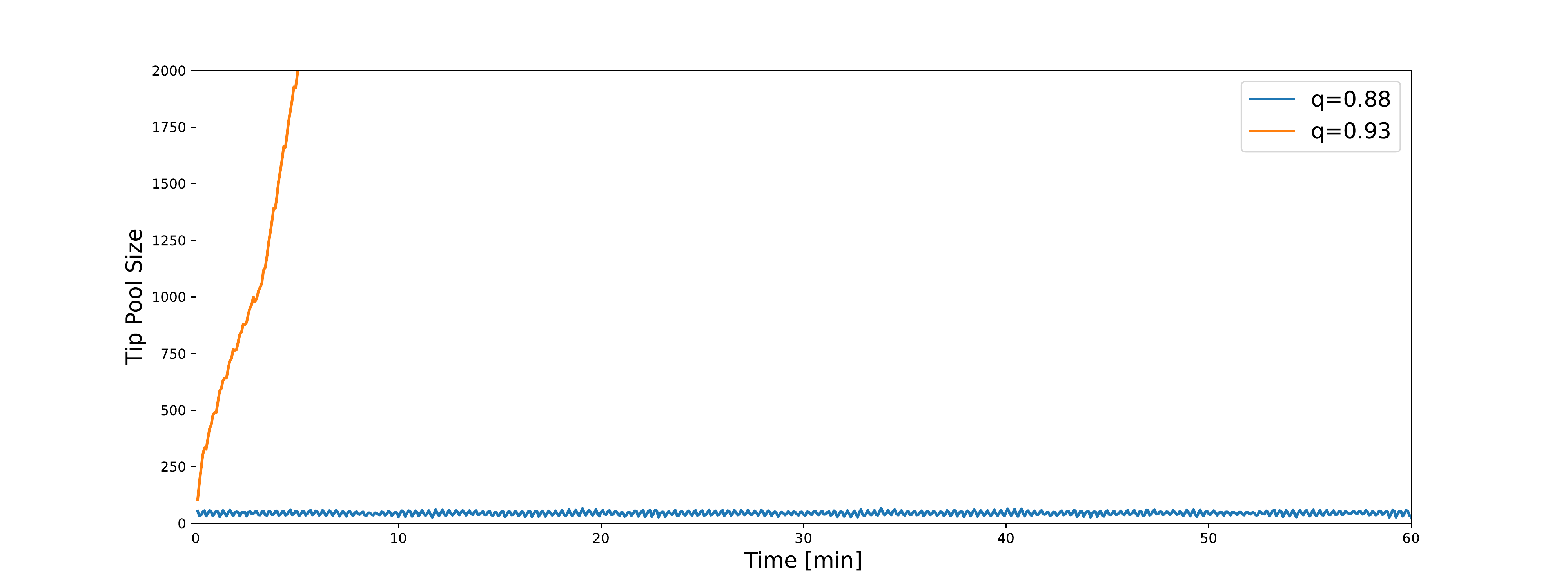}
\newline
d) $k = 16$

\caption{Average tip pool sizes for different $k$ and $q$ inflated during one hour of orphanage attack with no restriction on parents age. }
\label{fig:apdx-infinite-tips}
\end{figure}

\begin{figure}

\includegraphics[width=0.8\linewidth]{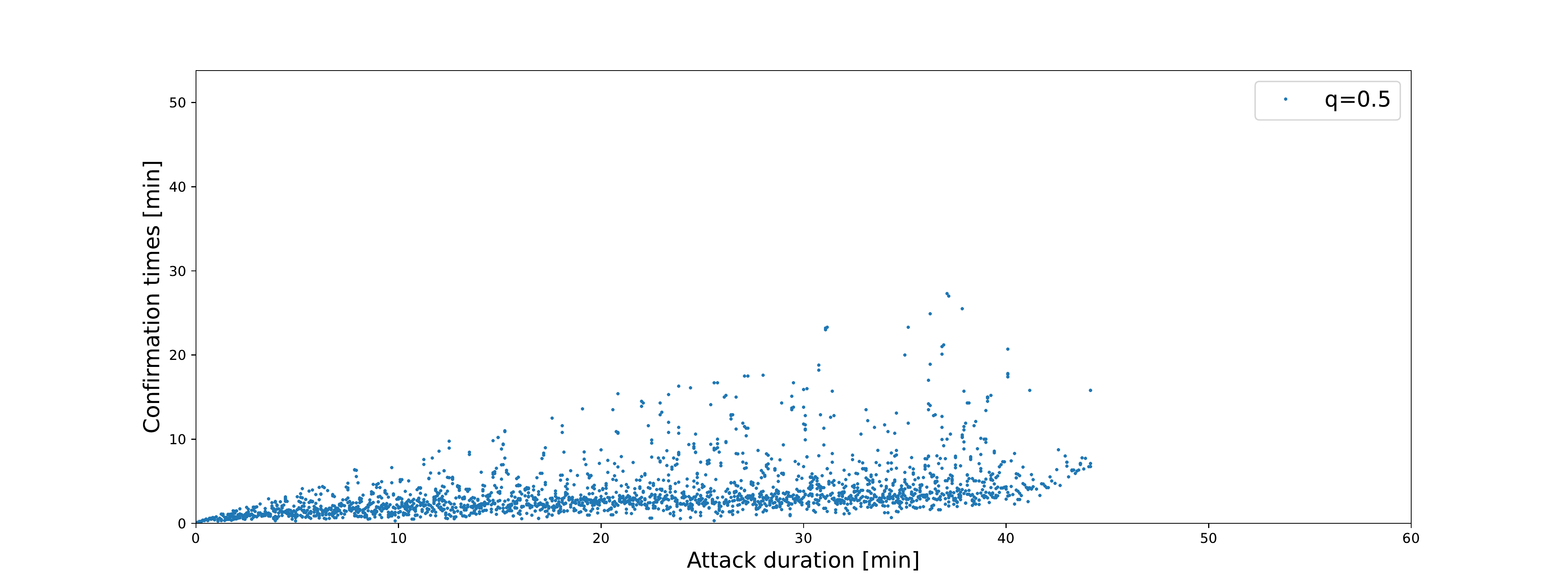}
\newline
a) $k = 2$

\includegraphics[width=0.8\linewidth]{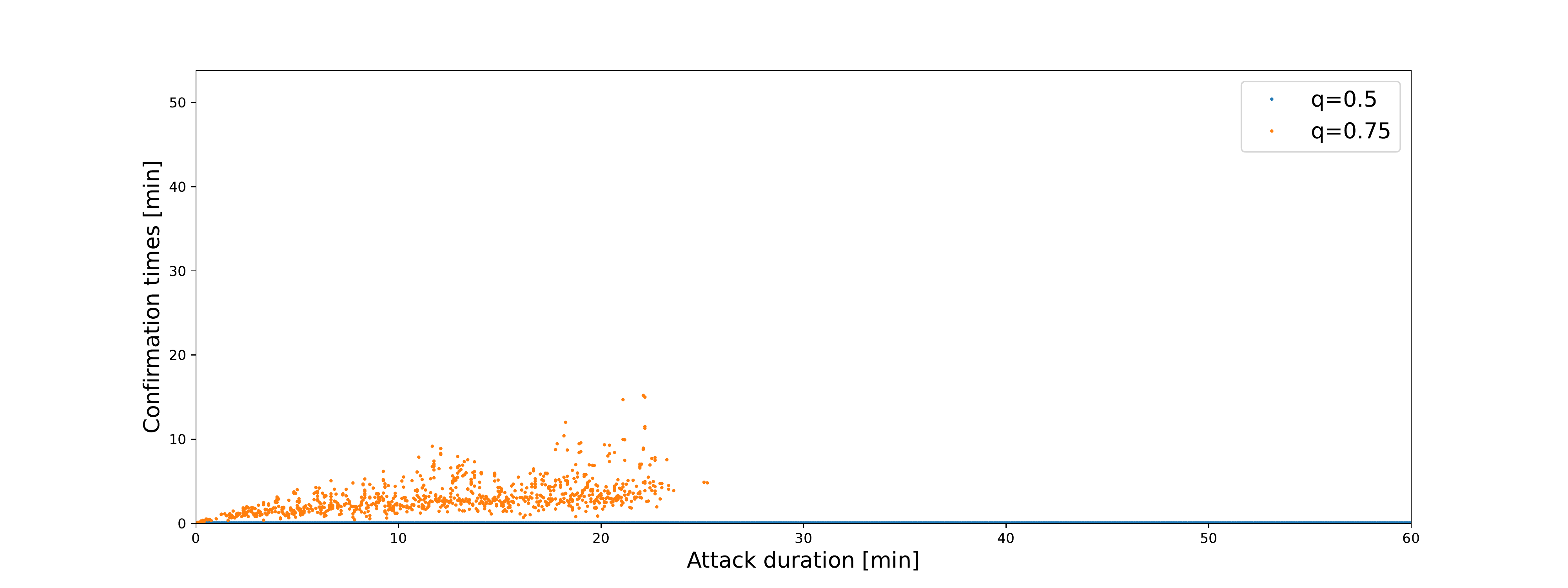}
\newline
b) $k = 4$

\includegraphics[width=0.8\linewidth]{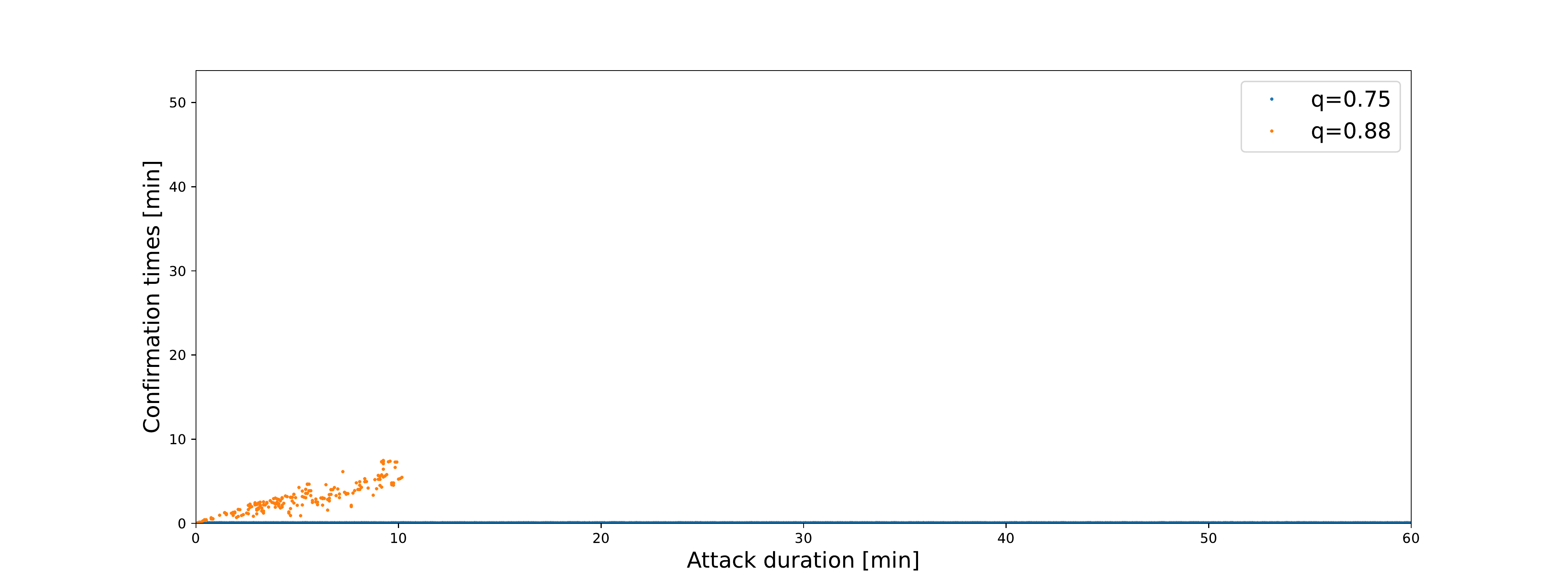}
\newline
c) $k = 8$

\includegraphics[width=0.8\linewidth]{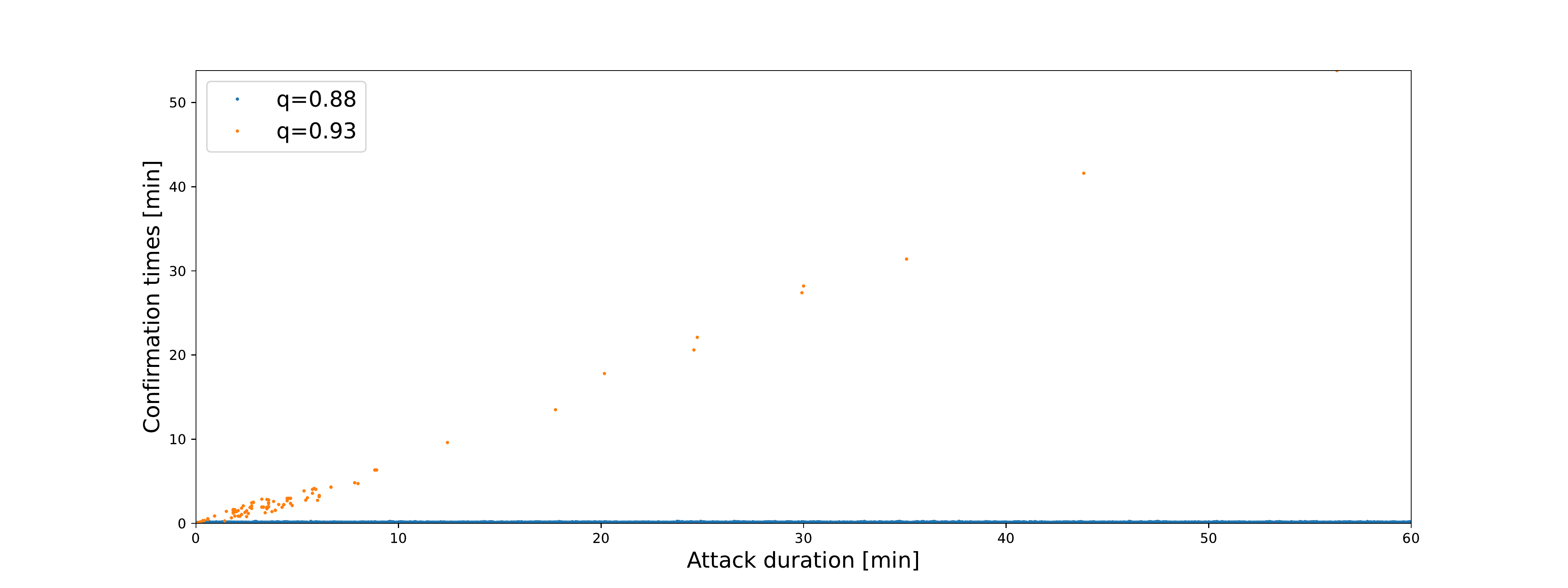}
\newline
d) $k = 16$

\caption{Finalization times for different $k$ and $q$ measured during one hour of orphanage attack with no restriction on parents age. }
\label{fig:apdx-infinite-conf}
\end{figure}

\end{document}